\documentclass{jfm}

\usepackage{graphicx}
\usepackage{newtxtext}
\usepackage{newtxmath}
\usepackage{natbib}
\usepackage{hyperref}
\usepackage{stmaryrd}
\usepackage{cancel}
\usepackage{amsmath}
\usepackage{comment}
\usepackage{url}

\DeclareMathOperator*{\argmin}{argmin} 

\hypersetup{
    colorlinks = true,
    urlcolor   = blue,
    citecolor  = black,
}

\newcommand{\RomanNumeralCaps}[1]

\title{Mean resolvent operator of a statistically steady flow}

\author{Colin Leclercq\aff{1}\corresp{\email{colin.leclercq@onera.fr}}, Denis Sipp\aff{1}
  }

\affiliation{\aff{1}ONERA, DAAA, Université Paris Saclay, 8 rue des Vertugadins, 92190 Meudon, France}

\begin{document}
\maketitle

\begin{abstract}
{This paper introduces a new operator relevant to input-output analysis of flows in a statistically steady regime far from the steady base flow: the \textit{mean resolvent} $\mathsfbi{R}_0$.} It is defined as the operator predicting, in the frequency domain, the mean linear response to forcing of the time-varying base flow. As such, it provides the statistically optimal linear time-invariant approximation of the input-output dynamics, {which may be useful, for instance, in flow control applications}. Theory is developed for the periodic case. The poles of the operator are shown to correspond to the Floquet exponents of the system, including purely imaginary poles at multiples of the fundamental frequency. In general, evaluating mean transfer functions from data requires averaging the response to many realizations of the same input. However, in the specific case of harmonic forcings, we show that the mean transfer functions may be identified without averaging: an observation referred to as `dynamic linearity' in the literature (Dahan et al., 2012). For incompressible flows in the \textit{weakly unsteady limit}, i.e. when amplification of perturbations by the unsteady part of the periodic Jacobian is small compared to amplification by the mean Jacobian, the mean resolvent $\mathsfbi{R}_0$ is well-approximated by the well-known resolvent operator about the mean-flow. Although the theory presented in this paper only extends to quasiperiodic flows, the definition of $\mathsfbi{R}_0$ remains meaningful for flows with continuous or mixed spectra, including turbulent flows. Numerical evidence supports the close connection between the two resolvent operators in quasiperiodic, chaotic and stochastic two-dimensional incompressible flows.

\end{abstract}

\begin{keywords}
Authors should not enter keywords on the manuscript, as these must be chosen by the author during the online submission process and will then be added during the typesetting process (see \href{https://www.cambridge.org/core/journals/journal-of-fluid-mechanics/information/list-of-keywords}{Keyword PDF} for the full list).  Other classifications will be added at the same time.
\end{keywords}

{\bf MSC Codes }  {\it(Optional)} Please enter your MSC Codes here

\section{Introduction}
\label{sec:intro}
\subsection{Existence of transfer functions for time-varying base flows ?}
In the early stages of `natural' (i.e. caused by small-amplitude two-dimensional Tollmien--Schlichting waves) laminar-turbulent transition of an incompressible flow on a smooth flat plate, the flow may be safely assumed to respond to perturbations in a linear time-invariant (LTI) fashion. Otherwise, the response of open shear flows to external excitations, even infinitesimal, is generally more complex. Turbulent flows certainly are not time-invariant indeed, but neither are idealistic two-dimensional flows like the ones behind a backward-facing step or in the wake of one or a few cylinders at moderate Reynolds numbers. Even at very low Reynolds number, unsteadiness may set in as a result of noise amplification or intrinsic instability. Hence, infinitesimal extrinsic perturbations do \textit{not} interact with a steady base flow but a time-evolving one. 

However, for feedback control using modern robust control techniques, an LTI model of the input-output dynamics is generally required. In the frequency domain, LTI dynamics indicates the existence of transfer functions, hence the question: how can we define transfer functions for time-varying, statistically steady base flows ?

\subsection{Option A: `dynamic linearity', i.e. using small-amplitude harmonic forcings}
Some authors have used harmonic forcings to identify transfer functions on simulations of statistically steady flows \citep{DA12,DA17,EV17} which are either laminar or turbulent, and possess either peaked or broadband spectra. For sufficiently low forcing amplitude, the authors evaluate a frequency response by taking the ratio of the complex amplitudes of the output and input at the forcing frequency (the concept of nonlinear transfer function \citep{NO08} is not relevant here due to the choice of low amplitude of the forcing). This observation hints at the existence of a transfer function in situations where it should not theoretically exist, given the time-varying nature of the base flow. The authors use the term `dynamic linearity' to characterize this interesting behaviour but no theoretical arguments seem currently available to understand it.

\subsection{Option B: Linearizing about the mean flow}
A second option consists in assuming that the linear response $\mathbf{u}$ to infinitesimal forcing $\mathbf{f}$ is characterized by the Jacobian matrix\footnote{{With the exception of appendix \ref{sec:Neumann}, we adopt in this paper a spatially-discrete framework where the incompressible velocity field is discretized over $N$ degrees of freedom.}} $\mathsfbi{J}_{\overline{\mathbf{U}}}$ evaluated about the mean flow $\overline{\mathbf{U}}$,  i.e. 
\begin{equation}
    \mathrm{d}_t\mathbf{u}=\mathsfbi{J}_{\overline{\mathbf{U}}}\mathbf{u}+\mathbf{f}.\label{eq:LTI}
\end{equation}
The resolvent operator about the mean flow 
\begin{equation}
    \mathsfbi{R}_{\overline{\mathbf{U}}}(s)=(s\mathsfbi{I}-\mathsfbi{J}_{\overline{\mathbf{U}}})^{-1}
\end{equation}
then describes an LTI input-output dynamics in the frequency domain ($s$ being the Laplace variable), and has been successfully used {for open-loop \citep{MO12,LU14,TO19,YE19,LI21} and} closed-loop \citep{LE19} control of statistically steady flows. However, such an input-output model remains only empirically validated so far, and there is no clear justification for its use, since the mean flow is not {an invariant solution} of the nonlinear unforced system. The use of such model is however motivated by a large body of literature devoted to linear analysis about a time-averaged mean flow, with clear predictive power. 

Indeed, modal analysis of $\mathsfbi{J}_{\overline{\mathbf{U}}}$ yields accurate predictions of both the dominant oscillation frequency in the wake of a cylinder \citep{HA97,PI02,BA06,SI07,MI08} and the associated coherent vortical structures. The growth rate of the leading eigenvalue, whose real part predicts the frequency, is nearly neutral, a property often called RZIF \citep{TU15,BE19,BE19b,BE21} for Real Zero Imaginary Frequency, in a context that is now broader than just cylinder wake flow. This powerful property has been used to build self-consistent models \citep{MA14,MA15,ME17,BE21} successfully mimicking the nonlinear flow with the simplest ingredients. {The idea of marginal stability of the mean flow is not recent \citep{MA56} but may only be justified under specific conditions, i.e. weak nonlinearity \citep{NO03,SI07}, monochromatic oscillations (\cite{ME13}, criterion (b) of \S 3.5; \cite{TU15}) or weak unsteadiness relative to the mean flow (\cite{ME13}; criterion (a) of \S 3.5).}

{In the case of linearly stable flows, or noise-amplifiers, the resolvent operator about the turbulent mean flow has been extensively used to predict second-order statistics \citep{JO01,JO10,MO14,BE16,ZA17,TO20} and in particular coherent flow structures characterized by spectral POD modes \citep{SE16a,SE16b,BE16,SC18,TO18}. The resolvent operator has also been used to obtain deterministic reduced-order models of unsteady flows from measurements of the mean flow and a few local probes \citep{GO16,BE17,SY19}. The input-output framework is also particularly fruitful for elucidating receptivity mechanisms through the analysis of the optimal forcing modes associated with the resolvent modes \citep{HW10,GA13b,SA15b,JE16,SE16b}.} 

{The intuition that coherent structures must extract their energy from the mean flow dates back to the contributions of \cite{LE90,BU93,DE06,CO09,PU09,MC10} among others. However, despite the unchallenged efficacy of resolvent and modal analyses about the mean flow for modelling purposes, a clear justification for these procedures is still lacking \citep{BE16,JO21}. As already said, the first difficulty in terms of interpretation arises from linearizing about a quantity, the mean flow, which is not an invariant solution of the governing equations. In the input-output framework of \cite{MC10}, the resolvent operator $\mathsfbi{R}_{\overline{\mathbf{U}}}$ is forced by a finite-amplitude internal forcing arising from nonlinear interactions of the perturbations. Although this framework is an exact reformulation of the Navier--Stokes equations, with no approximation involved, the procedure remains \textit{ad hoc} in the sense that alternative formulations may be obtained by decomposing the flow about any alternative reference state, for instance the steady base flow. One practical reason for the popularity of this particular formulation is the recurrent observation that for energetic frequencies of the flow, the matrix $\mathsfbi{R}_{\overline{\mathbf{U}}}(\mathrm{i}\omega)$ is often nearly rank 1 and the leading spectral POD mode is often well-approximated by the leading resolvent mode \citep{BE16,TO18}. The clear advantage in this situation is that coherent flow structures may be predicted without having to consider the unknown nonlinear forcing term.} 

{However, it is also increasingly clear that incorporating information about the nonlinear forcing into the linear operator, either in the form of an eddy viscosity \citep{IL18,MO19,SY19,MA19,PI21}, or through a low-rank state-feeback operator \citep{ZA17}, may significantly enhance its predictive power. For instance, the alignment between the leading resolvent mode and leading spectral POD mode may increase from 0.4 to nearly 0.9 by tuning the effective viscosity above the molecular viscosity \citep{PI21}. All these recent contributions confirm that $\mathsfbi{R}_{\overline{\mathbf{U}}}$ is a useful, yet suboptimal linear input-output representation. Worse still, \citep{KA20} recently showed that resolvent analysis about a mean flow is in general ill-posed as the poles of the operator depend on an arbitrary choice of formulation for the governing equations. In the case of a supersonic heated jet, discrepancies of up to $40\%$ were noted in the optimal gain at energetic frequencies, depending on the choice between the use of conservative versus primitive variables.} 

{All these observations motivate the search for a new input-output operator i) incorporating information about the nonlinear forcings, ii) which should be optimal in some sense and iii) which definition should be intrinsic. As we shall see, the new operator will help understand the `dynamic linearity' phenomenon and provide a physical interpretation to $\mathsfbi{R}_{\overline{\mathbf{U}}}$. Its poles will also exactly verify the RZIF property in the periodic case, whereas $\mathsfbi{R}_{\overline{\mathbf{U}}}$ only approximately does so.}

\subsection{Option C: Averaging the time-varying linear response}
\begin{figure}
\centering
    \includegraphics[width=0.7\textwidth]{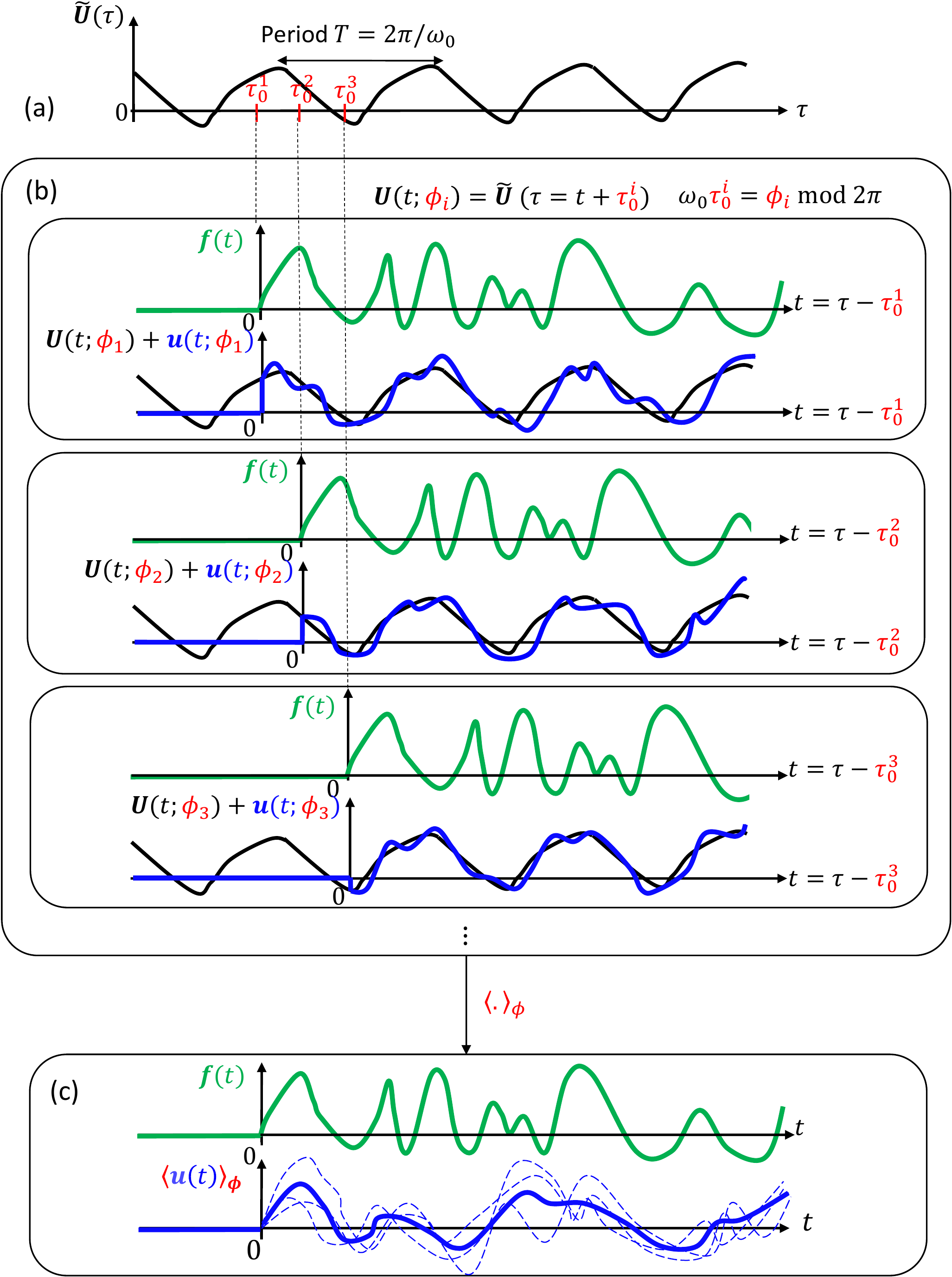}
    \caption{(a) Periodic base flow $\widetilde{\mathbf{U}}(\tau)$ for an arbitrarily defined origin of time $\tau=0$. (b) Linear responses $\mathbf{u}(t;\phi_i)$ to the same infinitesimal forcing $\mathbf{f}(t)$ started at different instants $\tau=\tau^i_0$. Like the underlying base flow $\mathbf{U}(t;\phi_i)={\widetilde{\mathbf{U}}(\tau=t+\tau^i_0)}$, the responses are parametrized by the phase $\phi=\phi_i$ of the periodic base flow at $t=0$, such that $\omega_0\tau_0=\phi\text{ mod }2\pi$. (c) The response to linear forcing being a distribution, the statistically optimal LTI model is obtained by computing the mean response $\langle \mathbf{u}(t)\rangle_\phi$.}
    \label{fig:schema_principe}
\end{figure}

In this paper, we will not make the \textit{ad hoc} hypothesis of linearity about the mean flow. We will only keep the assumption of linearity, which remains valid as long as the forcing amplitude is low enough, but now the base flow is {an exact time-varying solution of the Navier--Stokes equations}. Until \S \ref{sec:extension}, the base flow will be assumed to be periodic with period $T$ corresponding to a fundamental angular frequency $\omega_0:=2\pi/T$. We represent in figure \ref{fig:schema_principe}, the response $\mathbf{u}$ to an infinitesimal forcing signal $\mathbf{f}$ started at time $\tau_0$ relative to an arbitrary origin of time $\tau=0$ on the base flow trajectory {$\widetilde{\mathbf{U}}(\tau)$}. 
Because the base flow is not steady but periodic, the linear response to forcing is not unique but parametrized by the phase $0\leq \phi<2\pi$ of the limit cycle when the forcing starts. This key parameter for our analysis is defined such that 
\begin{equation}
    \omega_0\tau_0=\phi\text{ mod }2\pi.
\end{equation}
Indeed, the input-output dynamics is now characterized by a linear time-periodic (LTP) system
\begin{equation}
    \mathrm{d}_t\mathbf{u}=\mathsfbi{J}(t;\phi)\mathbf{u}+\mathbf{f},\quad \mathsfbi{J}(t+T;\phi)=\mathsfbi{J}(t;\phi),\label{eq:simple2}
\end{equation}
with a periodic Jacobian itself parametrized by the phase $\phi$ through the base flow $\mathbf{U}(t;\phi)={\widetilde{\mathbf{U}}(\tau=t+\tau_0)}$. For $\phi$ uniformly distributed in $[0,2\pi)$, there is in fact a distribution of timeseries $\mathbf{u}(t;\phi)$ corresponding to a given input $\mathbf{f}(t)$. The statistically optimal LTI approximation of this LTP system should predict a single output $\mathbf{v}(t)$ minimizing the mean error with respect to $\mathbf{u}(t;\phi)$, i.e. 
\begin{equation}
    \mathbf{v}(t)=\argmin_{\tilde{\mathbf{v}}(t)}\lim_{T'\to\infty}\dfrac{1}{T'}\int_0^{T'} \langle\|\mathbf{u}(t;\phi)-\tilde{\mathbf{v}}(t)\|^2\rangle_\phi\,\mathrm{d}t,
\end{equation}
where $\langle.\rangle_\phi$ denotes an ensemble average with respect to $\phi$. This output is obviously the mean response $\mathbf{v}(t)=\langle\mathbf{u}(t)\rangle_\phi$. We define the mean resolvent operator $\mathsfbi{R}_0$ (the notation will become clear in \S \ref{subsec:meanres}) as the operator predicting the mean response, in the frequency domain, for any given input
\begin{equation}
    \boxed{\text{mean resolvent }\mathsfbi{R}_0: \mathbf{f}(s)\mapsto\langle \mathbf{u}(s)\rangle_\phi}.\label{eq:def}
\end{equation}
This transfer function is well-defined because the relationship between the input $\mathbf{f}$ and the output $\langle \mathbf{u}\rangle_\phi$ is by construction causal, linear and time-invariant (i.e. independent of $\phi$).

\begin{figure}
    \centering
    \hspace{10mm}\includegraphics[width=0.75\textwidth]{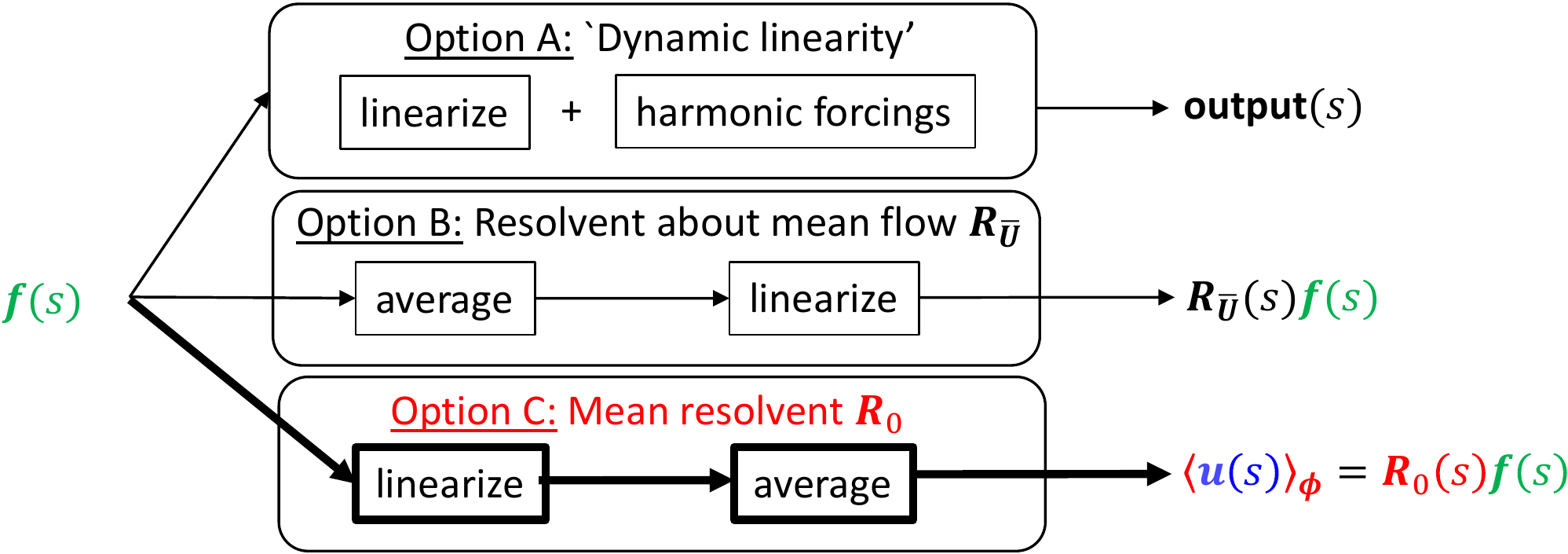}
    \caption{Three ways of defining a LTI operator for a periodic base flow.}
    \label{fig:3waysLTI}
\end{figure}

We see that in option C the order of operations is reversed compared to option B: we linearize then average instead of averaging then linearizing. In option A, linearization is involved as well, but there is no averaging at all. Instead, a very specific form of input is chosen, which is harmonic in time. The three options are summarized in figure \ref{fig:3waysLTI}.

\subsection{Goal and outline of the paper}
{The aim of this paper is to investigate option C and establish connections with options A and B. In particular, we will show that option A is a specific way to evaluate $\mathsfbi{R}_0$, while $\mathsfbi{R}_{\overline{\mathbf{U}}}$ is a (reduced-order) approximation of $\mathsfbi{R}_0$, under appropriate conditions.}

In \S \ref{sec:pinball100} and \S \ref{sec:theory}, we will continue to focus on incompressible periodic base flows. In \S \ref{sec:pinball100} we will perform a numerical experiment on the fluidic pinball in a periodic regime of oscillations at $Re=100$, in order to compare options B and C, namely mean resolvent versus resolvent about the mean flow. In \S \ref{sec:theory}, we will study the properties of the mean resolvent operator, with the goal to elucidate the observations made in \S \ref{sec:pinball100} and draw connections with the resolvent operator about the mean flow and `dynamic linearity'. Connections with the Koopman operator (\S \ref{subsec:Koopman}), the RZIF property (\S \ref{subsec:RZIF}) and model reduction (\S \ref{subsec:approxmeanres}) will also be made. In \S \ref{sec:extension} we perform the same numerical experiment as in \S \ref{sec:pinball100} but for more complex base flows, i.e. quasiperiodic, chaotic and stochastic. We will consider the fluidic pinball at $Re=110$ and $Re=120$ as well as the backward-facing step flow (with exogenous stochastic forcing) at $Re=500$. The goal is to stress the strong connection between mean resolvent and resolvent about the mean flow in all these dynamically distinct cases. No computations are done in a compressible case but in \S \ref{subsec:compress} we indicate the implications of our theoretical analysis in \S \ref{sec:theory} to such situations. A theoretical extension of \S \ref{sec:theory} is proposed for the quasiperiodic case in appendix \ref{subsec:QP}. We summarize our findings in \S \ref{sec:concl}.


\section{Resolvent about the mean flow versus mean resolvent: fluidic pinball at $ Re=100$}\label{sec:pinball100}

\begin{figure}
    \centering
    \begin{tabular}{cc}
     (a) \includegraphics[height=0.24\textwidth]{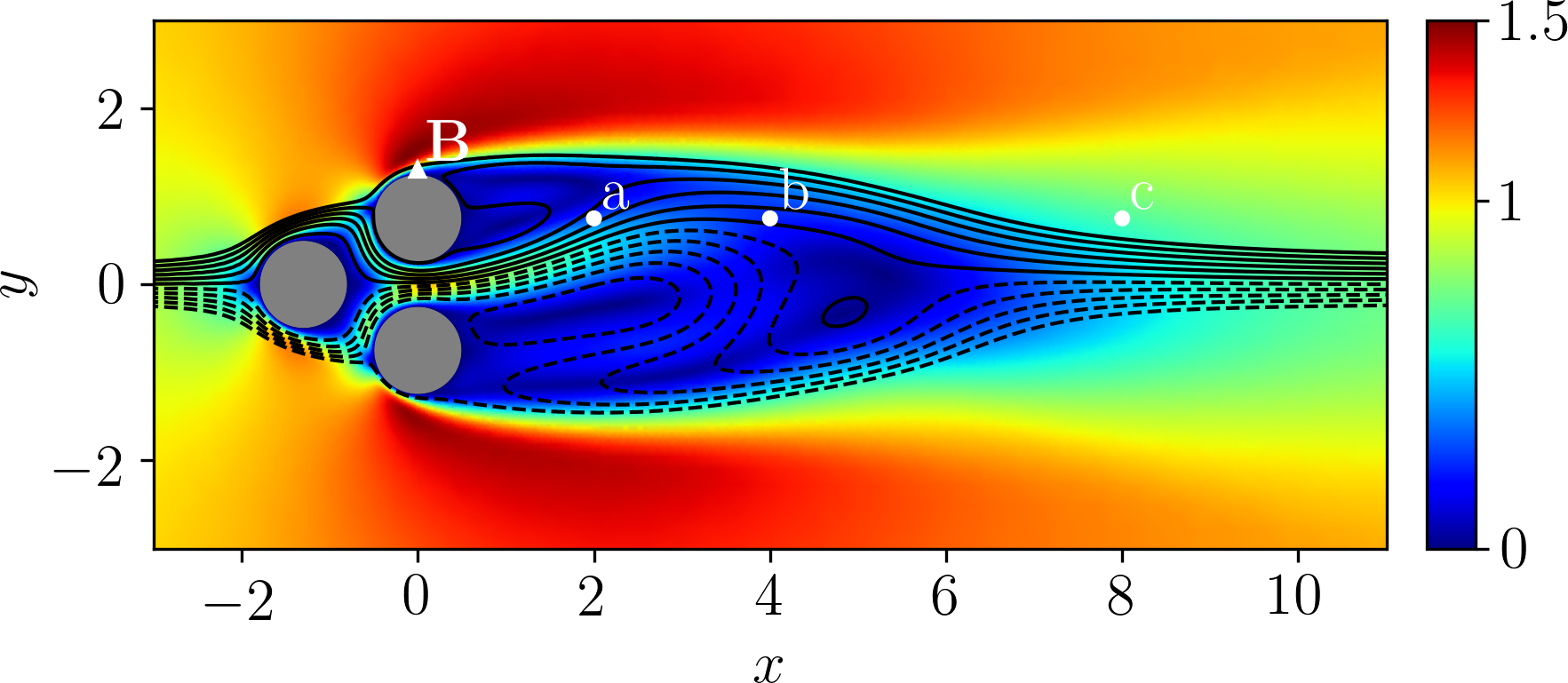}    &  (b) \includegraphics[height=0.24\textwidth]{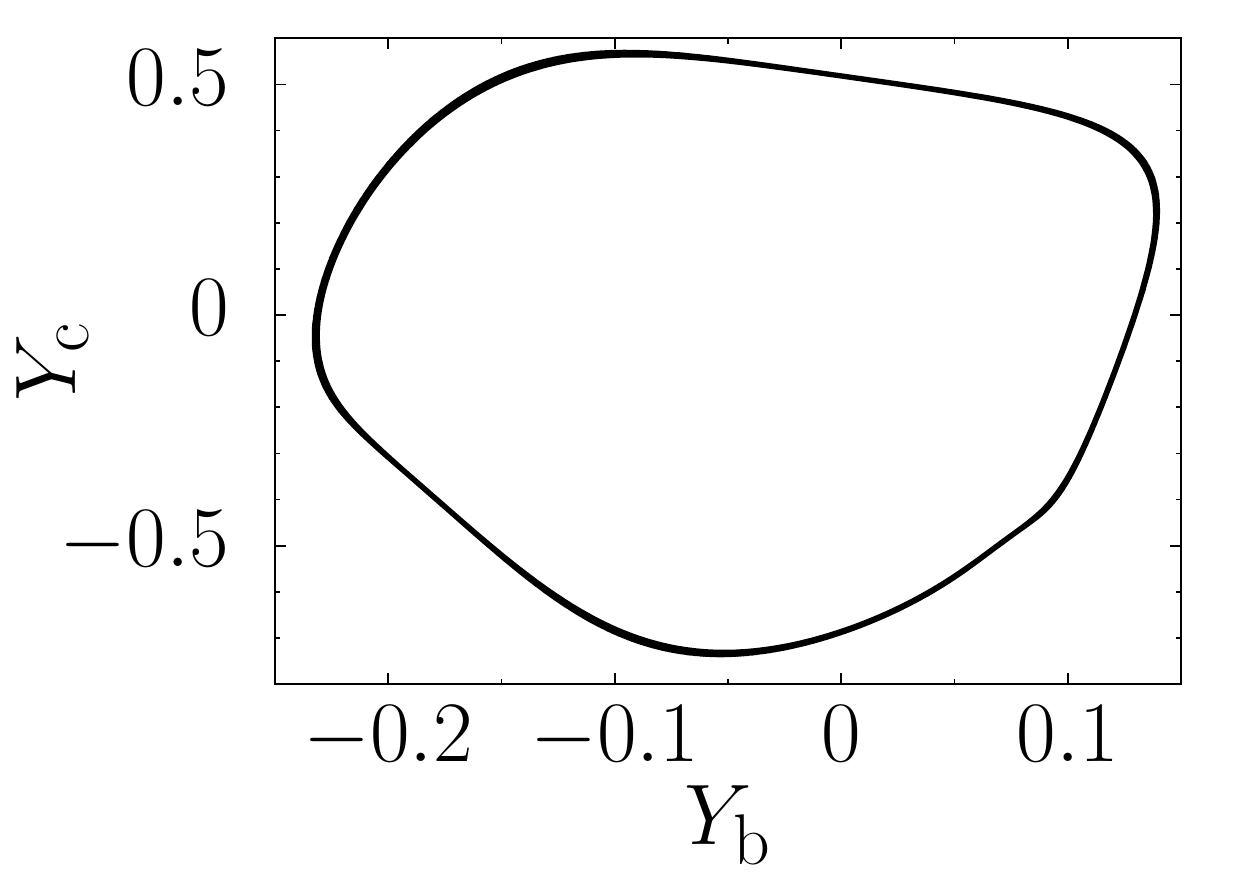}\\
    \end{tabular}
    \caption{Fluidic pinball at $Re=100$: periodic behaviour, (a) Mean velocity norm and streamlines and (b) phase portraits using $y$-velocity sensors $Y_\mathrm{b}$ and $Y_\mathrm{c}$.}
    \label{fig:fluidicpinball}
\end{figure}

In this section, we perform a numerical experiment on the case of the incompressible two-dimensional fluidic pinball in a periodic regime at $Re=100$. The goal is to compare transfer functions based on options B and C, i.e. average then linearize versus linearize then average. The precise configuration consists in the the numerical setting of \cite{DE20}. We define the unit length to be the diameter $D$ of each cylinder and the (convective) time unit to be $D/U_\infty$, where $U_\infty$ is the upstream velocity norm. The Reynolds number is defined as $Re:=U_\infty D/\nu$ where $\nu$ is the kinematic viscosity. The centers of the `front', `top' and `bottom' triangles are respectively located at $(x_f,y_f)=(-1.5\cos(\pi/6),0)$, $(x_t,y_t)=(0,0.75)$ and $(x_b,y_b)=(0,-0.75)$, hence forming an equilateral triangle of side length $1.5$. Instead of measuring the full input-output relation between $N$-dimensional inputs and outputs, we will focus in section \S \ref{sec:pinball100} on the single-input single-output (SISO) transfer between an actuator signal $u$ and a measurement $y$ such that
\begin{equation}
    \mathbf{f}(t)=\mathbf{B}u(t),\quad y(t;\phi)=\mathbf{C}^T\mathbf{u}(t;\phi).\label{eq:inout}
\end{equation}
The SISO viewpoint is more practical for the numerical experiment we perform, and it is also relevant in the context of flow control. More specifically, we will consider three probes $\mathbf{C}$ extracting the $y$-velocity component in the wake of the `top' cylinder at $y=y_t$ and respectively $x=2,4,8$ (white filled circles in figure \ref{fig:fluidicpinball}). Measurements of the periodic base flow will be denoted $Y=\mathbf{C}^T\mathbf{U}$. When necessary, the index `a',`b' or `c' will be used to specify which sensor we are referring to. The quantity $\mathbf{B}$ is a discretized Gaussian volume force field $\mathcal{B}(x,y;x_0,y_0,\sigma_x,\sigma_y)$ located at the `top' of the `top' cylinder
\begin{equation}
    \mathcal{B}(x,y;x_0,y_0,\sigma_x,\sigma_y)=\left[0,\exp\left({-\dfrac{(x-x_0)^2}{2\sigma_x^2}-\dfrac{(y-y_0)^2}{2\sigma_y^2}}\right)\right]^T,\label{eq:defB}
\end{equation}
with $\sigma_x=0.01$, $\sigma_y=0.1$ and the center $(x_0,y_0)$ of the Gaussian (white triangle in figure \ref{fig:fluidicpinball}) relative to the center $(x_t,y_t)$ of the cylinder being given by $(x_0-x_t,y_0-y_t)=(0,0.55)$.  

Numerical discretization details (computational domain, mesh, spatial and temporal schemes) are provided in \S \ref{sec:numdetails}.

\subsection{Unsteady base flow}
In figure \ref{fig:fluidicpinball}(a), we plot the streamlines and isocontours of velocity norm for the mean flow. We notice it is {asymmetric}, as noted by \cite{DE20}. In figure (b), we plot the phase portrait of the flow, using the vertical velocity probes `b' and `c': it is a closed curve, which confirms the periodic nature of the dynamics at $Re=100$. Fourier spectra are shown in figure \ref{fig:spectra_BF}(i) for the 3 sensors {(a,b,c)} located in the wake of the cylinders. These spectra confirm the periodic nature of the flow since they are discrete with a single fundamental frequency at $\omega_0=0.73$ and harmonics. Higher frequencies are more energetic in the far wake than in the near wake. Indeed, perturbations amplify while being convected, causing stronger nonlinear energy transfers downstream.

\subsection{Estimating mean transfer functions from input-output data}\label{subsec:paramphi}

The resolvent operator about the mean flow $\mathsfbi{R}_{\overline{\mathbf{U}}}$ and the mean resolvent operator $\mathsfbi{R}_0$ are both multiple-input-multiple-output (MIMO) transfer functions with $N$-dimensional inputs and outputs. SISO transfer functions between any time-invariant actuator-sensor pair $(\mathbf{C},\mathbf{B})$ may be derived from these operators using
\begin{equation}
    G_{\overline{\mathbf{U}}}(s):=\mathbf{C}^T\mathsfbi{R}_{\overline{\mathbf{U}}}(s)\mathbf{B},\quad\text{and}\quad\langle G(s)\rangle_\phi:=\mathbf{C}^T\mathsfbi{R}_0(s)\mathbf{B}.
\end{equation}
In order to estimate the mean transfer function $\langle G(s)\rangle_\phi$ from input-output data, we introduce `frequency response realizations'
\begin{equation}
    G(s;\phi,u):=\dfrac{y(s;\phi)}{u(s)},
\end{equation}
which are obtained by taking the ratio between the Laplace transforms of the input $u$ and the output $y$. The quantity $G$ is not directly associated with a transfer function since the linear response to forcing about a time-varying base flow is not time-invariant; hence the dependence on the phase $\phi$ and the input signal $u$. However, upon averaging with respect to $\phi$ (we recall that $\phi$ is assumed to be uniformly distributed in $[0,2\pi)$), the linear input-output dynamics becomes time-invariant and the dependence disappears. We then recover the mean transfer function based on the mean resolvent

\begin{equation}
\langle G(s)\rangle_\phi=\langle G(s;\phi,u)\rangle_\phi.
\end{equation}

For any complex frequency $s$ and input signal $u$, the variance $ \mathrm{Var}_\phi\,{G}(s;u)$ of the complex random variable $G(s;\phi,u)$ may be computed as
\begin{align}
    \mathrm{Var}_\phi\,{G}&:=\langle |G -\langle G \rangle_\phi|^2\rangle_\phi,\nonumber\\ 
    &= \langle |G|^2 \rangle_\phi - |\langle G \rangle_\phi|^2, 
\end{align}
and the ratio 
\begin{equation}
    \eta(s;u):=\dfrac{\sqrt{\mathrm{Var}_\phi\,{G}}}{|\langle G\rangle_\phi|}\label{eq:eta}
\end{equation}
may be interpreted as the inverse of a `signal-to-noise' ratio, allowing us to quantify the validity of the time-invariance hypothesis. The system is nearly LTI at a given frequency $s$ if $\eta\ll 1$ for any choice of $u$. {A geometric interpretation of $\eta$ will be proposed in figure \ref{fig:eta_show} of \S \ref{subsec:results}.}

The ratio $\eta$ is also useful to evaluate the convergence of the mean estimate $\langle G\rangle_{\phi,N_s}$ to the true mean $\langle G\rangle_{\phi}$, using $N_s\gg 1$ realizations of $G$. Since all realizations are independent and identically distributed random variables of variance $\mathrm{Var}_\phi\,G$, the mean estimate is, by the central limit theorem, a normal random variable of variance
\begin{equation}
    \mathrm{Var}_\phi\,\langle G\rangle_{\phi,N_s}=\dfrac{\mathrm{Var}_\phi\,G}{N_s}.\label{eq:varmean}
\end{equation} 
The convergence criterion $\sqrt{\mathrm{Var}_\phi\,\langle G\rangle_{\phi,N_s} }\ll |\langle G \rangle_\phi|$ for the mean estimate therefore translates into the condition $\eta \ll \sqrt{N_s}$ for the given choice of $u$ and $s$.

\subsection{Procedure}\label{sec:procedure}
The nonlinear simulation is initialized at $\tau=0$ with the steady base flow (fixed-point) perturbed by $\mathbf{B}$ (to quickly reach the limit-cycle). After an initial transient of more than 1000 convective time-units, the flow converges to a periodic regime of self-sustained oscillations.

This periodic base flow is then perturbed linearly using impulsive forcings $u(t)=\delta(t)$ by simply initializing the perturbation field with the actuator, i.e. $\mathbf{u}(0)=\mathbf{B}$. Other choices of input signals may be made for estimating mean transfer functions, which will be discussed later in \S \ref{subsec:identif} and \S \ref{sec:identifharm}. Following impulsive forcing, both nonlinear (for the periodic base-flow) and linear (for the perturbation) simulations are run for over more than 1000 convective time units. We compute $N_s=140$ impulse responses, by uniformly distributing the relative initial time $\tau_0$ over a period $T=8.64$ of the underlying limit-cycle, i.e. $\Delta \tau_0=T/N_s$. This amounts to uniformly distributing the initial phase $\phi$ over $[0,2\pi)$, as required. 

Note that since impulse responses never decay, we can only compute frequency responses shifted in the right-half plane, i.e. obtained for $s=\mathrm{i}\omega+\sigma$ with $\sigma>0$. This is indeed mandatory for convergence of the Laplace transform. The value of $\sigma=0.01$ is chosen such that the mean impulse response modulated by $\mathrm{e}^{-\sigma t}$ reaches negligible values towards the end of the time window of 1000 convective time units. The Fourier transform of the exponentially modulated signal, estimated via the discrete Fourier transform, then yields the Laplace transform of the impulse over the shifted imaginary axis. The frequency response over $\mathrm{i}\mathbb{R}$ (minus the singularities) may then be retrieved by analytic continuation, but this is not done here.

The present analysis may remind the reader of the work by \cite{YE19}, who performed resolvent analysis about the mean flow on a shifted imaginary axis with $\sigma>0$.

\subsection{Results}\label{subsec:results}

\begin{figure}
    \centering
    \includegraphics[width=1.\textwidth]{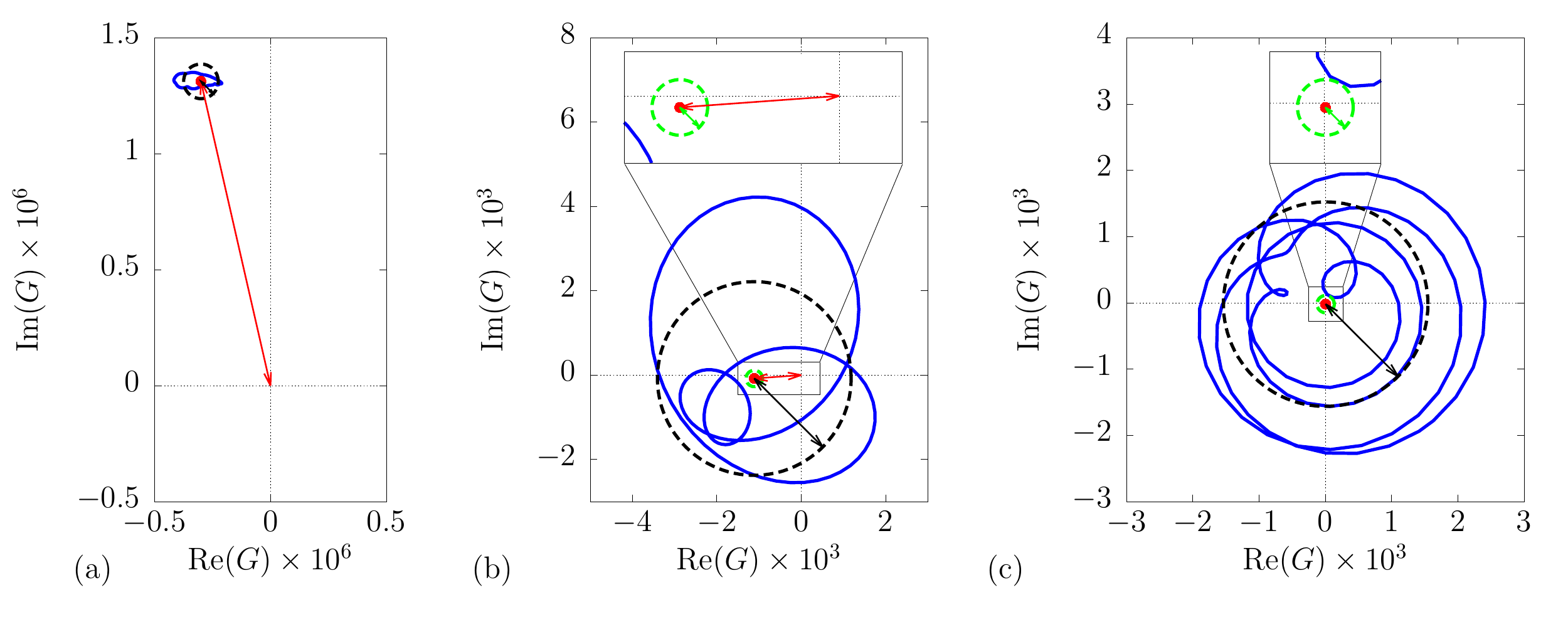}
    \caption{{Geometric interpretation of the ratio $\eta$ for the three cases (a) $\eta\ll 1$, (b) $1<\eta<\sqrt{N_s}$ and (c) $\sqrt{N_s}\ll \eta$ of table \ref{tab:eta}; see main text for description.}}
    \label{fig:eta_show}
\end{figure}

{First of all, a geometric interpretation of the ratio $\eta$ is proposed in figure \ref{fig:eta_show}, where three cases are considered: (a) $\eta\ll 1$, (b) $1\leq \eta<\sqrt{N_s}$ and (c) $\sqrt{N_s} \ll \eta$, corresponding to the $\left\{\text{probe,frequency}\right\}$ pairs of table \ref{tab:eta}. For a given value of $s$, the locus of $G(s;\phi)$ is represented by a closed blue curve in the complex plane as $\phi$ varies between 0 and $2\pi$. The red dot indicates the barycenter of the curve, which is the estimated mean transfer $\langle G\rangle_{\phi,N_s}$. The length of the red arrow approximates $|\langle G\rangle_\phi|$. There are two dashed circles centered on the red dot: a black one with radius approximating $\sqrt{\mathrm{Var}_\phi G}$ (length of black arrow), and a green one with radius equal to $\sqrt{\mathrm{Var}_\phi \langle G\rangle_{\phi,N_s}}$ (length of green arrow). In panel (a), the radius of the black circle is much smaller compared to the red arrow and the entire blue distribution may therefore be approximated by its red barycenter: the dynamics is quasi-time-invariant for that $\left\{\text{probe,frequency}\right\}$ pair. In panel (b), the black arrow is twice the size of the red arrow, hence replacing the entire blue distribution by the single red dot is a very crude approximation: the dynamics is not quasi-time-invariant. However, the green arrow is still much smaller than the red arrow (see inset), which means that the estimation of the mean transfer function is converged. But in panel (c), not only is the black arrow much larger than the red arrow (dynamics not time-invariant), but the green arrow is also much larger than the red arrow (see inset) so there are not enough samples to confidently estimate the mean transfer function. We finally recall that only the exact mean $\langle G\rangle_\phi$ is independent from the forcing signal, while all other quantities (blue curve and its variance, mean transfer estimate and its variance) depend on the specific choice of $u(t)$.}
\begin{table}
    \centering
\begin{tabular}{cc|c}
probe & frequency $s$ & $\eta$\\
  a  &  $0.01+8.00\mathrm{i}$ & $0.0554$\\
   a  & $0.01+0.813\mathrm{i}$ & $2.06$\\
   c & $0.01+4.00\mathrm{i}$ & $73.6$
\end{tabular}
    \caption{Values of the ratio $\eta$ evaluated from $N_s=140$ impulses for three $\left\{\text{probe,frequency}\right\}$ pairs.}
    \label{tab:eta}
\end{table}


\begin{figure}
    \includegraphics[width=1.0\textwidth]{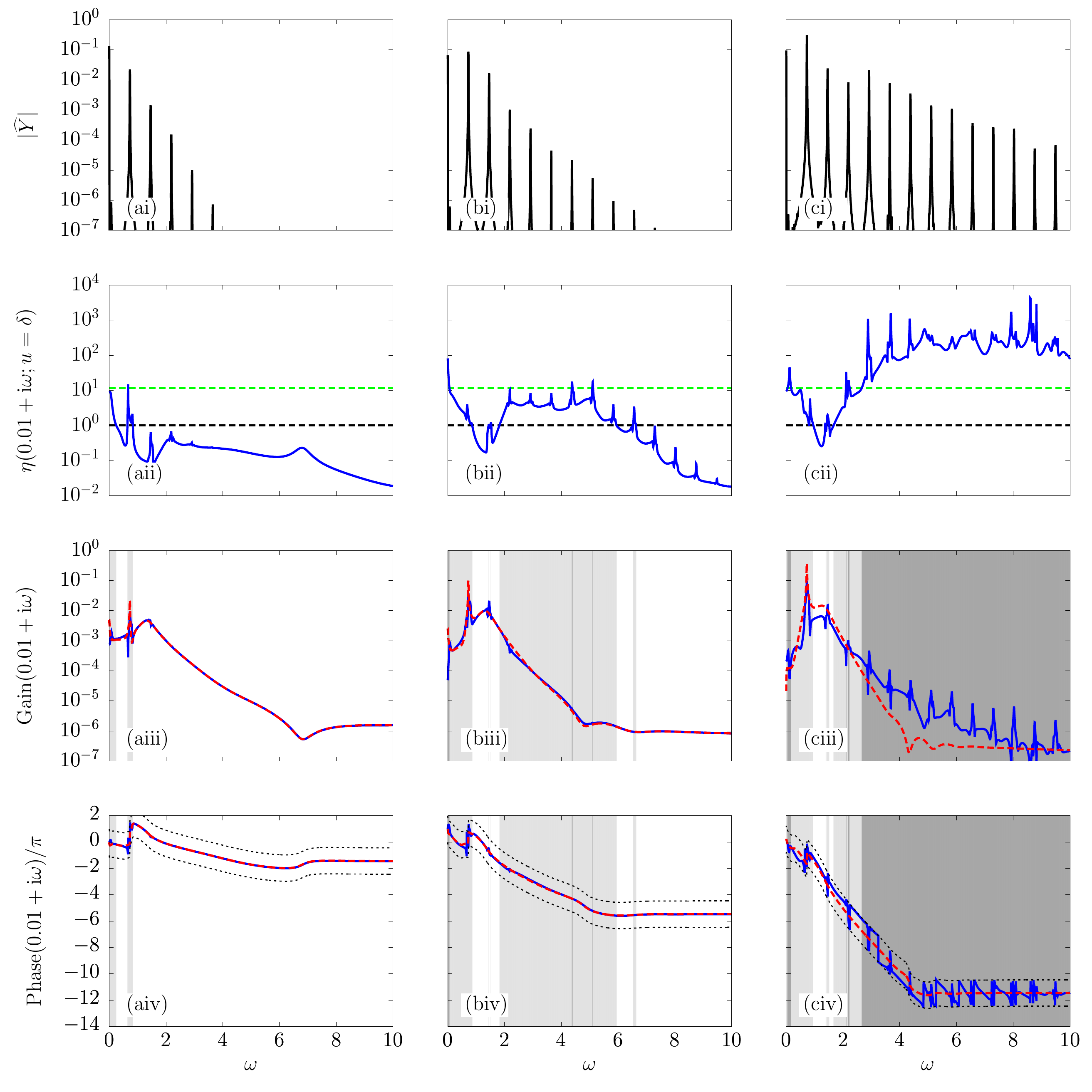}
    \caption{{Fluidic pinball in the periodic regime at $Re=100$. The labels (a,b,c) correspond to the three sensor positions. (i) Fourier spectrum of measurement $Y$ of the periodic base flow. The spectrum is discrete and no averaging is done so a single Hann window is applied over the entire signal of more than 1000 time units. (ii) Ratio $\eta$ (defined in (\ref{eq:eta})) for $N_s=140$ impulsive forcings $u(t)=\delta(t)$. The black dashed line indicates the threshold $\eta=1$ far below which the system is nearly LTI (with respect to $u=\delta$). The green dashed line indicates the threshold $\eta=\sqrt{N_s}$ far below which the estimate of the mean transfer function is converged. Gain (iii) and phase (iv) of mean frequency response $\langle G\rangle_{\phi,N_s}$ (solid blue) and frequency response about the mean flow $G_{\overline{\mathbf{U}}}$ (dashed red). Light-grey shading indicated frequency ranges where $1< \eta <\sqrt{Ns}$ and dark-grey shading corresponds to $\eta\geq \sqrt{N_s}$. In the phase plots, the two black-dotted curves indicate a shift of $+/-\pi$ with respect to the phase of $G_{\overline{\mathbf{U}}}$. For rows (ii)-(iii)-(iv), all quantities are evaluated on the shifted imaginary axis $\sigma+\mathrm{i}\omega$, with $\sigma=0.01$.}}
    \label{fig:spectra_BF}
\end{figure}

{We then proceed to a more systematic examination of $\eta$ for $u(t)=\delta(t)$ in row (ii) of figure \ref{fig:spectra_BF}, for the three probes {(a,b,c)} over the frequency range $0\leq \omega\leq 10$. The threshold $\eta=1$ is indicated with a dashed black line, while the ratio $\eta=\sqrt{N_s}$ is indicated with a green dashed line. The value of $\eta$ with respect to these two critical values is also indicated in the Bode diagrams of the mean transfer function $\langle G\rangle_\phi$ represented in rows (iii) and (iv), for the gain and phase respectively. No shading corresponds to a frequency range where $\eta< 1$, i.e. the dynamics is quasi-LTI, light-grey shading corresponds to $1\leq\eta<\sqrt{Ns}$, i.e. dynamics not LTI but mean estimate converged, and dark-grey shading corresponds to $\sqrt{N_s}\leq \eta$, i.e. mean estimate not even converged.}

{The first remark is that the ratio $\eta$ becomes larger as the probe moves downstream. As a result, converging the mean transfer estimate requires more samples downstream than upstream (see greater extent of dark grey regions). Correlatively, the frequency range of validity of the quasi-time-invariant region shrinks as the probe moves downstream. Both observations seem correlated with the fact that the unsteady fluctuations are more energetic downstream than upstream (see power spectra in row (i)). The ratio $\eta$ is also greater near resonance frequencies for the same reason: the base flow fluctuations are greater near these frequencies. It is interesting to note however that for all probe positions, the LTI approximation is valid in some frequency range near the maximum gain (but excluding resonances).}

{Whether $\eta$ is small compared to 1 or not, we always observe a qualitative agreement between the Bode diagrams of $G_{\overline{\mathbf{U}}}$ and $\langle G\rangle_{\phi}$, as long as the mean estimate is converged. For probe (a), the two diagrams are nearly undistinguisable from one another, and the deviation remains very small for probe (b) as well, except near resonant frequencies $k\omega_0$. For probe (c) though, the deviation becomes noticeable, even when $\eta<1$.} 

{Only the fundamental frequency leads to a resonance peak in $G_{\overline{\mathbf{U}}}$, whereas higher-order harmonics are also visible in $\langle G\rangle_\phi$. These extra peaks are indicative of the presence of poles at $s=\mathrm{i}k\omega_0$ in the mean transfer function, but the amplitude of the peaks rapidly decreases with $k$ and as the probe moves upstream.}



{In the theoretical section, we will attempt to elucidate the aforementioned observations: i) why are $\langle G\rangle_\phi$ and $G_{\overline{\mathbf{U}}}$ generally so similar, ii) why does the quality of the LTI approximation deteriorate downstream, iii) why are there poles at $s=\mathrm{i}k\omega_0$ in the mean transfer function and iv) why are higher-order resonances weaker and only noticeable downstream ?}

\section{Theory for incompressible periodic base flows}\label{sec:theory}
In this theoretical section, we come back to the more general MIMO viewpoint between a $N$-dimensional forcing $\mathbf{f}(t)$ and a $N$-dimensional response $\mathbf{u}(t;\phi)$. 
\subsection{Phase-dependent frequency response of LTP system}\label{subsec:Floquet}

Denote $\mathbf{\Phi}(t,t';\phi)$ the propagator from $t'$ to $t$ of the linear time-periodic system (\ref{eq:simple2}), such that
\begin{equation}
    \mathbf{u}(t;\phi)=\mathbf{\Phi}(t,0;\phi)\mathbf{u}_0+\int_{0}^t \mathbf{\Phi}(t,t';\phi)\mathbf{f}(t')\,\mathrm{d}t',\label{eq:response}
\end{equation}
where $\mathbf{u}_0$ is the initial condition on $\mathbf{u}$, regardless of $\phi$. By Floquet's theorem \citep{MA18}, we know that there exists a real $T$-periodic matrix $\mathsfbi{P}(t;\phi)$, invertible at all times $t$, and a constant complex matrix $\mathsfbi{K}$ such that $\mathbf{\Phi}(t,t';\phi)=\mathsfbi{P}(t;\phi)\mathrm{e}^{\mathsfbi{K}(t-t')}\mathsfbi{P}^{-1}(t';\phi)$. There are multiple possible determinations of the matrix $\mathsfbi{K}=(1/T)\mathrm{log}(\mathbf{\Phi}(T,0))$
depending on the definition of the logarithm. Assuming $\mathsfbi{K}$ to be diagonalizable $\mathsfbi{K}=\mathsfbi{S}\mathbf{\Lambda}\mathsfbi{S}^{-1}$
and choosing the principal determination for the logarithm, all the eigenvalues fall into the fundamental strip $-\omega_0/2<\mathrm{Im}(\lambda_i)\leq \omega_0/2, ~i=1,\dots,N$. These are called the principal Floquet exponents and they may be ranked in decreasing order of growth rate, i.e. $\mathrm{Re}(\lambda_1)\geq \mathrm{Re}(\lambda_2) \geq \dots\geq \mathrm{Re}(\lambda_N)$. Other determinations of the logarithm lead to eigenvalues $s_{ij}=\lambda_i+\mathrm{i}j\omega_0$ which are Floquet exponents as well, located in complementary strips $\omega_0(j-1/2)<\mathrm{Im}(s_{ij})\leq \omega_0(j+1/2)$ in the complex plane. The direct Floquet modes $\mathbf{v}^i$ are the columns of the $T$-periodic complex matrix $\mathsfbi{V}(t;\phi):=\mathsfbi{P}(t;\phi)\mathsfbi{S}$. The adjoint Floquet modes $\mathbf{w}^i$, such that $(\mathbf{w}^i,\mathbf{v}^j)=\delta_{ij}$ for the canonical inner product $(\mathbf{a},\mathbf{b})=1/T\int_0^T\mathbf{a}^H\mathbf{b}\,\mathrm{d}t$ on the space of periodic functions in $\mathbb{C}^N$ (where $^H$ denotes the complex transpose), are the columns of the matrix $\mathsfbi{W}(t;\phi)$ such that $\mathsfbi{W}^H\mathsfbi{V}=\mathsfbi{I}$ at all times (and phases $\phi$). Using this eigendecomposition, the propagator reads
\begin{equation}
    \mathbf{\Phi}(t,t';\phi)=\mathsfbi{V}(t;\phi)\mathrm{e}^{\mathbf{\Lambda} (t-t')} \mathsfbi{W}^H(t';\phi).\label{eq:propag}
\end{equation}
We may now expand the direct and adjoint Floquet modes as Fourier series. If the $j\textsuperscript{th}$ Fourier coefficient of $\mathsfbi{V}$ is proportional to $\mathrm{e}^{\mathrm{i}j\phi}$, then so is the $j\textsuperscript{th}$ Fourier coefficient of $\mathsfbi{W}$ since $\mathsfbi{W}^H\mathsfbi{V}=\mathsfbi{I}$ at all times. Therefore, we can write the following Fourier series
\begin{equation}
    \mathsfbi{V}(t;\phi)=\sum_j \widehat{\mathsfbi{V}}_j \mathrm{e}^{\mathrm{i}j(\omega_0 t+\phi)},\quad \mathsfbi{W}(t';\phi)=\sum_j \widehat{\mathsfbi{W}}_j \mathrm{e}^{\mathrm{i}j(\omega_0 t'+\phi)}. \label{eq:Four}
\end{equation}
Injecting (\ref{eq:propag}) and (\ref{eq:Four}) in (\ref{eq:response}) and assuming $\mathbf{u}_0=\boldsymbol{0}$, we have
\begin{align}
    \mathbf{u}(t;\phi)=&\int_{0}^t \mathsfbi{V}(t;\phi)\mathrm{e}^{\mathbf{\Lambda} (t-t')} \mathsfbi{W}^H(t';\phi)\mathbf{f}(t')\,\mathrm{d}t',\nonumber\\
    =&\int_{0}^t \left[\sum_j \widehat{\mathsfbi{V}}_j\mathrm{e}^{\mathrm{i}j(\omega_0t+\phi)}\right]\mathrm{e}^{\boldsymbol{\Lambda}(t-t')}\left[\sum_l \widehat{\mathsfbi{W}}^H_l\mathrm{e}^{-\mathrm{i}l(\omega_0t'+\phi)}\right] \mathbf{f}(t')\,\mathrm{d}t',\nonumber\\
    =&\sum_{j,l}\mathrm{e}^{\mathrm{i}(j-l)\phi}\int_{0}^t\left[ \widehat{\mathsfbi{V}}_j \mathrm{e}^{(\boldsymbol{\Lambda}+\mathrm{i}j\omega_0\mathsfbi{I})(t-t')}\widehat{\mathsfbi{W}}^H_l\right]\left[\mathrm{e}^{\mathrm{i}(j-l)\omega_0 t'}\mathbf{f}(t')\right]\,\mathrm{d}t',\nonumber\\
    =&\sum_{n}\mathrm{e}^{\mathrm{i}n\phi}\int_{0}^t\left[\sum_{j}\widehat{\mathsfbi{V}}_j \mathrm{e}^{(\boldsymbol{\Lambda}+\mathrm{i}j\omega_0\mathsfbi{I})(t-t')}\widehat{\mathsfbi{W}}^H_{j-n}\right]\left[\mathrm{e}^{\mathrm{i}n\omega_0 t'}\mathbf{f}(t')\right]\,\mathrm{d}t'.
\end{align}
We recognize convolution products of causal functions on the right-hand-side hence it is easy to take the Laplace transform, yielding the linear time-periodic input-output relation
\begin{equation}
\boxed{
    \mathbf{u}(s;\phi)=\sum_n \mathrm{e}^{\mathrm{i}n\phi}\mathsfbi{R}_n(s)\mathbf{f}(s-\mathrm{i}n\omega_0)},\label{eq:LTPIO}
\end{equation}
with the operators
\begin{equation}
\boxed{
    \mathsfbi{R}_n(s)=\sum_j \sum_{k=1}^N\dfrac{ \widehat{\mathbf{v}} ^k_j{(\widehat{\mathbf{w}}^k_{j-n})^H} }{s-(\lambda_k+\mathrm{i}j\omega_0)}},\label{eq:Floquetexpr}
\end{equation}
where $\widehat{\mathbf{v}}^k_j$ and $\widehat{\mathbf{w}}^k_j$ designate the $ k\textsuperscript{th}$ column of $ \widehat{\mathsfbi{V}}_j $ and $ \widehat{\mathsfbi{W}}_j $.
Unlike in LTI systems, the output at frequency $s$ depends on the input at an infinite number of frequencies $s-\mathrm{i}n\omega_0$. The output depends {explicitly} on the phase $\phi$ through the cross-frequency transfers $\mathrm{e}^{\mathrm{i}n\phi}\mathsfbi{R}_n(s)$ for $n\neq0$. The poles of the $\mathsfbi{R}_n$ are exactly the Floquet exponents of the LTP system. The adjoint Floquet modes characterize receptivity to forcing of the corresponding direct Floquet modes.

\subsection{Mean resolvent operator}\label{subsec:meanres}

We now seek the operator which predicts the mean output $\langle \mathbf{u}(s)\rangle_\phi$ for a given input $\mathbf{f}(s)$, by averaging (\ref{eq:LTPIO}) with respect to $\phi$. By doing so, all the cross-frequency transfers vanish and the only remaining transfer is $\mathsfbi{R}_0$
\begin{equation}
\boxed{
    \langle \mathbf{u}(s)\rangle_\phi=\mathsfbi{R}_0(s)\mathbf{f}(s)},
\end{equation}
which is consistent with our initial choice of notation for the mean resolvent in the introduction.

\subsubsection{Resonances at natural frequencies}\label{subsec:poles}
Next, we note that since the periodic base flow $\mathbf{U}(t)$ is self-sustained (no external forcing is required), there is a zero Floquet exponent in the LTP system with associated Floquet mode equal to $\mathrm{d}_t \mathbf{U}$ (see appendix \ref{sec:appendix}). If we restrict our attention to the case where the periodic base flow is linearly stable, then the zero Floquet exponent is the leading one, i.e. $\lambda_1=0$ and $\mathbf{v}^1=\mathrm{d}_t \mathbf{U}$\footnote{{In case of instability $\lambda_1>0$ but there is still a zero exponent $\lambda_j=0$ for some $j>1$.}}, so that $\widehat{\mathbf{v}}^1_j=\mathrm{i}j\omega_0\widehat{\mathbf{U}}_j$ where $\widehat{\mathbf{U}}_j$ denotes the $j^\textsuperscript{th}$ harmonic of the Fourier decomposition of $ \mathbf{U}(t) $. Therefore, using expression (\ref{eq:Floquetexpr}), the mean resolvent reads
\begin{equation}
\boxed{
    \mathsfbi{R}_0(s)=\sum_{j\neq0}\dfrac{ \mathrm{i}j\omega_0{\widehat{\mathbf{U}}}_j(\widehat{\mathbf{w}}^1_j)^H }{s-\mathrm{i}j\omega_0}+\sum_j\sum_{k=2}^N\dfrac{ \widehat{\mathbf{v}} ^k_j(\widehat{\mathbf{w}}^k_j)^H }{s-(\lambda_k+\mathrm{i}j\omega_0)}\label{eq:FloquetR0}}.
\end{equation}
We immediately see that there are purely imaginary poles at $s=\mathrm{i}j\omega_0$ as expected from the numerical experiment on the fluidic pinball. The coefficients associated to these poles in the expansion, also called the residuals
\begin{equation}
    \mathrm{Res}_{\mathrm{i}j\omega_0}\mathsfbi{R}_0(s)=\mathrm{i}j\omega_0{\widehat{\mathbf{U}}}_j(\widehat{\mathbf{w}}^1_j)^H,
\end{equation}
tend very rapidly to 0 in the operator norm as $|j|\to \infty$, because the $j^\textsuperscript{th}$ Fourier harmonic of both $\mathbf{U}$ and $\mathbf{w}^1$ are involved (the decay is $o(1/j^m)$ for any integer $m$ if we consider $\mathcal{C}^\infty$ periodic structures). This explains why resonance peaks caused by imaginary poles in figure \ref{fig:spectra_BF}(ii), become decreasingly visible as $|j|$ increases. The fact that the residuals depend on $\widehat{\mathbf{U}}_j$ also explains why resonances at high frequencies are however more visible downstream than upstream. Indeed, larger oscillation amplitude in the far wake causes more pronounced nonlinear energy transfers to higher harmonics (see base flow spectra in row (i) of figure \ref{fig:spectra_BF}), hence a better observability $|\mathbf{C}^T\widehat{\mathbf{U}}_j|$ of high-order harmonics $\widehat{\mathbf{U}}_j$ as the sensor moves downstream.

\subsubsection{System identification using frequency-rich inputs}\label{subsec:identif}
The upside of using frequency-rich signals for system identification is that it allows identifying the dynamics for multiple frequencies at once. The downside though, is that many realizations of the same input signal may be necessary to converge the mean response prior to identification, as we have seen in \S \ref{sec:pinball100} (in particular row (iv) in figure \ref{fig:spectra_BF}). The number of realizations necessary to converge the mean is very dependent on the input signal chosen. This dependence can be made explicit, using (\ref{eq:LTPIO}) and the definition of the variance:
\begin{equation}
    \mathrm{Var}_\phi \mathbf{u}(s)=\sum_{n\neq 0}\|\mathbf{f}(s-\mathrm{i}n\omega_0)\|_{\mathsfbi{R}_n^H\mathsfbi{R}_n}^2,\label{eq:var}
\end{equation}
where $\|\mathbf{a}\|_{\mathsfbi{M}}=(\mathbf{a}^H\mathsfbi{M}\mathbf{a})^{1/2}$ is the norm induced by the positive-definite matrix $\mathsfbi{M}$. A convenient choice of input signal is one which minimizes the variance, allowing for a minimal amount of realizations. Clearly, an input signal with a broadband spectrum like an impulse $\mathbf{f}(t)=\mathbf{B}\delta(t)$ (for which $\delta(s)=1$ for all $s$) is likely to generate a lot of output variance and is not necessarily a favourable choice for identification. An alternative to frequency-rich inputs is to use non-resonant harmonic forcings for identification, as we shall see in the next section.

Another advantage of using a broadband input signal for identification is that if the ratio $\eta$ is very small compared to 1 for such an input signal, then the ratio is likely to be very small for any input signal, hence the LTI approximation based on the mean response is likely to be meaningful. Expression (\ref{eq:var}) also explains why the quality of the LTI approximation deteriorates as the sensor moves downstream in figure \ref{fig:spectra_BF}(iv) for $ u(t)=\delta(t) $. Indeed,
\begin{equation}
    \mathrm{Var}_\phi G(s;u=\delta)=\sum_{n\neq 0} | \mathbf{C}^T\mathsfbi{R}_n(s)\mathbf{B}|^2,
\end{equation}
which, according to \eqref{eq:Floquetexpr}, increases downstream since the observability $|\mathbf{C}^T\widehat{\mathbf{v}}_j^k|$ of the Floquet modes probably grows downstream (this is obvious for the first Floquet mode since $\widehat{\mathbf{v}}^1_j=\mathrm{i}j\omega_0\widehat{\mathbf{U}}_j$).

\subsubsection{System identification using harmonic inputs: `dynamic linearity'}\label{sec:identifharm}
The Laplace transform of harmonic forcings of the form $\mathbf{f}(t)=\mathbf{B}\mathrm{e}^{\mathrm{i}\omega t}$ is given by
\begin{equation}
    \mathbf{f}(s)=\dfrac{\mathbf{B}}{s-\mathrm{i}\omega}.
\end{equation}
Plugging this expression into the input-output relation (\ref{eq:LTPIO}) leads to
\begin{equation}
    \mathbf{u}(s;\phi)=\sum_n \mathrm{e}^{\mathrm{i}n\phi}\dfrac{\mathsfbi{R}_n(s)\mathbf{B}}{s-\mathrm{i}(\omega+n\omega_0)}.\label{eq:dynlin}
\end{equation}
The temporal response may now be evaluated by inverting the Laplace transform. {For the linearly stable limit-cycle considered here,} the decaying Floquet exponents lead to a transient response while all the purely imaginary poles in the right-hand-side of (\ref{eq:dynlin}), either in the transfer operators or the forcing, lead to a permanent contribution. For non-resonant forcing frequencies $\omega \neq n\omega_0$, we have
\begin{align}
    \mathbf{u}(t;\phi)&=\mathrm{transient}\nonumber\\
    &+\mathsfbi{R}_0(\mathrm{i}\omega)\mathbf{B}\mathrm{e}^{\mathrm{i}\omega t}+\sum_{j\neq 0}\mathrm{i}j\omega_0\widehat{\mathbf{U}}_j(\widehat{\mathbf{w}}_j^1)^H\mathbf{B} \mathrm{e}^{\mathrm{i}j\omega_0t}\nonumber\\
   &+\sum_{n\neq 0}\mathrm{e}^{\mathrm{i}n\phi} \left[\mathsfbi{R}_n(\mathrm{i}(\omega+n\omega_0))\mathbf{B}
     \mathrm{e}^{\mathrm{i}(\omega+n\omega_0)t}+\sum_{j\neq 0}\mathrm{i}j\omega_0\widehat{\mathbf{U}}_j(\widehat{\mathbf{w}}_{j-n}^1)^H\mathbf{B} \mathrm{e}^{\mathrm{i}j\omega_0t}\right].\label{eq:phasedep}
\end{align}
The first line of (\ref{eq:phasedep}) collects the transient contributions, the second line gathers terms which are phase-independent, while the third line is phase-dependent and vanishes upon averaging. However, the point here is not to take an average but notice that the Fourier coefficient at the forcing frequency, which may be obtained using an harmonic average 
\begin{equation}
    \widehat{\mathbf{u}}(\omega;\phi)=\lim_{T'\to \infty}\dfrac{1}{T'}\int_0^{T'} \mathbf{u}(t;\phi)\mathrm{e}^{-\mathrm{i}\omega t}\,\mathrm{d}t
\end{equation}
is phase-independent and exactly given by the product of the input amplitude $\widehat{\mathbf{f}}(\omega)=\mathbf{B}$ by the mean resolvent operator evaluated at $s=\mathrm{i}\omega$:
\begin{equation}
    \boxed{\widehat{\mathbf{u}}(\omega)=\mathsfbi{R}_0(\mathrm{i}\omega)\widehat{\mathbf{f}}(\omega)
    }.
\end{equation}

In other terms, using small-amplitude harmonic forcings, it is possible to obtain frequency samples of the mean transfer function $\langle G\rangle_\phi=\mathbf{C}^T\mathsfbi{R}_0\mathbf{B}$ without the need to carry out an ensemble average over several realizations. This is in essence the `dynamic linearity' phenomenon described by \citep{DA12,DA17,EV17} and used therein for LTI system identification on time-varying base flows. 

In principle, using a superposition of harmonic forcings
\begin{equation}
    \mathbf{f}(t)=\mathbf{B}(A_1\mathrm{e}^{\mathrm{i}\omega_1 t}+A_2\mathrm{e}^{\mathrm{i}\omega_2 t}+\dots),
\end{equation}
it should even be possible to sample the mean transfer function at all input frequencies at once, using a single input realization, as long as $\omega_i-\omega_j\neq n\omega_0$ and $\omega_i\neq n\omega_0$ for any $i,j >0$.

\subsection{Connection with the Koopman operator}\label{subsec:Koopman}

The propagator from $t'=0$ to $t$ 
\begin{equation}
    \mathbf{\Phi}(t,t'=0;\phi)=\sum_j\sum_{k=1}^N \widehat{\mathbf{v}} ^k_j\mathrm{e}^{(\lambda_k+\mathrm{i}j\omega_0)t}\mathrm{e}^{\mathrm{i}j\phi}(\mathbf{w}^k(0;\phi))^H
\end{equation}
corresponds to the Koopman operator associated with the full-state observable $\mathbf{u}$ in the case of linear dynamics about the time-periodic base flow \citep{ME16}. The $\widehat{\mathbf{v}} ^k_j$ are the Koopman modes, the Floquet exponents $\lambda_k+\mathrm{i}j\omega_0$ are the Koopman eigenvalues, and $\mathbf{u}_0\mapsto\mathrm{e}^{\mathrm{i}j\phi}(\mathbf{w}^k(0;\phi))^H\mathbf{u}_0$ are the phase-dependent Koopman eigenfunctions.

Averaging the propagator with respect to $\phi$ leads to the mean propagator
\begin{equation}
    \langle\mathbf{\Phi}(t,0)\rangle_\phi=\sum_j\sum_{k=1}^N \widehat{\mathbf{v}} ^k_j\mathrm{e}^{(\lambda_k+\mathrm{i}j\omega_0)t}{(\widehat{\mathbf{w}}^k_j)^H}\label{eq:meanpropag}
\end{equation}
based on the phase-averaged Koopman eigenfunctions $\mathbf{u}_0\mapsto\langle \mathrm{e}^{\mathrm{i}j\phi}(\mathbf{w}^k(0;\phi))^H\mathbf{u}_0\rangle_\phi={(\widehat{\mathbf{w}}^k_j)}^H\mathbf{u}_0$.

By comparing equations (\ref{eq:meanpropag}) and (\ref{eq:Floquetexpr}) for $ n=0$, we see that the mean resolvent operator is the frequency domain representation of the mean propagator from $0$ to $t$, i.e. of the mean Koopman operator for the full-state observable of the LTP system
\begin{equation}
\boxed{
    \mathsfbi{R}_0(s)=\mathcal{L}[\langle \mathbf{\Phi}(t,0)\rangle_\phi]},
\end{equation}
where $\mathcal{L}[.]$ denotes the Laplace transform. 

\subsection{Connection with LTI dynamics about the mean flow}\label{subsec:connection}

\subsubsection{Connection with the harmonic transfer operator}
Evaluating (\ref{eq:LTPIO}) at various output frequencies $s+\mathrm{i}k\omega_0$, it is possible to introduce the harmonic transfer operator $\underline{\mathsfbi{H}}(s;\phi)$ such that
\begin{align}
    &\underbrace{\begin{pmatrix}\vdots \\\mathbf{u}(s-\mathrm{i}\omega_0)\\\mathbf{u}(s)\\ \mathbf{u}(s+\mathrm{i}\omega_0)\\\vdots\end{pmatrix}}_{\underline{\mathbf{u}}(s;\phi)}=\nonumber\\
    &\underbrace{\begin{pmatrix}
    \ddots & \ddots & \ddots & \ddots & \ddots\\
    \ddots & \mathsfbi{R}_{0}(s-\mathrm{i}\omega_0) & \mathsfbi{R}_{-1}(s-\mathrm{i}\omega_0)\mathrm{e}^{-\mathrm{i}\phi} & \mathsfbi{R}_{-2}(s-\mathrm{i}\omega_0)\mathrm{e}^{-2\mathrm{i}\phi} & \ddots\\
    \ddots & \mathsfbi{R}_{1}(s)\mathrm{e}^{\mathrm{i}\phi} & \mathsfbi{R}_{0}(s) & \mathsfbi{R}_{-1}(s)\mathrm{e}^{-\mathrm{i}\phi} & \ddots\\
    \ddots & \mathsfbi{R}_{2}(s+\mathrm{i}\omega_0)\mathrm{e}^{2\mathrm{i}\phi} & \mathsfbi{R}_{1}(s+\mathrm{i}\omega_0)\mathrm{e}^{\mathrm{i}\phi} & \mathsfbi{R}_{0}(s+\mathrm{i}\omega_0) & \ddots\\
     \ddots & \ddots & \ddots & \ddots & \ddots
    \end{pmatrix}}_{\underline{\mathsfbi{H}}(s;\phi)}
\underbrace{\begin{pmatrix}\vdots \\\mathbf{f}(s-\mathrm{i}\omega_0)\\\mathbf{f}(s)\\ \mathbf{f}(s+\mathrm{i}\omega_0)\\\vdots\end{pmatrix}}_{\underline{\mathbf{f}}(s)}.
\end{align}
The block $\underline{\mathsfbi{H}}_{jk}$ of this infinite matrix operator characterizes the transfer from the frequency $s+\mathrm{i}k\omega_0$ of the input $\mathbf{f}$ to the frequency $s+\mathrm{i}j\omega_0$ of the output $\mathbf{u}$ \citep{WE90,WE91,ZH02,ZH08}. The mean resolvent operator appears along the diagonal of the harmonic transfer operator, as it characterizes the phase-independent transfer from any frequency of the input to the same frequency at the output. In particular, we have
\begin{equation}
    \boxed{\mathsfbi{R}_0(s)=\underline{\mathsfbi{H}}_{00}(s)}.\label{eq:connectH}
\end{equation}
\subsubsection{The harmonic transfer operator as a feedback loop}\label{subsec:HFT}
Decompose the Jacobian as a mean $\overline{\mathsfbi{J}}$ and a periodic perturbation $\mathsfbi{J}'(t;\phi)$, which we expand as a Fourier series
\begin{equation}
    \mathsfbi{J}(t;\phi)=\overline{\mathsfbi{J}}+ \underbrace{\sum_{j} \widehat{\mathsfbi{J}}'_j \mathrm{e}^{\mathrm{i}j(\omega_0 t+{\phi})}}_{\mathsfbi{J}'(t;\phi)}.\label{eq:FourU}
\end{equation}
 Because $\mathsfbi{J}'(t;\phi)$ is real, the harmonics have {Hermitian} symmetry, i.e. $\widehat{\mathsfbi{J}}'_{-j}=\widehat{\mathsfbi{J}}'^{*}_{j}$, where $(.)^*$ denotes the complex conjugate (not the conjugate transpose $(.)^H$). Moreover, we also obviously have $\widehat{\mathsfbi{J}}'_0=\mathsfbi{0}$. Since we are considering the incompressible Navier--Stokes equations, the nonlinearity is quadratic hence the mean Jacobian is equal to the Jacobian operator about the mean flow
 \begin{equation}
     \overline{\mathsfbi{J}}=\mathsfbi{J}_{\overline{\mathbf{U}}}.\label{eq:JbarUbar}
 \end{equation}
 Assuming $\mathbf{u}^0=0$, taking the Laplace transform of (\ref{eq:simple2}) and plugging (\ref{eq:FourU})-(\ref{eq:JbarUbar}) yields
\begin{equation}
    s\mathbf{u}(s;\phi)=\mathsfbi{J}_{\overline{\mathbf{U}}}\mathbf{u}(s;\phi)+\sum_{j}\widehat{\mathsfbi{J}}'_j\mathrm{e}^{\mathrm{i}j\phi}\mathbf{u}(s-\mathrm{i}j\omega_0;\phi)+\mathbf{f}(s),
\end{equation}
which in turn leads, through harmonic balance \citep{KH02}, to an alternative form of the harmonic transfer operator, also referred to as the harmonic resolvent operator \citep{PA20,franceschini2022identification}
\begin{equation}
    \underline{\mathsfbi{H}}(s;\phi)=(\underline{\mathsfbi{D}}(s)-\underline{\mathsfbi{T}}(\phi))^{-1}.\label{eq:Halt}
\end{equation}

The infinite matrix $\underline{\mathsfbi{D}}(s)$ is block-diagonal, while $\underline{\mathsfbi{T}}(\phi)$ is block-Laurent \citep{KU15} with zero blocks on the diagonal:
\begin{align}
    \underline{\mathsfbi{D}}(s)&=
    \begin{pmatrix}
    \ddots & \ddots & \ddots & \ddots & \ddots\\
    \ddots & (s-\mathrm{i}\omega_0)\mathsfbi{I}-\mathsfbi{J}_{\overline{\mathbf{U}}} & \mathsfbi{0} & \mathsfbi{0} & \ddots\\
    \ddots & \mathsfbi{0} & s\mathsfbi{I}-\mathsfbi{J}_{\overline{\mathbf{U}}} & \mathsfbi{0} & \ddots\\
    \ddots& \mathsfbi{0} & \mathsfbi{0} & (s+\mathrm{i}\omega_0)\mathsfbi{I}-\mathsfbi{J}_{\overline{\mathbf{U}}} & \ddots\\
     \ddots & \ddots & \ddots & \ddots & \ddots
    \end{pmatrix},\\
    \underline{\mathsfbi{T}}(\phi)&=
    \begin{pmatrix}
    \ddots & \ddots & \ddots & \ddots & \ddots\\
    \ddots & \mathsfbi{0} & \widehat{\mathsfbi{J}}_1'^{*}\mathrm{e}^{-\mathrm{i}\phi} & \widehat{\mathsfbi{J}}_2'^{*}\mathrm{e}^{-2\mathrm{i}\phi} & \ddots\\
    \ddots & \widehat{\mathsfbi{J}}'_1\mathrm{e}^{\mathrm{i}\phi} & \mathsfbi{0} & \widehat{\mathsfbi{J}}_1'^{*}\mathrm{e}^{-\mathrm{i}\phi} & \ddots\\
    \ddots& \widehat{\mathsfbi{J}}'_2\mathrm{e}^{2\mathrm{i}\phi} & \widehat{\mathsfbi{J}}'_1\mathrm{e}^{\mathrm{i}\phi} & \mathsfbi{0} & \ddots\\
     \ddots & \ddots & \ddots & \ddots & \ddots
    \end{pmatrix}.
\end{align}

\begin{figure}
    \centering
    \includegraphics[height=0.25\textwidth]{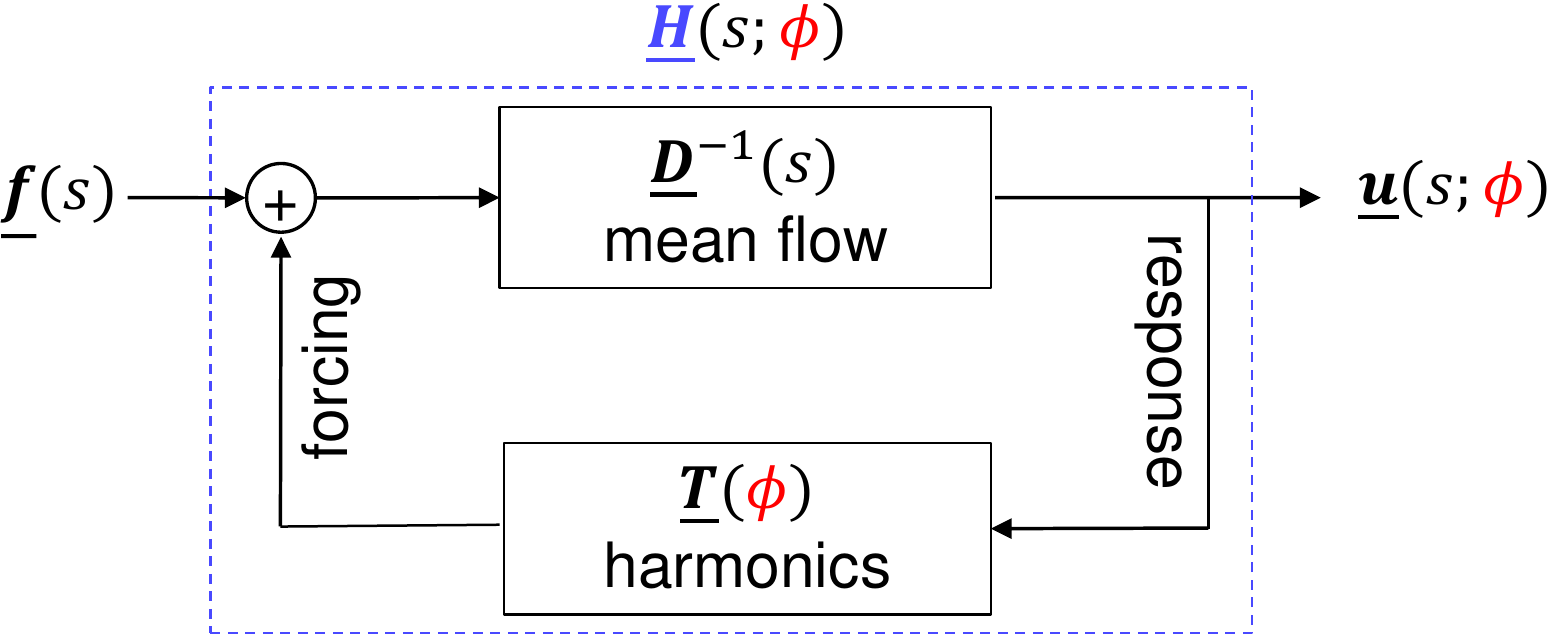}
    \caption{Block representation of the harmonic transfer operator $\underline{\mathsfbi{H}}(s;\phi)$ as a feedback loop between two blocks: $\underline{\mathsfbi{D}}^{-1}(s)$ accounts for interactions with the mean flow and $\underline{\mathsfbi{T}}(\phi)$ accounts for interactions with the harmonics of the periodic Jacobian.}
    \label{fig:diag}
\end{figure}

By recasting (\ref{eq:Halt}) in the form
\begin{equation}
    \underline{\mathsfbi{H}}(s;\phi)= (\underline{\mathsfbi{I}}-\underline{\mathsfbi{D}}^{-1}(s)\underline{\mathsfbi{T}}(\phi))^{-1}\underline{\mathsfbi{D}}^{-1}(s),\label{eq:loopH}
\end{equation}
the operator may be interpreted as a feedback loop between the two blocks $\underline{\mathsfbi{D}}^{-1}(s)$ and $\underline{\mathsfbi{T}}(\phi)$, as illustrated in figure \ref{fig:diag}.

A parallel may be drawn with the \textit{describing function methodology} \citep{GE68}, which applies to oscillating systems composed of an LTI block in feedback loop with a \textit{nonlinear time-invariant} block, instead of a \textit{linear time-varying} one as here. The describing function methodology may be used to obtain transfer functions parametrized by the forcing amplitude \citep{NO08}, whereas the present methodology yields transfer functions parametrized by the phase $\phi$. The two approaches are not mutually exclusive and one may extend the present framework by adding a nonlinear time-invariant block as well to consider the effect of forcing amplitude for harmonic inputs.

The operator $\underline{\mathsfbi{D}}^{-1}(s)$ is block-diagonal hence does not transfer energy from one frequency to another. Physically, this block represents interactions of the perturbation with the mean flow, and this process is independent of the phase $\phi$. This block takes a forcing at the input and delivers a velocity at the output. The feedback block does the opposite and corresponds to interactions with the fluctuating part of the periodic base flow. It contains only off-diagonal terms associated to the various harmonics of the periodic perturbation flow $\mathbf{U}'$, meaning that it can only transfer energy from one frequency to another, and this process is phase-dependent. This alternative interpretation of the harmonic transfer operator is key to understanding the connection of the mean resolvent operator to the resolvent operator about the mean flow, as we shall see next. 

\subsubsection{The resolvent about the mean flow approximates the mean resolvent}
 
The inverse $(\underline{\mathsfbi{I}}- \underline{\mathsfbi{D}}^{-1}(s)\,\underline{\mathsfbi{T}}(\phi))^{-1}$ in (\ref{eq:loopH}) may be expanded as a Neumann series $\sum_{k\geq 0}(\underline{\mathsfbi{D}}^{-1}(s)\,\underline{\mathsfbi{T}}(\phi))^k$ such that
\begin{equation}
    \underline{\mathsfbi{H}}(s;\phi)=\underbrace{\underline{\mathsfbi{D}}^{-1}(s)}_{\underline{\mathsfbi{H}}^0(s)}+\underbrace{\underline{\mathsfbi{D}}^{-1}(s)\underline{\mathsfbi{T}}(\phi)\underline{\mathsfbi{D}}^{-1}(s)}_{\underline{\mathsfbi{H}}^1(s;\phi)}+\underbrace{\underline{\mathsfbi{D}}^{-1}(s)\underline{\mathsfbi{T}}(\phi)\underline{\mathsfbi{D}}^{-1}(s)\underline{\mathsfbi{T}}(\phi)\underline{\mathsfbi{D}}^{-1}(s)}_{\underline{\mathsfbi{H}}^2(s;\phi)}+\dots.\label{eq:expH}
\end{equation}
{The series converges if $\|\underline{\mathsfbi{D}}^{-1}(s)\,\underline{\mathsfbi{T}}(\phi)\|_2<1$\footnote{{Here $\|.\|_2$ denotes the spectral norm, i.e. maximum singular value, not the $H_2$ norm.}}. Assume $\mathrm{Im}(s)=:\sigma>\sigma_\mathrm{max}$ where $\sigma_\mathrm{max}\geq 0$ is the maximum growth rate among the poles of $\underline{\mathsfbi{D}}^{-1}$ and $\underline{\mathsfbi{H}}$, then the series converges if a) base flow unsteadiness is sufficiently weak or b) $\sigma$ is sufficiently large (see appendix \ref{subsec:harmres}). Each contribution $\underline{\mathsfbi{H}}^i(s;\phi)$ corresponds to going $i$ times around the loop, i.e. interacting $i$ times with the unsteady part of the base flow.}

Expansion (\ref{eq:expH}) on the harmonic transfer operator leads to a similar expansion on the mean resolvent operator since the latter operator is a sub-block of the former according to (\ref{eq:connectH}), i.e. 
\begin{equation}
    \mathsfbi{R}_0=\underline{\mathsfbi{H}}^0_{00}+\underline{\mathsfbi{H}}^1_{00}+\underline{\mathsfbi{H}}^2_{00}+\dots.\label{eq:expansionMR}
\end{equation} 
This expansion may be interpreted in the following way: $\underline{\mathsfbi{H}}^i_{00}$ collects contributions of $\underline{\mathsfbi{H}}$ which deliver an output at the same frequency than the input, after interacting $i$ times with the fluctuating part of the base flow. Using the general terms
\begin{align}
    \underline{\mathsfbi{D}}_{jk}^{-1}(s)&:=\mathsfbi{R}_{\overline{\mathbf{U}}}(s+\mathrm{i}j\omega_0)\delta_{jk},\label{eq:D}\\
    \underline{\mathsfbi{T}}_{jk}(\phi)&:=\widehat{\mathsfbi{J}}'_{j-k}\mathrm{e}^{\mathrm{i}(j-k)\phi},\label{eq:T}
\end{align}
we find
\begin{align}
    \underline{\mathsfbi{H}}^0_{00}&=\mathsfbi{R}_{\overline{\mathbf{U}}},\label{eq:R00}\\
    \underline{\mathsfbi{H}}^1_{00}&=\mathsfbi{0},\label{eq:R01}\\
    \underline{\mathsfbi{H}}^2_{00}&=\sum_j \mathsfbi{R}_{\overline{\mathbf{U}}}(s) \widehat{\mathsfbi{J}}'^{*}_{j}\mathsfbi{R}_{\overline{\mathbf{U}}}(s+\mathrm{i}j\omega_0)\widehat{\mathsfbi{J}}'_{j}\mathsfbi{R}_{\overline{\mathbf{U}}}(s),\label{eq:R02}\\
    \underline{\mathsfbi{H}}^3_{00}&=\sum_{j,k}\mathsfbi{R}_{\overline{\mathbf{U}}}(s)\widehat{\mathsfbi{J}}'^{*}_{j}\mathsfbi{R}_{\overline{\mathbf{U}}}(s+\mathrm{i}j\omega_0) \widehat{\mathsfbi{J}}'_{j-k}\mathsfbi{R}_{\overline{\mathbf{U}}}(s+\mathrm{i}k\omega_0)\widehat{\mathsfbi{J}}'_{k}\mathsfbi{R}_{\overline{\mathbf{U}}}(s),\label{eq:R03}\\
    &\vdots\nonumber
\end{align}
The order 0 term is equal to the resolvent operator about the mean flow, which corresponds to interacting 0 times with the fluctuating Jacobian. The order 1 term is exactly 0, because it is impossible to interact just once with the fluctuating Jacobian and come out of the loop at the same frequency than the input. The first correction to the resolvent operator about the mean flow then arises at order 2, when successive interactions with $\widehat{\mathsfbi{J}}'_k$ and its complex conjugate $\widehat{\mathsfbi{J}}'^{*}_{j}=\widehat{\mathsfbi{J}}'_{-j}$ occur, allowing the output to be at the same frequency than the input. 

{Expansion (\ref{eq:expansionMR}) may be used to bound the absolute and relative differences between $\mathsfbi{R}_0$ and $\mathsfbi{R}_{\overline{\mathbf{U}}}$. Introduce the following system norm
\begin{equation}
    \|\mathsfbi{G}\|_{\infty,\sigma}:=\sup_{\omega\in\mathbb{R}}\|\mathsfbi{G}(\sigma+\mathrm{i}\omega)\|_2,\label{eq:defnorm}
\end{equation}
and, for $\sigma>\sigma_\mathrm{max}$, the small parameter
\begin{equation}
    \epsilon_\sigma:=\sum_k\|\mathsfbi{R}_{\overline{\mathbf{U}}}\widehat{\mathsfbi{J}}'_k\|_{\infty,\sigma},\label{eq:defsig}
\end{equation}
characterizing the ($\sigma$-dependent) loop gain associated with the feedback interconnection of $\underline{\mathsfbi{D}}^{-1}(s)$ and $\underline{\mathsfbi{T}}(\phi)$, averaged with respect to $\phi$.} 
 {Using (\ref{eq:expansionMR}) and the definition (\ref{eq:defsig}) of $\epsilon_\sigma$, we may write
\begin{align}
    \|\mathsfbi{R}_0-\mathsfbi{R}_{\overline{\mathbf{U}}}\|_2(s)&\leq\|\mathsfbi{R}_{\overline{\mathbf{U}}}\|_2(s) (\epsilon_\sigma^2 + \epsilon_\sigma^3+\dots),\nonumber\\
    &= \|\mathsfbi{R}_{\overline{\mathbf{U}}}\|_2(s)\dfrac{\epsilon_\sigma^2}{1- \epsilon_\sigma}.
    \label{eq:ineq}
\end{align}
The geometric series $\sum_{k=0}^\infty \epsilon_\sigma^k$ converges to $(1-\epsilon_\sigma)^{-1}$ if and only if $\epsilon_\sigma<1$, which occurs if either a) base flow unsteadiness is sufficiently weak or b) $\sigma$ is large enough (see appendix \ref{subsec:meanres}). Since $\|\mathsfbi{R}_{\overline{\mathbf{U}}}\|_2(s)\to 0$ as either $\sigma\to\infty$ at fixed $\omega$ or $|\omega|\to \infty$ at fixed $\sigma$, the absolute difference in (\ref{eq:ineq}) also tends to zero in these limits, i.e. $\|\mathsfbi{R}_0-\mathsfbi{R}_{\overline{\mathbf{U}}}\|_2(s)\to 0$. Moreover, the relative error at fixed $s$ 
\begin{equation}
\boxed{
    \dfrac{\|\mathsfbi{R}_0-\mathsfbi{R}_{\overline{\mathbf{U}}}\|_2(s)}{\|\mathsfbi{R}_{\overline{\mathbf{U}}}\|_2(s)}=O(\epsilon_\sigma^2)},\label{eq:order2}
\end{equation}
is \textit{order 2}, hence vanishes very quickly with the small parameter $\epsilon_\sigma$. As already said, there are two independent ways to make $\epsilon_\sigma$ small. Taking large $\sigma$ (option b) allows for low relative difference between the two operators even in strongly unsteady base flows. But in this case, using an LTI approximation of the input-output dynamics becomes insufficient (see ratio $\eta$ introduced in \S \ref{subsec:paramphi}). Therefore, we argue that the key reason why $\mathsfbi{R}_{\overline{\mathbf{U}}}$ is a physically relevant operator in incompressible flows, is because it approximates $\mathsfbi{R}_0$ when base flow unsteadiness is weak (option a). The fact that the relative error is order 2 with respect to base flow unsteadiness may explain the robust agreement observed between the two operators. In the case of the fluidic pinball (see Bode diagrams in figure\ref{fig:spectra_BF}(iii,iv)), agreement was observed for a low value of $\sigma=0.01$, indicating small amplification of the perturbations by the unsteady part of the base flow.} 

{It is interesting to note that the small correction to $\mathsfbi{R}_{\overline{\mathbf{U}}}$ involves the unsteady part of the Jacobian operator, which indirectly incorporates information about the endogenous nonlinear forcings of \cite{MC10} into the linear operator $\mathsfbi{R}_0$. Indeed, the fluctuating part of the Jacobian is aware of the harmonic balance between the various Fourier components of the nonlinear base flow. In this regard, the present work may be seen as an extension of the recent contributions seeking to incorporate information about nonlinear forcings into the linear operator through either a turbulent viscosity \citep{MO19,PI21} or a state-feedback operator \citep{ZA17}.  
}

\subsection{Reduced-order models of the mean resolvent}\label{subsec:approxmeanres}

Another way to write the mean resolvent is in the form
\begin{equation}
\boxed{
    \mathsfbi{R}_0(s)=\underline{\mathsfbi{C}}(s\underline{\mathsfbi{I}}-\underline{\boldsymbol{\Lambda}})^{-1}\underline{\mathsfbi{B}}},\label{eq:IO}
\end{equation}
where
\begin{align}
    \underline{\mathsfbi{C}}&=[\dots,\widehat{\mathsfbi{V}}_{-1},\widehat{\mathsfbi{V}}_0,\widehat{\mathsfbi{V}}_{1}\dots],\\
    \underline{\mathsfbi{B}}&=[\dots,\widehat{\mathsfbi{W}}_{-1},\widehat{\mathsfbi{W}}_0,\widehat{\mathsfbi{W}}_{1}\dots]^H,\\
    \underline{\mathsfbi{I}}_{jk}&=\mathsfbi{I}\delta_{jk},\\
    \underline{\boldsymbol{\Lambda}}_{jk}&=(\boldsymbol{\Lambda}+\mathrm{i}j\omega_0\mathsfbi{I})\delta_{jk}.
\end{align}
The input (resp. output) matrix $\underline{\mathsfbi{B}}$ (resp. $\underline{\mathsfbi{C}}$) has $N$ columns (resp. {rows}) and an infinite number of rows (resp. columns), while the diagonal state matrix $\underline{\boldsymbol{\Lambda}}$ is infinite-dimensional. Expression (\ref{eq:IO}) is associated to a state-space representation 
\begin{align}
   \mathrm{d}_t \underline{\mathbf{x}} &=  \underline{\boldsymbol{\Lambda}}\,\underline{\mathbf{x}} + \underline{\mathsfbi{B}}\,\mathbf{f},\\
   \langle\mathbf{u}\rangle_\phi &= \underline{\mathsfbi{C}}\, \underline{\mathbf{x}},
\end{align}
where the internal state $\underline{\mathbf{x}}=[\dots,\mathbf{x}_{-1},\mathbf{x}_0,\mathbf{x}_1,\dots]^T$ is an infinite column vector, even though the input $\mathbf{f}$ and the output $\langle\mathbf{u}\rangle_\phi$ both have a finite dimension $N$. {However, for large enough $|j|$ the residuals of the poles $\lambda_k+\mathrm{i}j\omega_0$ in $\underline{\boldsymbol{\Lambda}}$ become negligible, as they involve the Fourier coefficients $\widehat{\mathsfbi{W}}_j$ in $\underline{\mathsfbi{B}}$ and $\widehat{\mathsfbi{V}}_j$ in $\underline{\mathsfbi{C}}$ (for $\mathcal{C}^\infty$ functions, the decay is $o(1/j^m)$ for any positive integer $m$). Hence the infinite-dimensional state vector $\underline{\mathbf{x}}$ may be projected onto a finite-dimensional one by simply eliminating high-frequency poles based on some cutoff value for the residual norm.}

The resolvent operator about the mean flow approximates $\mathsfbi{R}_0$ and it only has $N$ poles: it may therefore be interpreted as a reduced-order model of order $N$ of the mean resolvent. Even though the approximation is good for weakly unsteady flows, the operator $\mathsfbi{R}_{\overline{\mathbf{U}}}$ does not take into account the fluctuating part of the Jacobian, hence the model may not be optimal for the fixed order $N$. Moreover, there is no reason a priori to choose a model order equal to $N$: in model reduction, the order is fixed by the minimal number of poles necessary to capture the input-output dynamics up to a given precision. The appropriate order may therefore be chosen smaller or greater than $N$ depending on the desired precision.


\subsection{The `RZIF' property}\label{subsec:RZIF}
We recall that $\lambda_1=0$ (see \S \ref{subsec:poles} and appendix \ref{sec:appendix}) therefore $s=\mathrm{i}j\omega_0$ are Koopman eigenvalues of the mean propagator satisfying exactly the so-called RZIF property originally discussed in the context of mean flow stability analysis \citep{TU15}. Since $\mathbf{v}^1=\mathrm{d}_t\mathbf{U}$, the associated Koopman modes $\widehat{\mathbf{v}}_j^1=\mathrm{i}j\omega_0\widehat{\mathbf{U}}_j$ are parallel to the Fourier components $\widehat{\mathbf{U}}_j$ of the periodic base flow. This is only approximately true in the case of eigenvectors of the mean Jacobian associated with RZIF eigenvalues, unless the oscillations of the periodic base flow are monochromatic \citep{ME13,TU15}, the dynamics is weakly non-linear \citep{NO03,SI07} or weakly unsteady \cite{ME13}. {Therefore, the proposed framework appears to be more generic than linear analysis about a mean flow, as the poles of $\mathsfbi{R}_0$ and associated modes exactly satisfy the RZIF property, while this is only approximately true for $\mathsfbi{R}_{\overline{\mathbf{U}}}$.}


Finally, we add that the RZIF property is also verified for unstable periodic base flows, but in this case the zero Floquet exponent does not correspond to the leading one, i.e. $\lambda_1\neq 0$, which has a strictly positive growth rate (see appendix \ref{sec:appendix}). 

\section{Towards more complex base-flows}\label{sec:extension}
The previous sections were only concerned with periodic base flows, but the definition (\ref{eq:def}) of the mean resolvent $\mathsfbi{R}_0$ may be easily generalized to any other statistically steady base flow. Instead of taking the mean output with respect to the phase $\phi$, we may take more generally the mean output with respect to the relative time $\tau_0$ at which the forcing signal is released:
\begin{equation}
\boxed{
    \text{mean resolvent } \mathsfbi{R}_0: \mathbf{f}(s)\mapsto \langle\mathbf{u}(s)\rangle_{\tau_0}}.
\end{equation}
For such a definition to hold, we assume that the single trajectory $\widetilde{\mathbf{U}}(\tau)$ of the unsteady base flow, about which we linearize the dynamics, covers the entire dynamical attractor. Otherwise, an average over multiple base flow realizations also needs to be performed.

 We start by performing similar numerical experiments as in \S \ref{sec:pinball100}, but for more complex incompressible two-dimensional base flows. The numerical procedure explained in \S \ref{sec:procedure} is repeated, but now varying the time $\tau_0$ at which the impulse is triggered, rather than the phase $\phi$, which is no longer defined. The definitions introduced in \S \ref{subsec:paramphi} are extended by replacing $\langle.\rangle_\phi$ with $\langle.\rangle_{\tau_0}$, since now `frequency response realizations' are parametrized by $\tau_0$.
 
 In \S \ref{sec:fluidic110120}, we consider the fluidic pinball in the quasiperiodic and chaotic regimes by increasing the Reynolds number to respectively $Re=110$ and $Re=120$ \citep{DE20}. For the quasiperiodic case, an extension of the periodic theory is possible and reported in appendix \ref{subsec:QP}. Next, in \S \ref{sec:BFS}, we consider the stochastic flow past a backward-facing step at $Re=500$, as in \cite{HE12}. Two values are considered for the variance $\sigma_w$ of the stochastic forcing needed to sustain unsteadiness in the base flow. 
 
 By considering these extra cases, we cover all possible signs of the maximum Lyapunov exponent (MLE), which is the maximal growth rate associated with the linearized dynamics about a statistically steady base flow. The MLE may be estimated from the mean impulse response according to
\begin{equation}
    \mathrm{MLE} = \lim_{t\to\infty}\dfrac{1}{t}\ln \dfrac{\|\langle\mathbf{u}(t)\rangle_{\tau_0}\|}{\|\langle \mathbf{u}(0)\rangle_{\tau_0}\|}.\label{eq:MLE}
\end{equation}
In the periodic case, it is equivalent to take the ensemble average with respect to $\tau_0$ or $\phi$ and the MLE corresponds to $\mathrm{Re}(\lambda_1)=0$. In the quasiperiodic case, we also have $\mathrm{MLE}=0$, for the chaotic regime $\mathrm{MLE}>0$, while for the stochastic regime $\mathrm{MLE}<0$ (results of these computations are reported in appendix \ref{sec:MLE}). The goal is to provide numerical evidence of the strong connection between the mean resolvent and the resolvent about the mean flow in all these cases, even though our theory only applies to the periodic and quasi-periodic cases. 

Finally, in \S \ref{subsec:compress}, we discuss implications of our previous theoretical analysis to the case of compressible flows.

 \subsection{Quasi-periodic and chaotic flows: fluidic pinball at $ Re=110$ and $ Re=120$}\label{sec:fluidic110120}
 The numerical setup is identical to that of \S \ref{sec:pinball100} (with more details in appendix \ref{sec:numdetails}) for $Re=100$.
 \subsubsection{Unsteady base flow}
\begin{figure}
    \centering
    \begin{tabular}{cc}
      (ai) \includegraphics[height=0.24\textwidth]{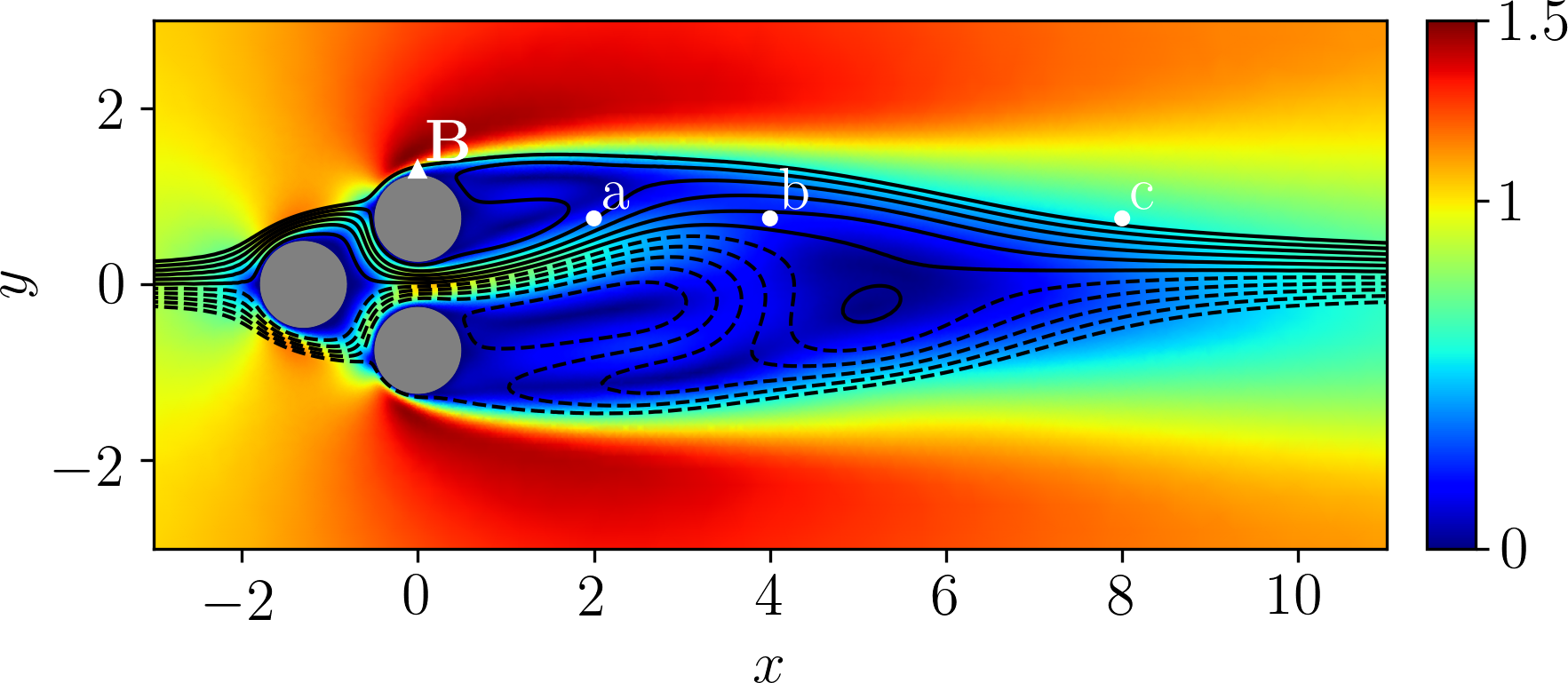}    &  (bi) \includegraphics[height=0.24\textwidth]{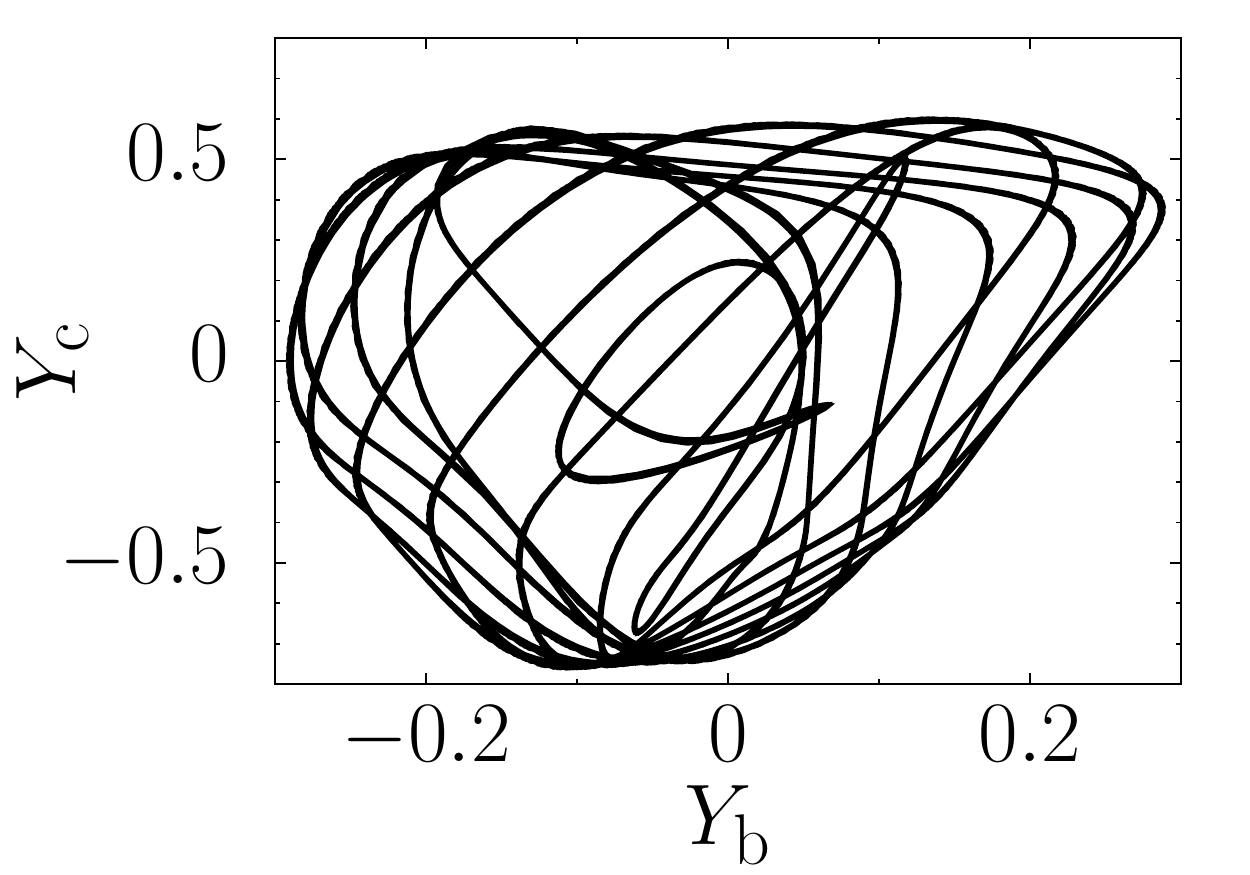}\\
       (aii) \includegraphics[height=0.24\textwidth]{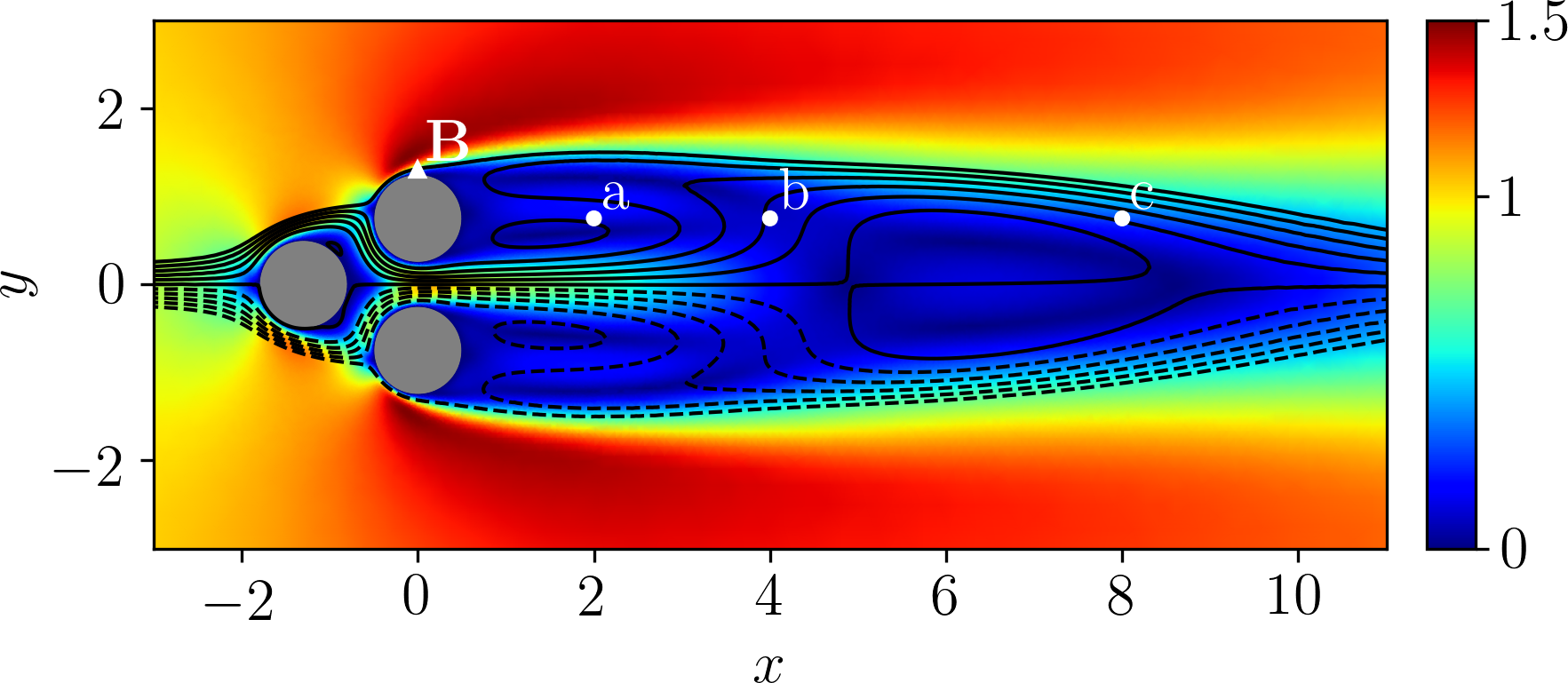}    &  (bii) \includegraphics[height=0.24\textwidth]{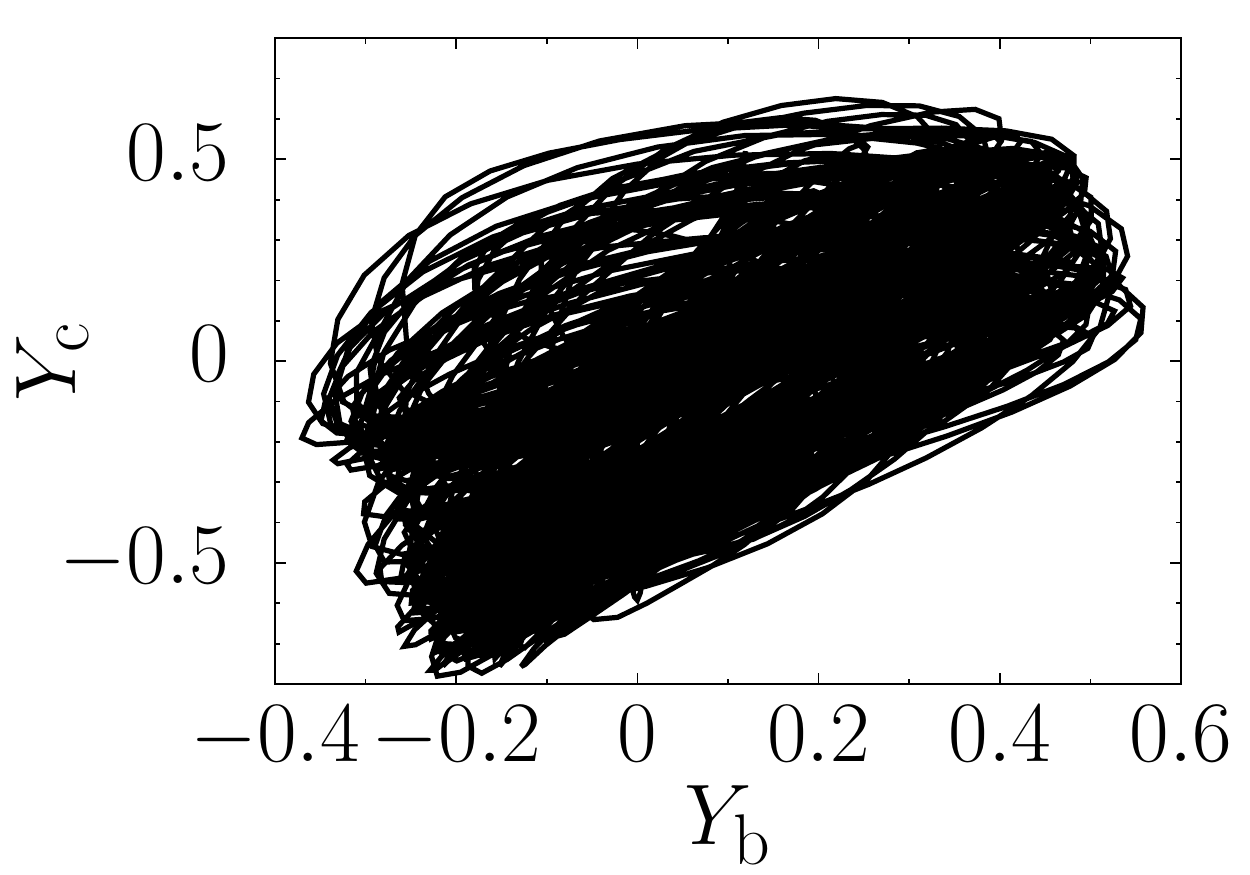}
    \end{tabular}
    \caption{Fluidic pinball in different regimes: (i) quasiperiodic at $Re=110$ and (ii) chaotic at $Re=120$. (a) Mean velocity norm and streamlines and (b) phase portraits using $y$-velocity sensors $Y_\mathrm{b}$ and $Y_\mathrm{c}$.}
    \label{fig:fluidicpinball_V2}
\end{figure}

 In figure \ref{fig:fluidicpinball_V2}(a), we plot the streamlines and velocity norm isocontours for the mean flow in the (i) quasiperiodic case and the (ii) chaotic case. We notice that the mean flow is {asymmetric} in the quasi-periodic case, but symetric in the chaotic case. The recirculation zone is more extended downstream in the latter case. 
 
 In figure \ref{fig:fluidicpinball_V2}(b), we plot the phase portrait of the flow, using the vertical velocity probes 2 and 3. The amplitude of the fluctuations increases with the Reynolds number. The phase portrait is well-structured in the quasiperiodic case but it is impossible to infer the dimensionality of the torus from this two-dimensional projection of the attractor. The complex structure of the phase portrait in figure \ref{fig:fluidicpinball_V2}(bii) is typical of chaotic dynamics. 
 
\begin{figure}
    \centering
    \includegraphics[width=1.0\textwidth]{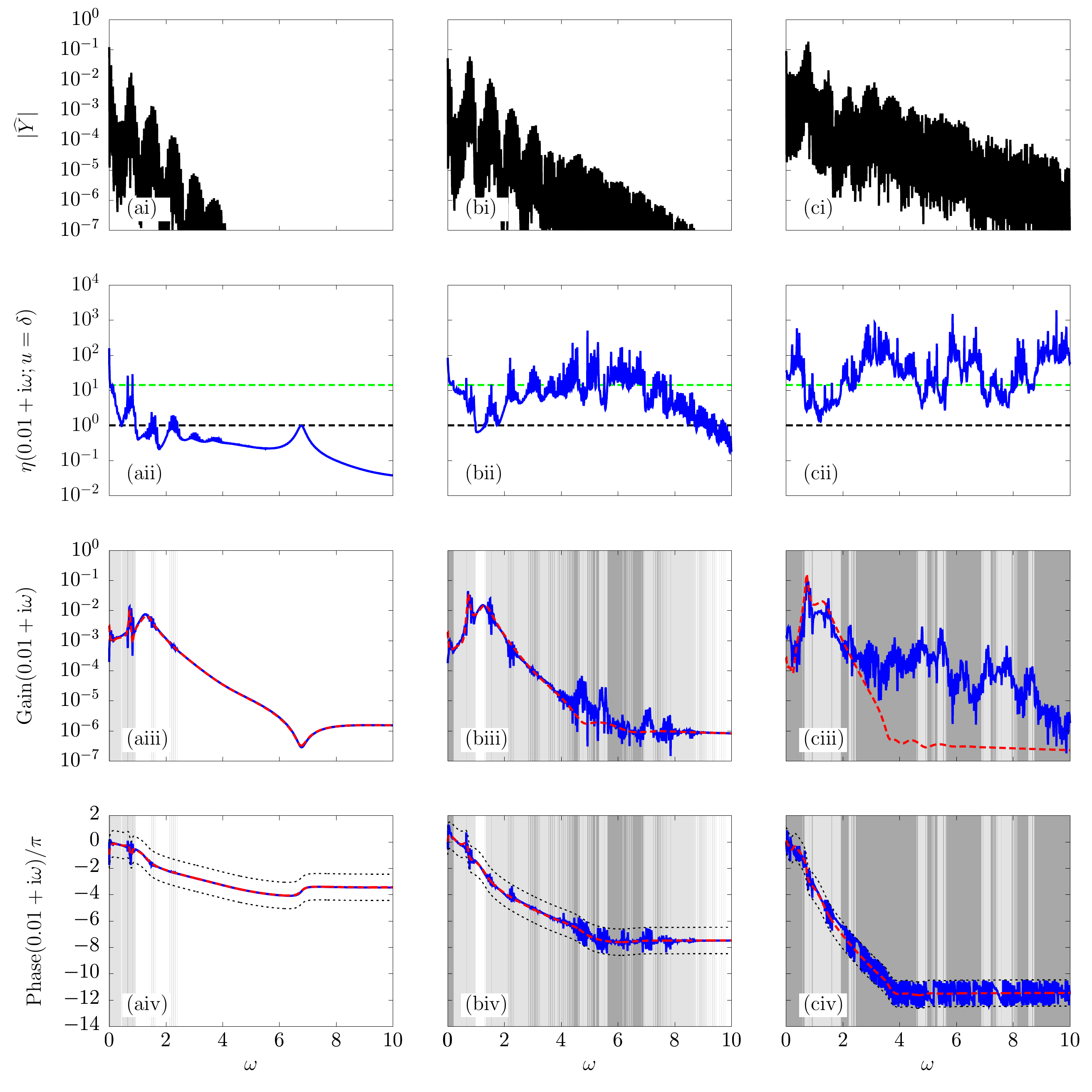}
    \caption{{Fluidic pinball in the quasiperiodic regime at $Re=110$. The labels (a,b,c) correspond to the three sensor positions. (i) Fourier spectrum of measurement the unsteady base flow $Y$. The spectrum is discrete and no averaging is done so a single Hann window is applied over the entire signal of more than 1000 time units. (ii) Ratio $\eta$ (defined in (\ref{eq:eta})) for $N_s=204$ impulsive forcings $u(t)=\delta(t)$. The black dashed line indicates the threshold $\eta=1$ far below which the system is nearly LTI (with respect to $u=\delta$). The green dashed line indicates the threshold $\eta=\sqrt{N_s}$ far below which the estimate of the mean transfer function is converged. Gain (iii) and phase (iv) of mean frequency response estimate $\langle G\rangle_{\tau_0,N_s}$ (solid blue) and frequency response about the mean flow $G_{\overline{\mathbf{U}}}$ (dashed red). Light-grey shading indicated frequency ranges where $1< \eta <\sqrt{N_s}$ and dark-grey shading corresponds to $\eta\geq \sqrt{N_s}$. In the phase plots, the two black-dotted curves indicate a shift of $+/-\pi$ with respect to the phase of $G_{\overline{\mathbf{U}}}$. For rows (ii)-(iii)-(iv), all quantities are evaluated on the shifted imaginary axis $\sigma+\mathrm{i}\omega$, with $\sigma=0.01$.}}
    \label{fig:res_pinball110}
\end{figure}

\begin{figure}
    \centering
    \includegraphics[width=1.0\textwidth]{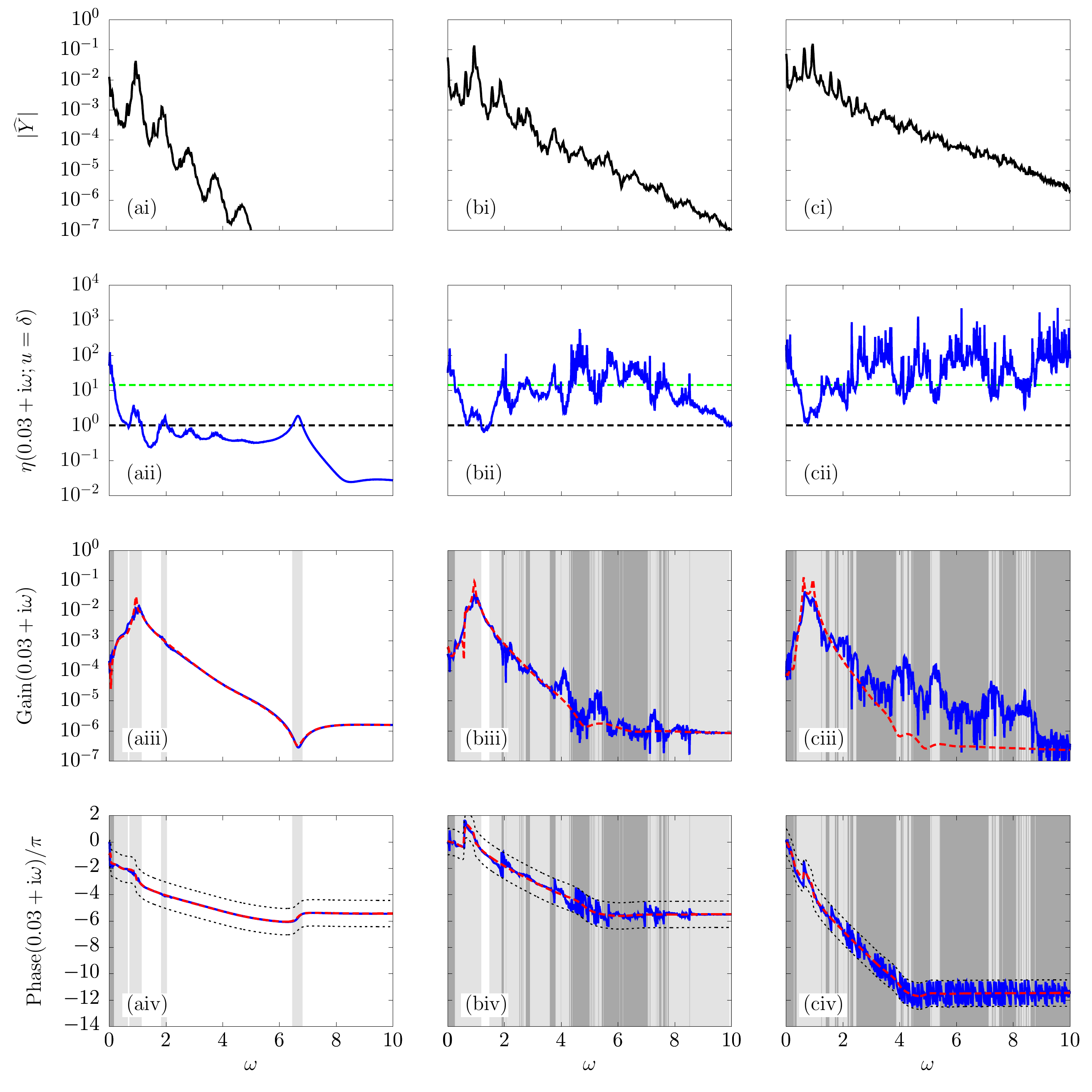}
    \caption{Same caption as figure \ref{fig:res_pinball110} for the chaotic regime at $Re=120$, with $N_s=204$ samples and $\sigma=0.03$.}
    \label{fig:res_pinball120}
\end{figure}

 Fourier spectra are shown in row (i) of figures \ref{fig:res_pinball110} and \ref{fig:res_pinball120} for each flow at each probe location (a,b,c). These confirm the qualitatively different dynamics of the flows: the spectrum is discrete in the quasiperiodic case, while it is continuous in the chaotic case, despite the apparent presence of peaks. In the latter case, they possess an intrinsic non-zero bandwidth, unlike in the periodic and quasi-periodic cases, where the `thickness' of the peak only arises from the estimation error of the discrete Fourier transform. It is not directly possible to infer the number of basic incommensurate frequencies in the quasiperiodic case by examining the spectra, but there clearly is energy in a much greater number of frequencies compared to the periodic case (see figure \ref{fig:spectra_BF}(i) for comparison).
 
\subsubsection{Procedure}
In both the quasi-periodic and chaotic cases, $N_s=204$ impulse responses are computed over more than 1000 convective time units, each starting 2 convective time units after the preceeding one, i.e. $\Delta \tau_0=2$. We assumed that this procedure allowed to sample a representative part of these attractors with a single realization of $\widetilde{\mathbf{U}}(\tau)$. We recall that $\sigma>\mathrm{MLE}$ is necessary for convergence of the Laplace transform of the linear impulse response. In the quasi-periodic case, the shifted frequency response is evaluated for $\sigma=0.01$ since $\mathrm{MLE}=0$. For the chaotic case, we choose $\sigma=0.03$ since $\mathrm{MLE}=0.02$ (see appendix \ref{sec:MLE}). 

\subsubsection{Results}
{Results are reported respectively in figures \ref{fig:res_pinball110} and \ref{fig:res_pinball120} for $Re=110$ and 120, and they are qualitatively very similar to the periodic case in figure \ref{fig:spectra_BF}. The ratio $\eta$ increases as the probe moves downstream, and as a consequence the assumption of time-invariant input-output dynamics is quite poor for probes (b) and (c), whereas it is quite good for probe (a). Unlike for $Re=100$, the ratio $\eta$ is never smaller than 1 for probe (c) at $Re=110$ and 120, even when the mean estimate is converged. The convergence of the mean estimate is also slower downstream than upstream, and the number of samples $N_s=204$ is insufficient at almost all frequencies for probe (c) using impulsive forcings $u(t)=\delta(t)$.}

{Good overall agreement between the Bode diagrams of $G_{\overline{\mathbf{U}}}$ and $\langle G\rangle_{\tau_0}$ is observed for the two flow regimes. For probe (a), the Bode diagrams of the two transfer functions are nearly undistinguishable from one another in both cases, except in the vicinity of some energetic peaks in the power spectrum of the base flow (see row (i)). For some high enough frequencies $\omega>4$, there are also noticeable differences between the two transfer functions at probes (b) and (c), even when $\eta<\sqrt{N_s}$. However, mean estimate convergence is only ensured when $\eta\ll\sqrt{N_s}$ so it is difficult to conclude on these deviations.}


In the quasiperiodic case, we notice the presence of resonance peaks in the mean transfer functions, at natural frequencies of the unsteady base flow, which are not visible in $G_{\overline{\mathbf{U}}}$. Resonances at high frequencies are not visible in the near wake and only manifest at sensors {(b)} and {(c)} in the far wake. The observations made for the quasi-periodic case are similar to the periodic case and the same justification may be provided in both cases (see appendix \ref{subsec:QP} for an extension of the theory developed in \S \ref{sec:theory} to the quasiperiodic case). It is interesting to note that the close agreement between $\langle G\rangle_{\tau_0}$ and $G_{\overline{\mathbf{U}}}$ carries over to the chaotic regime, even though we do not have a theory for base flows with a continuous power spectrum.

 \subsection{Stochastic flow: backward-facing step at $Re=500$}\label{sec:BFS}
 
 {The numerical setup and code for the nonlinear flow are identical to that of \cite{HE12,SI16}. The mesh has $63,902$ triangles, $32,792$ vertices and $N=258,970$ velocity degrees of freedom. The length unit corresponds to the height $h$ of the step, while the (convective) time unit corresponds to $h/U_\infty$. The time step is $\mathrm{d}t=0.002$. The main difference with the fluidic pinball is that the steady base flow is linearly stable hence requires constant forcing in order to reach a statistically stationary unsteady state. The forcing is chosen to be of the form $\mathbf{F}(t)=\mathbf{B}_w w(t)$ with $w$ a Gaussian-white noise and $\mathbf{B}_w$ a discretized Gaussian volume force $\mathcal{B}(x,y;x_0,y_0,\sigma_x,\sigma_y)$ with $(x_0,y_0)=(-0.5,0.25)$ and $\sigma_x=\sigma_y=0.1$ (see equation (\ref{eq:defB}) for the definition of $\mathcal{B}$ and white squares centered at $(x_0,y_0)$ in figures \ref{fig:bfs} (a,b)). Two noise variances are considered: $\sigma_w=\sqrt{10}$ and $\sigma_w=10$, respectively representing weakly and strongly nonlinear dynamics as in \cite{HE12}.} 
 
 {Linear perturbations about the time-varying base flow are solved in a similar fashion to the case of the fluidic pinball (see appendix \ref{sec:numdetails}). Note that the forcing $\mathbf{F}$ produces the nonlinear base flow $\widetilde{\mathbf{U}}$ and should not be confused with $\mathbf{f}=\mathbf{B}u$, which produces the linear response $\mathbf{u}$ about $\widetilde{\mathbf{U}}$.} The actuator $\mathbf{B}$ is a discretized version of $\mathcal{B}$ with $(x_0,y_0)=(-0.05,0.01)$ and $(\sigma_x,\sigma_y)=(0.01,0.1)$ (see white triangles in figures \ref{fig:bfs} (a,b)). We again use three probes to monitor the flow response: two $y$-velocity probes in the shear layer at respectively $(2.5,0)$ and $(5,0)$, and a friction sensor at $12.5\leq x\leq 12.9$ on the lower wall, behind reattachment (see white dots and rectangle in figures \ref{fig:bfs}(a,b)). The perturbation and base flow measurements will also be denoted $y=\mathbf{C}^T\mathbf{u}$ and $Y=\mathbf{C}^T\mathbf{U}$ with index `a',`b' or `c' when necessary.
 
 \subsubsection{Unsteady base flow}
\begin{figure}
    (a)\includegraphics[width=0.9\textwidth]{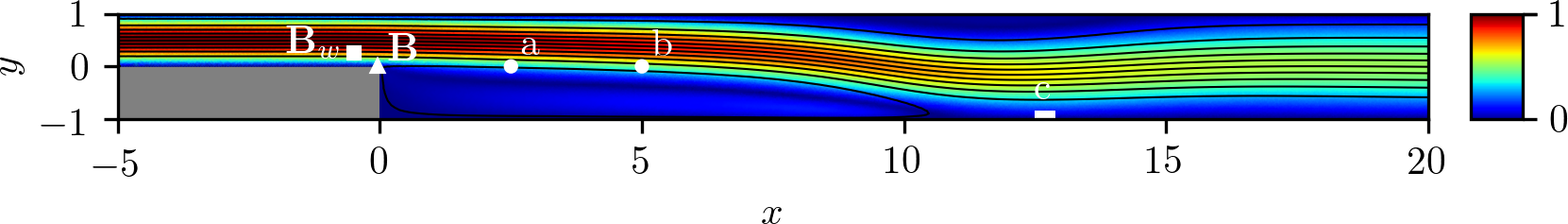}\\
    (b)\includegraphics[width=0.9\textwidth]{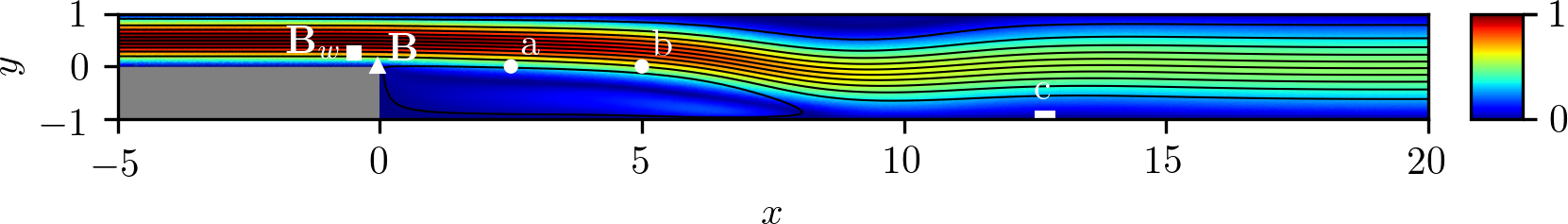}\\
    (c)\includegraphics[width=0.96\textwidth]{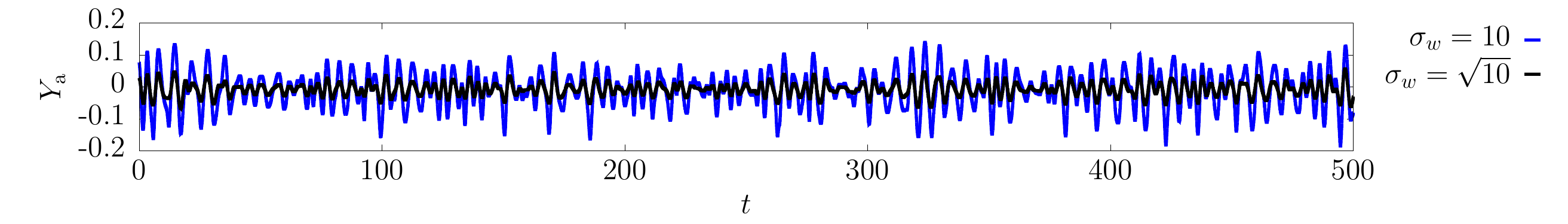}
    \caption{Isocontours of mean velocity norm and streamlines for the backward-facing step flow at $Re=500$: (a) $\sigma_w=\sqrt{10}$, (b) $\sigma_w=10$, where $\sigma_w$ is the rms of the amplitude $w(t)$ multiplying the external noise field $\mathbf{B}_w$ (white square), such that $\mathbf{F}(t)=\mathbf{B}_w w(t)$. The actuator is $\mathbf{B}$ (white triangle), $y$-velocity probes are placed at locations `a' and `b' in the shear layer (white circles), while a friction measurement (white rectangle) is taken at `c'. (c) Timeseries of $Y_\mathrm{a}$ for $\sigma_w=\sqrt{10}$ (dashed black line) and $\sigma_w=10$ (blue solid line).}
    \label{fig:bfs}
\end{figure}

 The effect of the noise variance on the mean flow is quite visible in figure \ref{fig:bfs}(a) for $\sigma_w=\sqrt{10}$ and \ref{fig:bfs}(b) for $\sigma_w=10$. For stronger perturbations, the recirculation region shrinks, with a reattachement point at $x=10.7$ in the first case and $x=8.3$ in the second case. Timeseries of sensor (a) are also shown in figure \ref{fig:bfs}(c): the signals are clearly stochastic and the fluctuation amplitude is consistent with $\sigma_w$. The Fourier spectra are shown in row (i) of figures \ref{fig:res_steplow} and \ref{fig:res_stephigh} for the two values of $\sigma_w$ and the three probes {(a,b,c)}. The spectra are broadband and there is more perturbation energy at higher $\sigma_w$, as expected. Also, perturbation energy increases further downstream, from sensor (a) to (c). At sensor (a), the spectrum for $\sigma_w=10$ is roughly equal to that for $\sigma_w=\sqrt{10}$, multiplied by a frequency-independent factor of $\sqrt{10}$, which is indicative of quasi-linear behaviour. The signature of strong nonlinear effects is visible downstream, as the spectra for both values of $\sigma_w$ no longer have the same overall shape: saturation occurs around $\omega \approx 1$, where linear amplification mechanisms dominate.
 
 \subsubsection{Procedure}
 Again, we use a single realization of $\widetilde{\mathbf{U}}(\tau)$ corresponding to a single random timeseries $w(\tau)$ but vary the relative time $\tau_0$ of linear impulsive forcing to obtain various impulse responses. The nonlinear flow is initialized with the steady base solution (fixed point), and after a transient of more than 150 time units, it is considered statistically stationary and linear impulses are then performed. The time lag $\Delta \tau_0$ between two impulses is of 2 convective time units, as in the quasiperiodic and chaotic flows. We run $N_s=308$ linearized impulse responses, over 220 time units only. Indeed, while impulse responses never decay for the fluidic pinball, since $\mathrm{MLE}\geq 0$, they do for the backward-facing step flow since $\text{MLE}<0$ (see appendix \ref{sec:MLE}) and reach negligible amplitude over that period of time, making shorter integration possible. Since $\mathrm{MLE}<0$, the Laplace transform may be evaluated directly on the imaginary axis.

\subsubsection{Results}
\begin{figure}
    \includegraphics[width=1.0\textwidth]{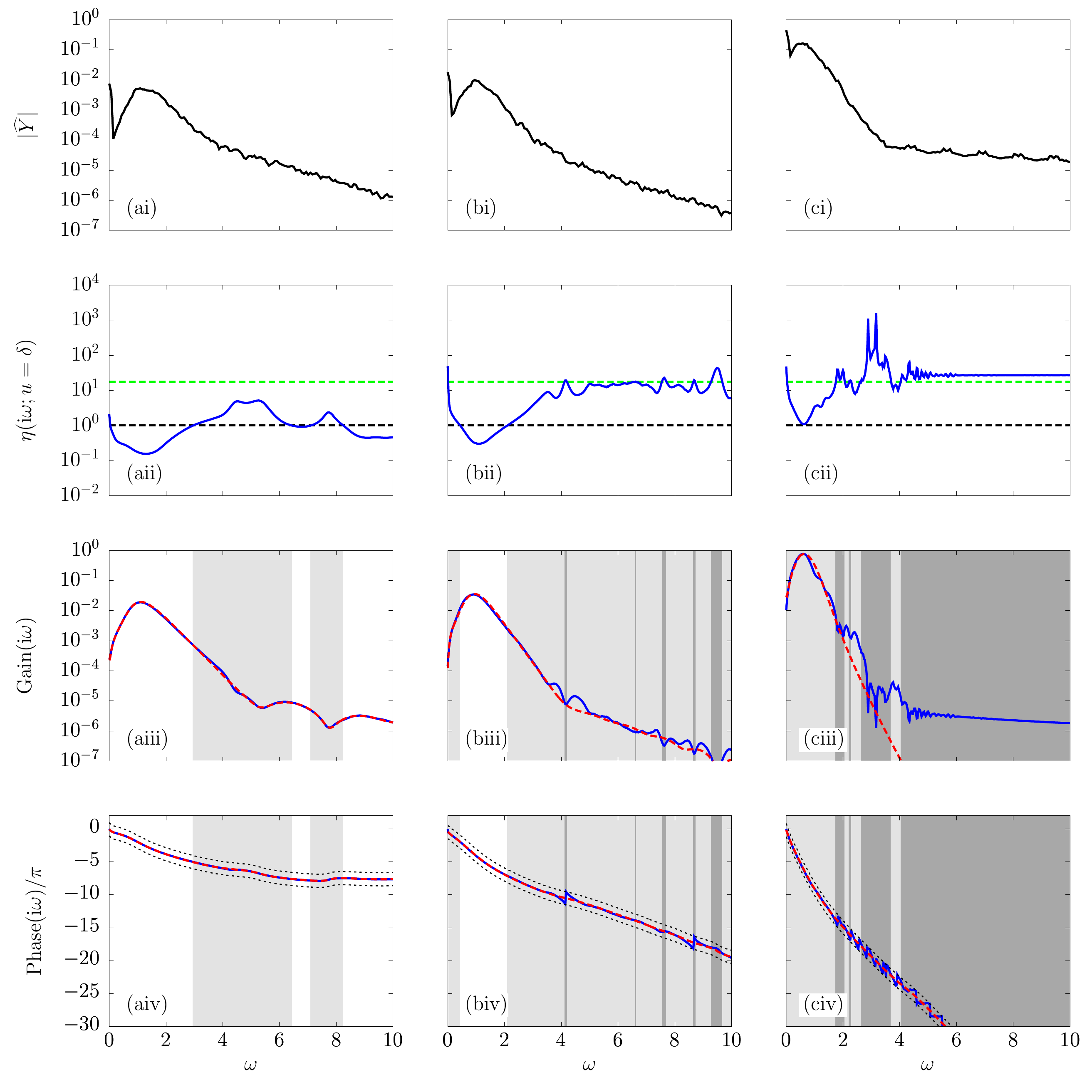}
    \caption{Same caption as figure \ref{fig:spectra_BF} for the backward-facing step flow at $\sigma_w=\sqrt{10}$ (low-noise case), with $N_s=308$ samples. Unlike for the fluidic pinball, the frequency response is on the imaginary axis, with no shift since $\mathrm{MLE}<0$. The continuous power spectrum is obtained by averaging over Hann windows of 86 convective times each, with $50\%$ overlap, for a total signal length of 860 time units.}
    \label{fig:res_steplow}
\end{figure}

\begin{figure}
    \includegraphics[width=1.0\textwidth]{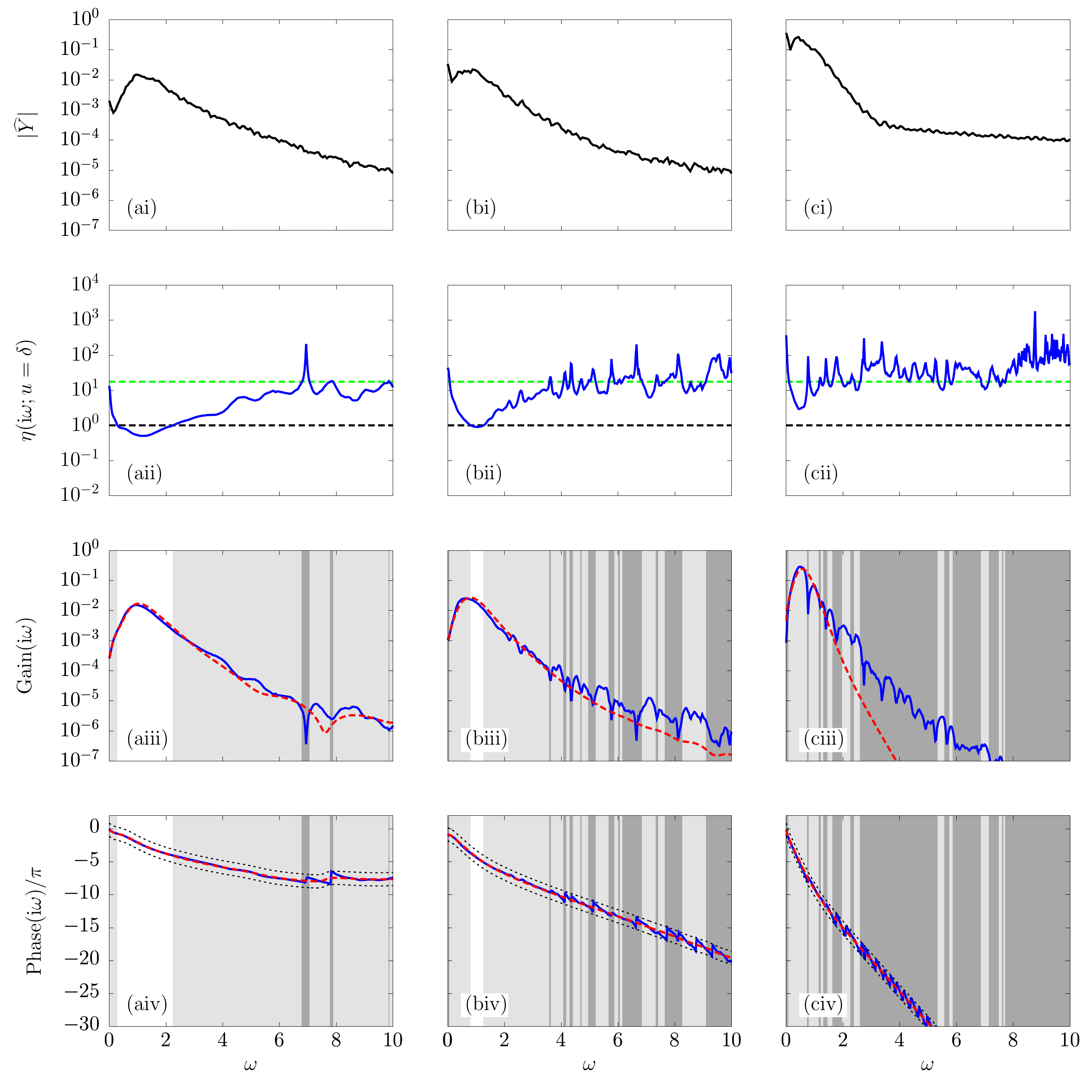}
    \caption{Same caption as figure \ref{fig:res_steplow} for the high-noise case $\sigma_w=10$, with $N_s=308$ samples.}
    \label{fig:res_stephigh}
\end{figure}

{Results for the stochastic case are presented in figures \ref{fig:res_steplow} and \ref{fig:res_stephigh}, respectively corresponding to low $\sigma_w=\sqrt{10}$ and high $\sigma_w=10$ noise variance. Interestingly, the results are qualitatively very similar to the fluidic pinball, even though the nature of the dynamics is quite different. The main difference is that there are no longer sharp variations in the gain and phase at very specific frequencies, in accordance with the absence of resonance peak in the base flow spectrum (see row (i)).}

{The value of $\eta$ increases downstream, slowing down the convergence of the mean estimate and progressively invalidating the time-invariant approximation. This is correlated to spatial perturbation growth in the streamwise direction. Similarly, increasing the noise variance $\sigma_w$ has a negative impact on $\eta$, as unsteady perturbations become stronger. For probes (a) and (b), there is a region where the time-invariant approximation is satisfactory, i.e. $\eta<1$, which is located around the maximum gain, for both values of $\sigma_w$. The frequency of maximum linear amplification of $u$ coincides with that of maximum nonlinear amplification of the white noise $w$ (see base flow power spectrum in row (i)).}



\subsection{Strongly compressible flows}\label{subsec:compress}
It was recently pointed out by \cite{KA20} that resolvent analysis about the mean flow is ill-posed in the case of strongly compressible flows, where large discrepancies in the optimal gains are observed depending on the formulation of the Navier--Stokes equations, whether in primitive or conservative variables. The inconsistency was attributed to the fact that the mean flow eigenvalues are not conserved under a nonlinear coordinate change. Numerical experiments have not been carried out here for compressible cases, but we wish to highlight the potential of the mean resolvent to resolve this issue, since we showed that the poles of the operator correspond to Floquet exponents (and Koopman eigenvalues) of the system in the LTP case (see \S \ref{subsec:Floquet} and \S \ref{subsec:Koopman}), which are invariant under a nonlinear coordinate change (see appendix \ref{sec:invariance}). 

Another point we want to stress is the assumption (\ref{eq:JbarUbar}) we made early on in \S \ref{subsec:connection} that the mean Jacobian is equal to the Jacobian operator about the mean flow. This is no longer true for cubic nonlinearities found in the compressible Navier--Stokes equations written either in primitive or conservative formulation. However, we may still use expansion (\ref{eq:R00})--(\ref{eq:R03}) if we replace $\mathsfbi{R}_{\overline{\mathbf{U}}}$ with
\begin{equation}
    \mathsfbi{R}_{\overline{\mathsfbi{J}}}(s)=(s\mathsfbi{I}-\overline{\mathsfbi{J}})^{-1},
\end{equation}
which we may call the resolvent operator about the mean Jacobian. Alternatively, a quadratic formulation of the compressible Navier--Stokes equations may be used \citep{VI98,IO00}, even though this form may not easily handle shock discontinuities due to its non-conservative form.

\section{Conclusions and outlook}\label{sec:concl}
{This paper is concerned with the definition of a time-invariant operator best characterizing linear input-output behaviour in statistically steady flows. Rather than making the \textit{ad hoc} assumption that the governing equations should be linearized about a time-invariant mean flow, we force the time-varying tangent system about unsteady trajectories on the attractor and collect the responses to the same input signal for various realizations of the unsteady base flow (this is done by varying the relative time $\tau_0$ at which the input is triggered on a single base flow trajectory). We then consider the ensemble average of responses produced by the given input to obtain a mean transfer function. By considering two-dimensional incompressible configurations, the fluidic pinball and the backward-facing step flow, we investigated four possible dynamical regimes: periodic, quasi-periodic, chaotic and stochastic. Inverting the order between the linearization and averaging steps lead to interesting findings.}

{First of all, our framework allows to quantify the validity of the time-invariant hypothesis for a given input signal and probe, frequency by frequency. The ratio $\eta$ between the standard deviation and the module of the mean quantifies the uncertainty associated with the mean frequency response and may be used for robust controller design. In all flows considered, the frequency response based on $\mathsfbi{R}_{\overline{\mathbf{U}}}$ appears to approximate the mean frequency response very well, but the two objects are not identical. In particular, for the periodic case, extra resonance peaks at exact multiples of the fundamental frequency(ies) are visible in the mean transfer function.} 

{We were able to explain these observations in the periodic and quasiperiodic cases, by averaging the harmonic transfer operator \citep{WE90,WE91,PA20,franceschini2022identification} with respect to the relative initial time of forcing $\tau_0$, or simply phase $\omega_0 \tau_0=\phi\,\text{mod}\,2\pi$ in the periodic case.
Mean transfer functions based on the resulting \textit{mean resolvent operator} $\mathsfbi{R}_0$, may be identified from input-output data without the need for averaging over several input realizations if harmonic forcings are used, even though the linearized system is not time-invariant. This justifies the principle of `dynamic linearity' put forward by \cite{DA12,DA17,EV17} in the context of flow control.}

{We then showed that the poles of $\mathsfbi{R}_0$ correspond to (quasi-)Floquet exponents of the LT(Q)P system, which may also be interpreted as Koopman eigenvalues. As a result, these poles are invariant under a nonlinear coordinate change, i.e. do not depend on an arbitrary choice of formulation for the Navier--Stokes equation. This strong property may help solve the ambiguity of classical resolvent analysis about a mean flow recently noted by \cite{KA20}, which is particularly noticeable in strongly compressible flows. There is also a set of marginally stable poles at multiples of the fundamental frequency, which causes the resonance peaks in the mean transfer functions. This so-called `RZIF' property \citep{TU15}, which is only approximate when considering the poles of $\mathsfbi{R}_{\overline{\mathbf{U}}}$, appears to be exact when considering the poles of $\mathsfbi{R}_0$.}

{Next, we further investigated the connection between $\mathsfbi{R}_{\overline{\mathbf{U}}}$ and $\mathsfbi{R}_0$ and found that the former operator approximates the latter in the incompressible case (the link is lost for cubic nonlinearities) within the \textit{weakly unsteady limit}, where amplification by the unsteady part of the base flow is small compared to amplification due to the mean flow. The relative difference between the two operators also vanishes for large positive growth rates (and the absolute difference vanishes for large frequencies). There are however two key differences between the two operators. One is that $\mathsfbi{R}_0$ indirectly incorporates information about the endogenous nonlinear forcing term of \cite{MC10}, which are not taken into account by $\mathsfbi{R}_{\overline{\mathbf{U}}}$, unless turbulent viscosity \citep{MO19,PI21} or state-feedback \citep{ZA17} is introduced. The missing information is embedded in the fluctuating part of the Jacobian, which is aware of the nonlinear equilibrium between the various Fourier components of the base flow. The other difference is that $\mathsfbi{R}_0$ has an infinite-dimensional internal state, while $\mathsfbi{R}_{\overline{\mathbf{U}}}$ has a finite-dimensional one. However, only a finite number of poles have a significant contribution to $\mathsfbi{R}_0$, explaining why a finite-dimensional approximation of this operator, in terms of internal state, is possible.}

{The definition of a new resolvent operator $\mathsfbi{R}_0$ is intended to extend that of the usual resolvent operator $\mathsfbi{R}_{\overline{\mathbf{U}}}$ about the mean flow, therefore, any analysis which may be done with $\mathsfbi{R}_{\overline{\mathbf{U}}}$ (reduced-order modelling, data-assimilation of second-order statistics, input-output analysis, flow control, etc.) may, in principle, also be done with (improved) $\mathsfbi{R}_0$, using yet to be defined numerical methods. For instance, a compelling prospect would be to compute the singular modes of $\mathsfbi{R}_0$ and compare them with the spectral POD modes to see if they align better than the singular modes of $\mathsfbi{R}_{\overline{\mathbf{U}}}$. It is hoped that the new operator may help understand and overcome the observed shortcomings of the classical resolvent operator, i.e. modelling of nonlinear forcings and compressibility effects. However, it is important to note that the present paper does not directly address turbulent nor compressible flows, and future work needs to be done to confirm the relevance of $\mathsfbi{R}_0$ to these flows.} 

{From a theoretical standpoint, an interesting perspective would be to extend the present formalism to the case of continuous or mixed spectra characteristic of chaotic, stochastic and turbulent flows. Koopman operator theory may be key to this endeavour \citep{VC19}. From a methodological standpoint, a key question would be to find an optimal way to compress the infinite state of the mean resolvent operator into a finite one. Since the mean resolvent is related to the mean Koopman operator, this problem may perhaps be tackled using extensions of the DMD method \citep{SC10,WI15,HE21}. Recent papers enhancing the predictive power of $\mathsfbi{R}_{\overline{\mathbf{U}}}$ using optimization techniques \citep{ZA17,PI21} may possibly bear connections with this compression problem.}

\backsection[Declaration of interests]{The authors report no conflict of interest.}

\backsection[Author ORCIDs]{Colin Leclercq, https://orcid.org/0000-0002-8262-0697; Denis Sipp, https://orcid.org/0000-0002-2808-3886}

\appendix

\section{Fluidic pinball: discretization and numerical methods} \label{sec:numdetails}

As in \cite{DE20}, the computational domain extends over $-6\leq x\leq 20$ and $-6\leq y\leq 6$, with the same Dirichlet boundary condition $\widetilde{\mathbf{U}}=(1,0)$ at the inflow $x=-6$, `top' $y=6$ and `bottom' $y=-6$ boundary conditions. A standard outflow boundary condition $\widetilde{P}\mathbf{n}-Re^{-1}\nabla \widetilde{\mathbf{U}}\cdot \mathbf{n}=\mathbf{0}$ is used at $x=20$, which is similar to the stress-free boundary condition of \cite{DE20}. The boundary condition on the cylinders is obviously no-slip and impermeability, i.e. $\widetilde{\mathbf{U}}=\mathbf{0}$. The freely available software FreeFem++ \citep{MR3043640} is used to time-march the incompressible Navier--Stokes equations discretized on Taylor--Hood finite elements; $P2$ for velocity components and forcings, and $P1$ for pressure. The mesh comprises $66,668$ triangles, $33,739$ vertices and the total number of velocity degrees of freedom, including both components, is $N=268,296$. 

\begin{figure}
    \centering
    \includegraphics[width=1.0\textwidth]{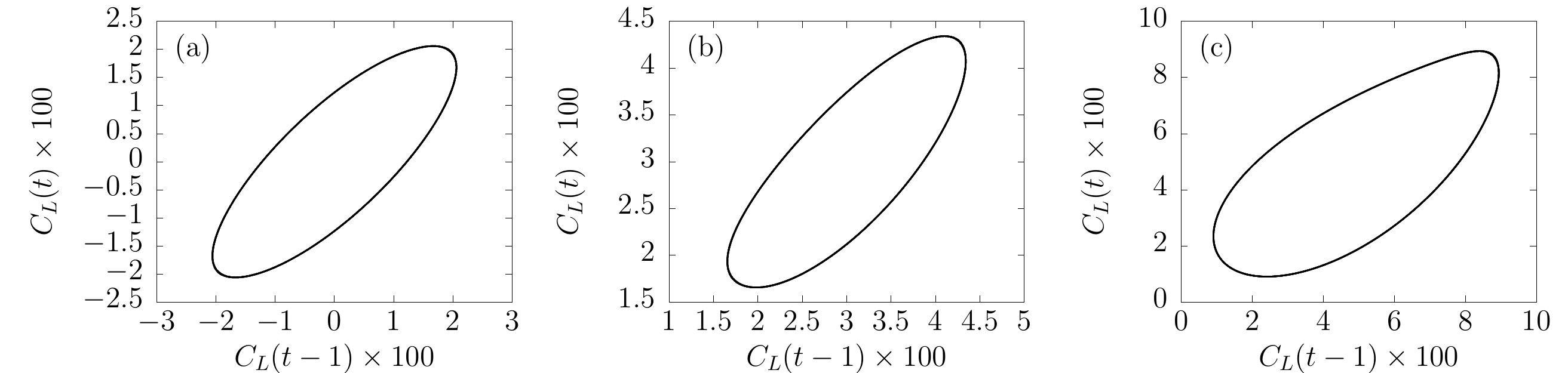}
    \caption{Time-delay embedded plot of lift coefficient for the periodic regimes at (a) $Re=68$, (b) $Re=75$ and (c) $Re=100$.}
    \label{fig:validation_Cl}
\end{figure}

In order to validate spatial resolution, we produce 4 mesh $M1-M4$ of variable refinement, and compute the lift and drag coefficients for the {asymmetric} steady base flow at $Re=75$ and $Re=100$. The coefficients are based on the total force by unit length exerted on all three cylinders normalized by $1/2\rho U_\infty^2 D$, where $\rho$ denotes density. The results are given in table \ref{tab:resol} and justify our choice of mesh M2 for the present study. Note that the values provided here do not match with those of \cite{DE20} as the authors did not include the viscous contribution to the forces (private communication). For completeness, we provide in figure \ref{fig:validation_Cl} the temporal evolution of the lift coefficient in the periodic regimes at $Re=68,75,100$ analyzed by \cite{DE20}, using the usual definition of the coefficients. This may be useful for validation purposes in future studies on this configuration.

\begin{figure}
    \centering
    \includegraphics[width=1.\textwidth]{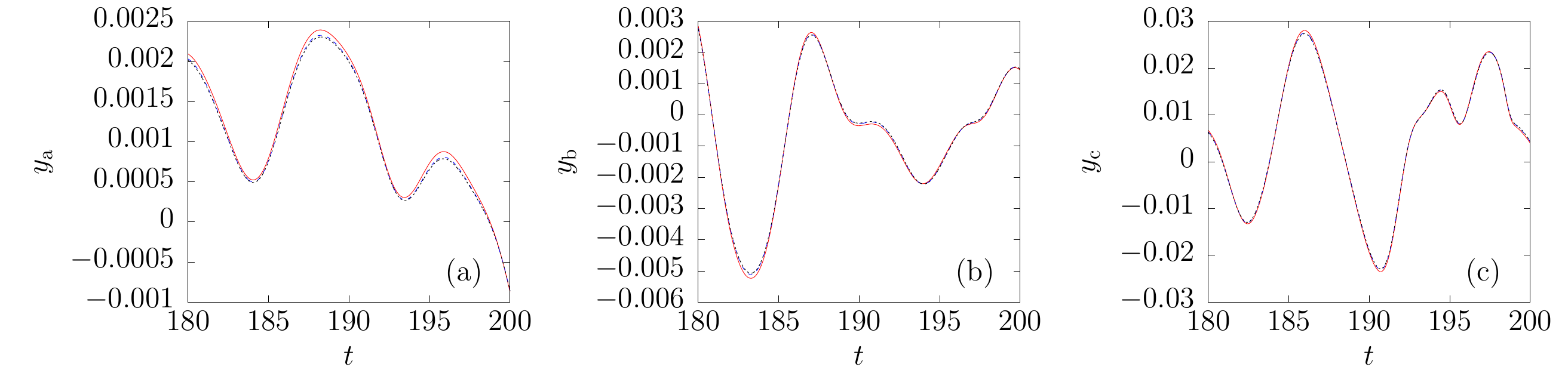}
    \caption{Realization of linear impulse response about quasiperiodic flow at $Re=110$ for $\mathrm{d}t=0.01$ (solid red), $\mathrm{d}t=0.005$ (dashed blue) and $\mathrm{d}t=0.0025$ (dotted black), on all 3 probes (a,b,c).}
    \label{fig:validationdt}
\end{figure}

The solvers are based on a sequential open-source code without time-splitting ($\mathrm{https://github.com/denissipp/AMR\_Sipp\_Schmid\_2016}$; \cite{SI16}). Code parallelization was found unnecessary for this study, which requires multiple realizations of impulsive forcings which can be run in parallel indeed. The nonlinear problem is solved in perturbative form with respect to the time-invariant base flow $\widetilde{\mathbf{U}}_b$ (i.e. fixed-point), after computing the latter using a Newton method. The time-stepping is semi-implicit: the diffusion term and interaction of the perturbation $\widetilde{\mathbf{U}}'':=\widetilde{\mathbf{U}}-\widetilde{\mathbf{U}}_b$ with $\widetilde{\mathbf{U}}_b$ are implicit, while the interaction of $\widetilde{\mathbf{U}}''$ with itself is explicit (Adams--Bashforth). The linear non-autonomous problem about the time-varying base flow is also solved semi-implicitely: the diffusion term and interaction of the perturbation $\mathbf{u}$ with $\widetilde{\mathbf{U}}_b$ are implicit, while the interaction of $\mathbf{u}$ with $\widetilde{\mathbf{U}}''$ is explicit (Adams--Bashforth as well). The time-stepping scheme is second-order for both problems, however we found the latter to be more sensitive to step size than the former. Temporal resolution tests were carried out for linear impulses about the quasiperiodic flow at $Re=110$. Figure \ref{fig:validationdt} shows a portion of an impulse response for three values of $\mathrm{d}t=0.01,0.05,0.0025$, indicating convergence for the lower value of $\mathrm{d}t=0.0025$ on all three probes. Finally, the linearized time-stepping code was validated by computing the impulse response about the time-averaged mean flow. The corresponding `frequency response realization' (shifted in the right-half plane, as specified in \S \ref{sec:procedure}) was obtained by taking the Laplace transform of the impulse response and then compared with frequential samples computed using the resolvent operator about the mean flow (on the same shifted axis).

\begin{table}
\centering
\begin{tabular}{lcccc}
&   $M1$ &    $M2$ &   $M3$ &   $M4$\\\hline
Nb. of triangles & 19628 & 66668 & 78538 & 102965\\
Nb. of vertices & 10041 & 33739 & 39725 & 52005\\\hline
$C_L$ {asymmetric} base flow $Re=75$  &  0.0383305  &  0.0381093 &   0.0381902  &  0.0381794\\
$C_L$ {asymmetric} base flow $Re=100$  &  0.0610811  &  0.0608825 &   0.0609652 &   0.0609484\\\hline
$C_D$ {asymmetric} base flow $Re=75$  &  3.91464  & 3.91629  &   3.91598  &  3.91634\\
$C_D$ {asymmetric} base flow $Re=100$  &  3.66636  &  3.66776 &   3.66721 & 3.66758   
\end{tabular}
\caption{Resolution tests on lift and drag coefficients $C_L$ and $C_D$ for asymmetric base flow at $Re=75$ and $Re=100$.}\label{tab:resol}
\end{table}

\section{Marginal Floquet exponents for self-sustained periodic base flow}\label{sec:appendix}
We find it useful to include a proof of this classical result. The periodic base flow $\mathbf{U}$ is a solution of the unforced discretized incompressible Navier--Stokes equations projected onto the space of solenoidal velocity fields
\begin{equation}
    \mathrm{d}_t\mathbf{U}=\mathbf{F}(\mathbf{U}),
\end{equation}
and by differentiating with respect to time
\begin{equation}
    \mathrm{d}_t (\mathrm{d}_t\mathbf{U})=\underbrace{\mathrm{\nabla}_{\mathbf{U}(t)}\mathbf{F}}_{\mathsfbi{J}(t)}\mathrm{d}_t\mathbf{U}.
\end{equation}
Since $\mathrm{d}_t\mathbf{U}$ is a solution of the unforced LTP system, its evolution from $t'$ to $t$ is governed by the propagator 
\begin{align}
    \mathrm{d}_t\mathbf{U}(t)&=\mathbf{\Phi}(t,t')\mathrm{d}_t\mathbf{U}(t'),\nonumber\\
    &=\mathsfbi{V}(t)\mathrm{e}^{\mathbf{\Lambda}(t-t')}\mathsfbi{W}^H(t')\mathrm{d}_t\mathbf{U}(t').
\end{align}
For $t'=t-T$, using the periodicity of $\mathrm{d}_t\mathbf{U}(t)$ and the Floquet modes, we have
\begin{equation}
    (\mathsfbi{I}-\mathrm{e}^{\mathbf{\Lambda} t})\mathsfbi{W}^H(t)\mathrm{d}_t\mathbf{U}(t)=0
\end{equation}
Clearly, $\mathrm{d}_t\mathbf{U}(t)$ is a Floquet mode associated with the fundamental Floquet exponent $0$ (and copies $\mathrm{i}j\omega_0$ in complementary strips). Since the periodic base flow is stable and Floquet exponents are ranked in decreasing order of growth rate, we have $\lambda_1=0$ and $\mathbf{v}^1=\mathrm{d}_t\mathbf{U}$. Note that if the base flow is an unstable periodic orbit of the unforced system, then there is still a zero Floquet exponent, but it will no longer be the leading one since the maximum Lyapunov exponent has to be strictly positive.

\section{Convergence of operator expansions}\label{sec:Neumann}
{In this appendix, we consider operators acting on continuous functions of space and replace fonts accordingly; for instance, the Jacobian matrix $\widehat{\mathsfbi{J}}_i'$ is replaced by the operator $\widehat{\mathcal{J}}_i'$. For simplicity, we will consider the specific case of spatially periodic $\mathcal{C}^\infty$ functions in an infinite one-dimensional domain with period $L$, such that 
\begin{align}
    \mathcal{R}_{\overline{U}}(s)&=\left[s\mathcal{I}-(-\overline{U}\partial_x+Re^{-1}\partial^2_{xx})\right]^{-1},\\
    \widehat{\mathcal{J}}'_i&=-\widehat{U}_i'\partial_x,
\end{align}
where $\overline{U}$ and $\widehat{U}_i'$ are scalar constants (homogeneous base flow along $x$) and $\mathcal{I}$ is the identity. The fundamental wavenumber $2\pi/L$ is denoted $k_L$. In this functional space, Fourier modes $\{e_n:x\mapsto \mathrm{e}^{\mathrm{i}n k_L x},n\in\mathbb{Z}\}$ form an orthonormal basis for the canonical inner product $(a,b)_{\mathrm{H}}:=\frac{1}{L}\int_{0}^L a^*(x)b(x)\,\mathrm{d}x$.} 
\subsection{Harmonic transfer operator}\label{subsec:harmres}
{The continuous-in-space-version of series (\ref{eq:expH}) converges if and only if the spectral radius $\rho(\underline{\mathcal{L}})$ of the operator $ \underline{\mathcal{L}}(s;\phi):=\underline{\mathcal{D}}^{-1}(s)\underline{\mathcal{T}}(\phi)$ is strictly less than one \citep{SU76}. The operator $\underline{\mathcal{L}}$ is an infinite-matrix of operators $\underline{\mathcal{L}}=(\underline{\mathcal{L}}_{jl})_{j,l\in\mathbb{Z}}$, where $\underline{\mathcal{L}}_{jl}=\mathcal{R}_{\overline{U}}(s+\mathrm{i}j\omega_0)\widehat{\mathcal{J}}'_{j-l}\mathrm{e}^{\mathrm{i}(j-l)\phi}$. Introduce $\underline{e}^n$, the infinite-diagonal-matrix of operators such that $\underline{e}_{j,l}^n=e_n\delta_{jl}$. The columns of this matrix are orthonormal with respect to the inner product $[\underline{a},\underline{b}]=\sum_j (\underline{a}_j,\underline{b}_j)_\mathrm{H}$. The operator $\underline{\mathcal{L}}$ is said to be block-diagonal because it leaves invariant the subspaces $V_n$ generated by the columns of $\underline{e}^n$ (and these subspaces are in a direct sum). More specifically, for every $n$, we may represent the operator $\underline{\mathcal{L}}|_{V_n}$ as an infinite-matrix $\underline{\mathsfbi{L}}^n$, i.e. $\underline{\mathcal{L}}\underline{e}^n=\underline{e}^n\underline{\mathsfbi{L}}^n$, where (for values of $s$ which do not cancel the denominator; it suffices that $\sigma=\mathrm{Im}(s)>0$)
\begin{equation}
    \underline{\mathsfbi{L}}^n_{jl}=-\dfrac{\widehat{U}_{j-l}'\mathrm{e}^{\mathrm{i}(j-l)\phi}\mathrm{i}nk_L}{s+\mathrm{i}j\omega_0+\overline{U}\mathrm{i}nk_L+Re^{-1}n^2k^2_L},
\end{equation}
therefore the operator norm
\begin{equation}
    \|\underline{\mathcal{L}}\|=\sup_n \|\underline{\mathsfbi{L}}^n\|
\end{equation}
(see p. 30 of \cite{CO90}). Moreover $\rho(\underline{\mathcal{L}})\leq \|\underline{\mathcal{L}}\|$ so it is sufficient to show that $\sup_n \|\underline{\mathsfbi{L}}^n\|<1$ in appropriate conditions. For any given $n$, we have $\underline{\mathsfbi{L}}^n=\mathrm{i}nk_L\,\underline{\mathsfbi{r}}^n(s)\underline{\mathsfbi{u}}'$, where
\begin{align}
    \underline{\mathsfbi{r}}^n_{jl}(s)&=(s+\mathrm{i}j\omega_0+\overline{U}\mathrm{i}nk_L+Re^{-1}n^2k^2_L)^{-1}\delta_{jl},\\
    \underline{\mathsfbi{u}}'_{jl}&=\widehat{U}'_{j-l}\mathrm{e}^{\mathrm{i}(j-l)\phi},
\end{align}
are respectively a diagonal matrix and a Laurent matrix of Fourier coefficients of $U'(t;\phi)$. The operator norm being submultiplicative, we have $\|\mathsfbi{L}^n\|\leq |n|k_L\|\underline{\mathsfbi{r}}^n\|\|\underline{\mathsfbi{u}}'\|$. For $\sigma>0$, we have $\|\underline{\mathsfbi{r}}^n(s)\|\leq(\sigma+Re^{-1}n^2k_L^2)^{-1}$. Moreover, $\|\underline{\mathsfbi{u}}'\|\leq \|U'\|_\infty=\max_t |U'(t)|$ (see chapter III in \cite{GO03}), therefore
\begin{align}
    \|{\underline{\mathcal{L}}(s)}\|&\leq k_L\|U'\|_\infty \max_n \dfrac{n}{\sigma+Re^{-1}n^2k_L^2},\\
    &\leq \dfrac{\|U'\|_\infty}{2k_L}\sqrt{\dfrac{Re}{\sigma}}\label{eq:C8},
\end{align}
(evaluating the maximum over the set of real numbers). Interestingly, there are two independent ways to make the norm of $\mathcal{L}(s)$ small: either a) weak base flow unsteadiness $\|U'\|_\infty$ or b) large $\sigma$.}

\begin{figure}
    \centering
    (a)\includegraphics[width=0.45\textwidth]{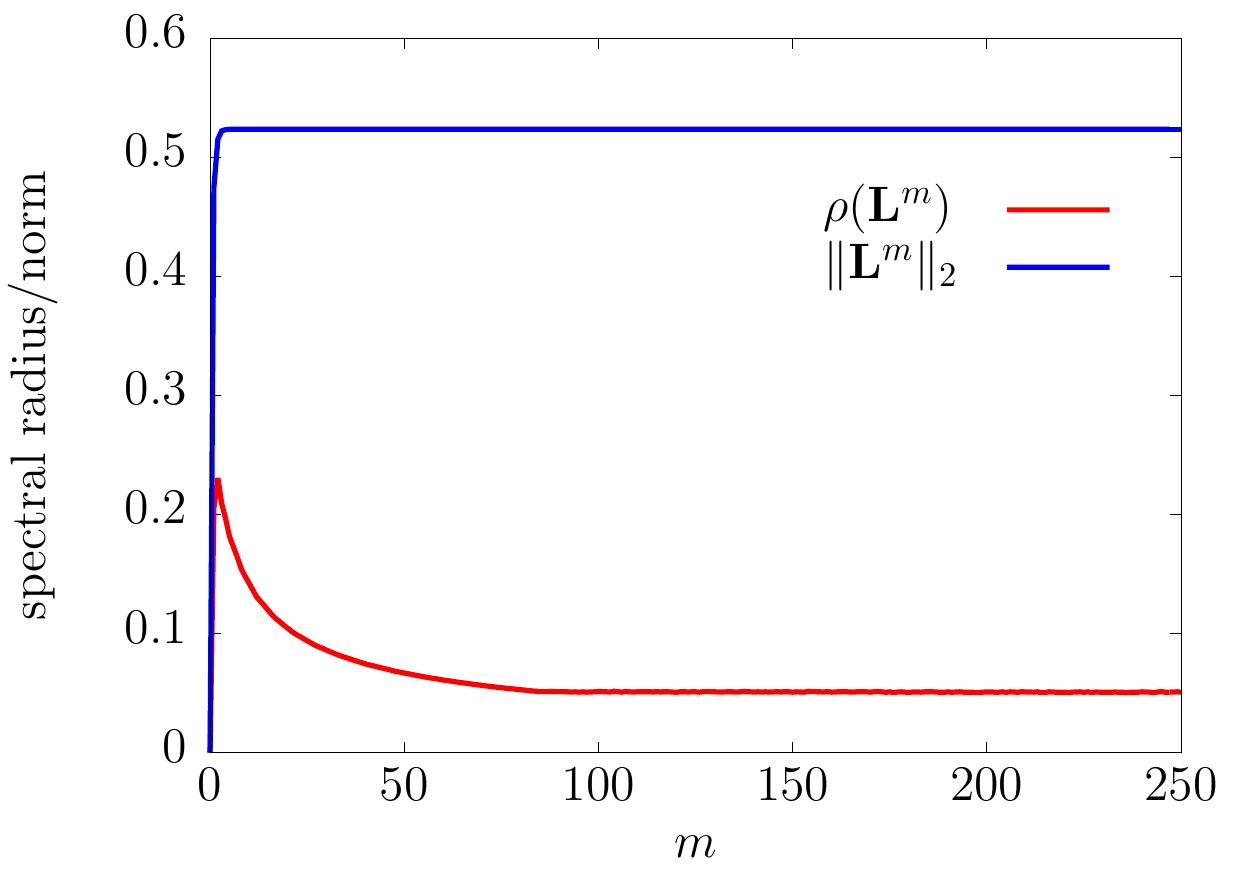} \hfill (b)\includegraphics[width=0.45\textwidth]{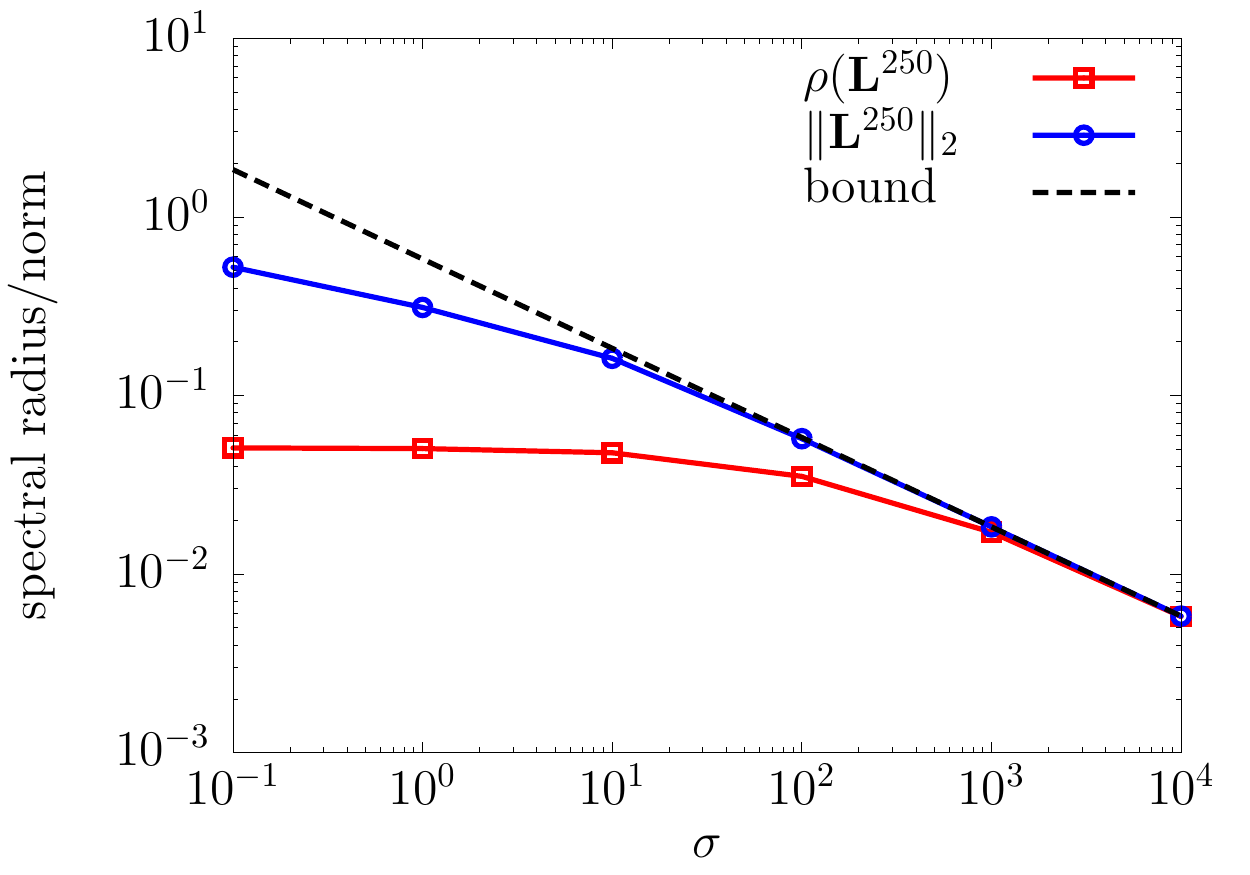}
    \caption{{(a) Evolution of the spectral norm/radius of the finite-dimensional approximation $\mathsfbi{L}^m$ of $\underline{\mathcal{L}}(s)$ with $m$ for $s=0.1+\mathrm{i}$. (b) Evolution of the spectral norm/radius of $\mathsfbi{L}^{250}$ for $s=\sigma+\mathrm{i}$ versus $\sigma$, with bound from equation (\ref{eq:C8}).}}
    \label{fig:convergence}
\end{figure}

{To check the bound (\ref{eq:C8}) numerically, we choose $U'(t;\phi):=2\sum_{j=1}^\infty \mathrm{e}^{-j}\cos[j(\omega_0 t+\phi)]$ such that $\widehat{U}'_j=\mathrm{e}^{-|j|}$ and $\|U'\|_\infty=2/(\mathrm{e}-1)\approx 1.16$. We also choose $Re=k_L=\overline{U}=1$ and $\omega_0=\pi$. For each value of $n$, we define the finite-dimensional approximation $\mathsfbi{L}^{n,m}:=(\underline{\mathsfbi{L}}^n)_{-m\leq i,j\leq m}$ of $\underline{\mathsfbi{L}}^n$. We may then approximate the operator $\underline{\mathcal{L}}$ by the finite block-diagonal matrix $\mathsfbi{L}^{m}:=\mathrm{blkdiag}[\mathsfbi{L}^{-m,m},\dots,\mathsfbi{L}^{0,m},\dots,\mathsfbi{L}^{m,m}]$. For such finite-dimensional matrices, we simply have $\rho(\mathsfbi{L}^m)=\max_n\rho(\mathsfbi{L}^{n,m})$ and $\|\mathsfbi{L}^m\|_2=\max_n\|\mathsfbi{L}^{n,m}\|_2$. The results are plotted in figure \ref{fig:convergence}: in panel (a), we fix $s=0.1+\mathrm{1i}$ and vary $0\leq m \leq 250$ while in panel (b) we fix $m=250$ and $\omega=1$ and vary $-1\leq \log_{10}(\sigma)\leq 4$. Panel (a) confirms that $m=250$ is large enough to converge the spectral radius and norm of $\underline{\mathcal{L}}(s)$ for the $s$ considered (we verified that this the case for all values of $\sigma$ in panel (b) as well). Panel (b) confirms expression (\ref{eq:C8}) for the bound and the decrease of the spectral norm as $O(\sigma^{-1/2})$, which guarantees that the Neumann series (\ref{eq:expH}) converges at any $\omega$ for large enough $\sigma>0$.} 

\subsection{Mean resolvent operator}\label{subsec:meanres}
{Since the mean resolvent is a sub-block of the harmonic transfer operator, the convergence of the Neumann series (\ref{eq:expH}) implies the convergence of expansion (\ref{eq:expansionMR}) for the mean resolvent. Moreover, similarly to (\ref{eq:C8}), it may be shown that for any integer $j$
\begin{equation}
    \|\mathcal{R}_{\overline{U}}\mathcal{J}'_j\|_{\infty,\sigma}\leq \dfrac{|\widehat{U}_j'|}{2k_L}\sqrt{\dfrac{Re}{\sigma}},
\end{equation}
where $\|.\|_{\infty,\sigma}$ is the maximum value of the operator norm over all real frequencies $\omega$, for a given value of $\sigma>0$. Since $U'(t)$ is $\mathcal{C}^\infty$, $\widehat{U}'_j=o(1/j^m)$ for any integer $m$, and in particular for $m\geq 2$, therefore the quantity $\sum_j |\widehat{U}'_j|$ is well-defined and so is $\epsilon_\sigma:=\sum_j\|\mathcal{R}_{\overline{U}}\widehat{\mathcal{J}}'_j\|_{\infty,\sigma}$ (continuous version of definition (\ref{eq:defsig}).} 

{Again, it is interesting to note that there are two independent ways to make $\epsilon_\sigma$ small. One is a) by reducing $\sum_j |\widehat{U}'_j|$, which is a measure of base flow unsteadiness, the other b) is by increasing $\sigma$.} 



\section{Extension of theory to quasiperiodic flows}\label{subsec:QP}

The theory presented in \S \ref{sec:theory} appears to extend to quasi-periodic flows. Indeed, $m$-torii possess $m$ basic incommensurate frequencies $\boldsymbol{\Omega}=[\omega_0,\omega_1,\dots,\omega_{m-1}]^T$, hence their discrete Fourier spectrum is indexed by $\mathbb{Z}^m$ instead of $\mathbb{Z}$. By simply introducing a bijection $\mathbf{b}$ from $\mathbb{Z}$ to $\mathbb{Z}^m$ such that $\mathbf{b}(0)=(0,\dots,0)$, it seems that we can reuse the same formalism as in \S \ref{subsec:HFT}. Using this bijection, the Fourier expansion of the quasi-periodic Jacobian reads
\begin{equation}
    \mathsfbi{J}(t;\tau_0)=\mathsfbi{J}_{\overline{\mathbf{U}}}+\sum_{k} \widehat{\mathsfbi{J}}'_{k}\mathrm{e}^{\mathrm{i}\mathbf{b}(k)\cdot \boldsymbol{\Omega}(t+\tau_0)},\label{eq:QP}
\end{equation}
and the random phase $\phi \in [0,2\pi)$ is now replaced with a random relative initial time $\tau_0$ at which point the linear forcing is switched on. The hatted quantities are now defined as harmonic averages \citep{AR17}
\begin{equation}
    \widehat{\mathsfbi{J}}'_k:=\lim_{T\to\infty} \frac{1}{T}\int_{0}^T \mathsfbi{J}'(t;\tau_0) \mathrm{e}^{-\mathrm{i}\mathbf{b}(k)\cdot \boldsymbol{\Omega} (t+\tau_0)}\,\mathrm{d}t.
\end{equation}
In general, $\mathbf{b}$ is not odd, i.e. $\mathbf{b}(k)\neq -\mathbf{b}(-k)$ for $k\neq 0$, hence we do not have Hermitian symmetry, i.e. $\widehat{\mathsfbi{J}}'_{-k}\neq \widehat{\mathsfbi{J}}'^{*}_k$. However, since $\mathsfbi{J}'$ is real, for all $k\neq 0$ there must exist $p$ such that $\widehat{\mathsfbi{J}}'_{p}= \widehat{\mathsfbi{J}}^{'*}_k$. We now provide an example of a bijection from $\mathbb{Z}$ to $\mathbb{Z}^m$. Cantor's pairing function $\pi(\alpha_1,\alpha_2):=\frac{1}{2}(\alpha_1+\alpha_2)(1+\alpha_1+\alpha_2)+\alpha_2$ defines a well-known bijection from $\mathbb{N}^2$ to $\mathbb{N}$. It may be generalized by recurrence over $m\geq 2$, using $\pi_m(\alpha_1,\dots,\alpha_m):=\pi(\pi_{m-1}(\alpha_1,\dots,\alpha_{m-1}),\alpha_m)$ and $\pi_1:=\mathrm{Id}$; the $\pi_m$ are called Cantor tuple functions and represent bijections from $\mathbb{N}^m$ to $\mathbb{N}$. A well-known bijection from $\mathbb{Z}$ to $\mathbb{N}$ is provided by the function $f$ such that $f(\alpha)=2\alpha$ if $\alpha\geq0$ and $f(\alpha)=-2\alpha-1$ otherwise. By composing $f$ and $\pi_m$, a bijection $\mathbf{b}$ from $\mathbb{Z}$ to $\mathbb{Z}^m$ may be formed, which satisfies the condition $\mathbf{b}(0)=(0,\dots,0)$.

Using this bijection, we recover an equivalent of the harmonic transfer operator in the quasi-periodic regime. The block $\underline{\mathsfbi{H}}_{jk}$ maps the input $\mathbf{f}$ at frequency $s+\mathrm{i}\mathbf{b}(k)\cdot\boldsymbol{\Omega}$ to the output $\mathbf{u}$ at frequency $s+\mathrm{i}\mathbf{b}(j)\cdot\boldsymbol{\Omega}$. For $j,k\in \mathbb{Z}$, the general terms of the two infinite block-matrices $\underline{\mathsfbi{D}}(s)$ and $\underline{\mathsfbi{T}}(\phi)$ simply change to
\begin{align}
    \underline{\mathsfbi{D}}_{jk}(s)&:=[(s+\mathrm{i}\mathbf{b}(j)\cdot\boldsymbol{\Omega})\mathsfbi{I}-\mathsfbi{J}_{\overline{\mathbf{U}}}]\delta_{jk},\\
    \underline{\mathsfbi{T}}_{jk}(\tau_0)&:=\widehat{\mathsfbi{J}}'_{j-k}\mathrm{e}^{\mathrm{i}(\mathbf{b}(j)-\mathbf{b}(k))\cdot \boldsymbol{\Omega} \tau_0}.\label{eq:Tbis}
\end{align}
 We conclude, as in the periodic case, that the resolvent operator $\mathsfbi{R}_{\overline{\mathbf{U}}}$ about the mean Jacobian approximates the mean resolvent operator $\mathsfbi{R}_0$ and that corrective terms are of order 2 with respect to $\|\mathsfbi{J}'\|_\mathcal{W}$.

Following \cite{ME20}, an equivalent to Floquet's theorem in the quasi-periodic regime, due to \cite{SE78}, may also be formulated: under appropriate conditions, there exists a quasi-periodic transformation $\mathsfbi{P}(t;\tau_0)$ and a constant matrix $\mathsfbi{K}$ such that the system (\ref{eq:simple2}) admits a propagator of the form $\mathbf{\Phi}(t,t';\tau_0)=\mathsfbi{P}(t;\tau_0)\mathrm{e}^{\mathsfbi{K}(t-t')}\mathsfbi{P}^{-1}(t';\tau_0)$. Direct and adjoint quasi-Floquet modes $\mathsfbi{V}(t;\tau_0)$ and $\mathsfbi{W}(t;\tau_0)$ may be defined in the same way as before, but they are now quasi-periodic instead of periodic, and adjoint with respect to the inner product $(\mathbf{a},\mathbf{b}):=\lim_{T\to\infty}(1/T)\int_0^T \mathbf{a}^H\cdot \mathbf{b}\,\mathrm{d}t$. The propagator may then be written as
\begin{equation}
    \mathbf{\Phi}(t,t';\tau_0)=\mathsfbi{V}(t;\tau_0)\mathrm{e}^{\mathbf{\Lambda} (t-t')} \mathsfbi{W}^H(t';\tau_0).\label{eq:quasipropag}
\end{equation}
with
\begin{equation}
    \mathsfbi{V}(t;\tau_0)=\sum_{k} \widehat{\mathsfbi{V}}_{k}\mathrm{e}^{\mathrm{i}\mathbf{b}(k)\cdot \boldsymbol{\Omega}(t+\tau_0)},\quad     \mathsfbi{W}(t;\tau_0)=\sum_{k} \widehat{\mathsfbi{W}}_{k}\mathrm{e}^{\mathrm{i}\mathbf{b}(k)\cdot \boldsymbol{\Omega}(t+\tau_0)}\label{eq:FourQP}
\end{equation}
and $\boldsymbol{\Lambda}$ a diagonal matrix of quasi-Floquet exponents. We therefore arrive at an expression similar to (\ref{eq:LTPIO})
\begin{equation}
    \mathbf{u}(s;\phi)=\sum_n \mathrm{e}^{\mathrm{i}\mathbf{b}(n)\cdot \boldsymbol{\Omega}\tau_0}\mathsfbi{R}_n(s)\mathbf{f}(s-\mathrm{i}\mathbf{b}(n)\cdot\boldsymbol{\Omega}),
\end{equation}
with the operators
\begin{equation}
    \mathsfbi{R}_n(s)=\sum_j\sum_{k=1}^N\dfrac{ \widehat{\mathbf{v}}^k_j(\widehat{\mathbf{w}}^k_{j-n})^H }{s-(\lambda_k+\mathrm{i}\mathbf{b}(j)\cdot\boldsymbol{\Omega})}.
\end{equation}
Just as before 
\begin{equation}
    \langle \mathbf{u}(s)\rangle_{\tau_0}=\mathsfbi{R}_0(s)\mathbf{f}(s).
\end{equation}

\section{Maximum Lyapunov exponent in various configurations} \label{sec:MLE}
\begin{figure}
    \centering
    \includegraphics[width=1.0\textwidth]{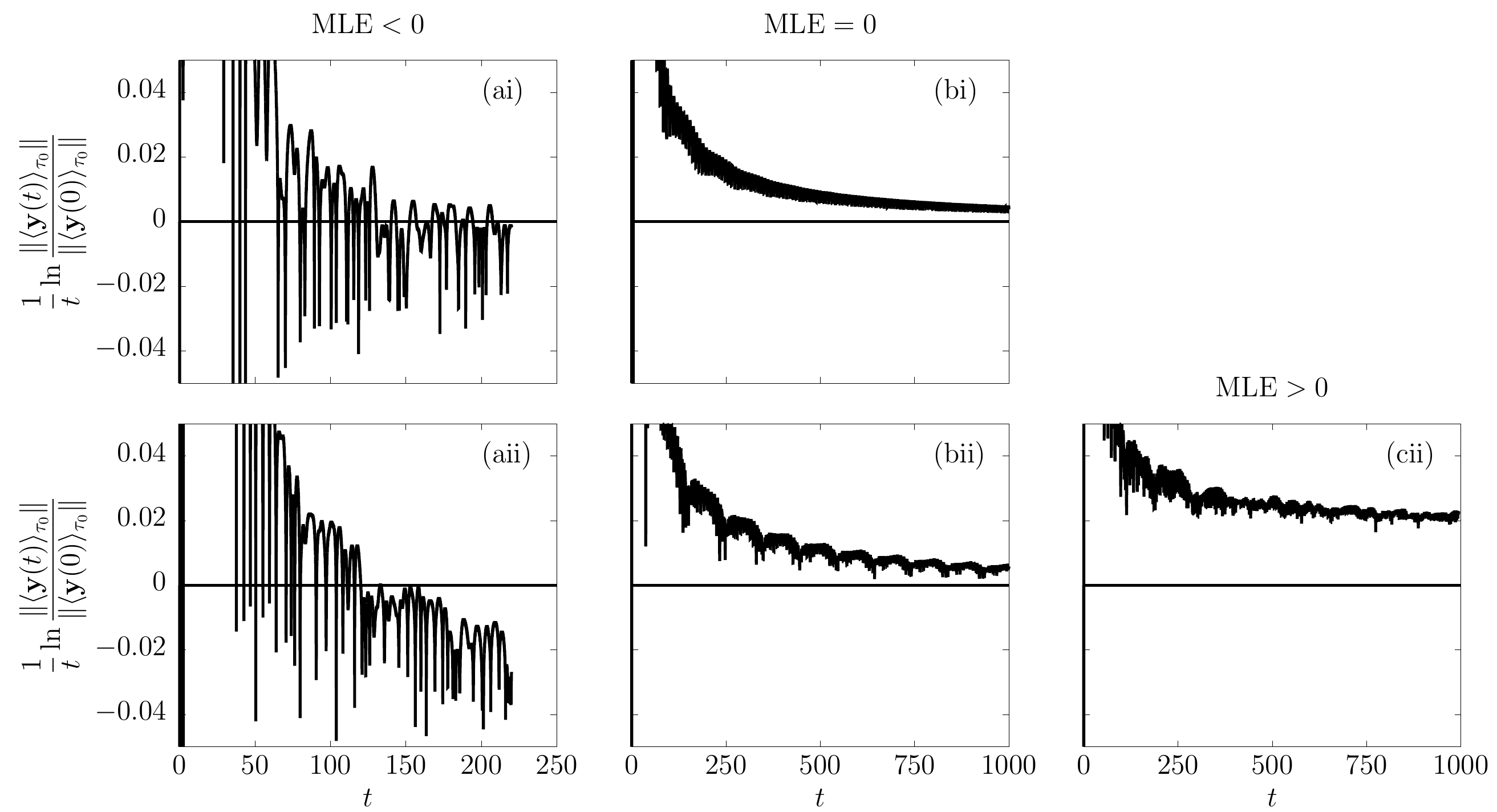}
    \caption{Estimation of maximum Lyapunov exponent (MLE) for the 5 simulations, based on the mean impulse response; (a) backward-facing step flow for (ai) $\sigma_w=\sqrt{10}$ and (aii) $\sigma_w=10$, fluidic pinball at (bi) $Re=100$, (bii) $Re=110$ and (cii) $Re=120$.}
    \label{fig:MLE}
\end{figure}

The maximum Lyapunov (\ref{eq:MLE}) is evaluated from the partial-state measurement $\mathbf{y}=[y_\mathrm{a},y_\mathrm{b},y_\mathrm{c}]^T$ as
\begin{equation}
    \mathrm{MLE} = \lim_{t\to\infty}\dfrac{1}{t}\ln \dfrac{\|\langle\mathbf{y}(t)\rangle_{\tau_0}\|}{\|\langle \mathbf{y}(0)\rangle_{\tau_0}\|}.
\end{equation}
In the periodic case, taking an ensemble average with respect to $\tau_0$ is equivalent to taking an ensemble average with respect to $\phi$. Results are shown in figure \ref{fig:MLE} for the five cases considered. Even though the timeseries are not long enough to reach full convergence of the maximum Lyapunov exponent ($\mathrm{MLE}$), they clearly demonstrate the three sought behaviours: (a) $\mathrm{MLE}<0$, (b) $\mathrm{MLE}=0$, (c) $\mathrm{MLE}>0$. The noise amplifier has negative MLE which depends on the noise intensity: the larger $\sigma_w$, the larger $|\mathrm{MLE}|$. Both the periodic and quasiperiodic regimes have null MLE. In the quasiperiodic case, the number of zero Lyapunov exponents provides the dimensionality of the torus \citep{OT15}, i.e. the number of basic incommensurate frequencies, but here we only computed the maximum one. In the chaotic case (fluidic pinball at $Re=120$), we find $\mathrm{MLE} \approx 0.02$.

\section{Invariance of the Floquet exponents to a coordinate change}\label{sec:invariance}

Consider the bijective nonlinear coordinate change $\mathbf{h}$ on the base flow
\begin{equation}
    \mathbf{U}_{\mathbf{h}}=\mathbf{h}(\mathbf{U}),
\end{equation}
then the invertible periodic matrix 
\begin{equation}
    \mathsfbi{N}(t;\phi)=\nabla_{\mathbf{U}(t;\phi)}\mathbf{h}
\end{equation}
characterizes the change of coordinate of the linear variables
\begin{equation}
    \mathbf{u}_{\mathbf{h}}=\mathsfbi{N}(t;\phi)\mathbf{u},\quad \mathbf{f}_{\mathbf{h}}=\mathsfbi{N}(t;\phi)\mathbf{f}.
\end{equation}
The propagator of the LTP system for the new input/output variables is simply given by
\begin{equation}
    {\mathbf{\Phi}_{\mathbf{h}}(t,t';\phi)=\mathsfbi{V}_{\mathbf{h}}(t;\phi)\mathrm{e}^{\mathbf{\Lambda}(t-t')}\mathsfbi{W}_{\mathbf{h}}^{H}(t';\phi)}
\end{equation}
where direct/adjoint Floquet modes change basis
\begin{equation}
    \mathsfbi{V}_\mathbf{h}(t;\phi)=\mathsfbi{N}(t;\phi)\mathsfbi{V}(t;\phi),\quad \mathsfbi{W}_\mathbf{h}(t';\phi)=\mathsfbi{N}(t';\phi)\mathsfbi{W}(t';\phi)
\end{equation}
but Floquet exponents in the fundamental strip remain characterized by the same matrix $\mathbf{\Lambda}$.

\bibliographystyle{jfm}
\bibliography{biblio}

\begin{thebibliography}{82}
\expandafter\ifx\csname natexlab\endcsname\relax\def\natexlab#1{#1}\fi
\def\au#1{#1} \def\ed#1{#1} \def\yr#1{#1}\def\at#1{#1}\def\jt#1{\textit{#1}}
  \def\bt#1{#1}\def\bvol#1{\textbf{#1}} \def\vol#1{#1} \def\pg#1{#1}
  \def\publ#1{#1}\def\arxiv#1{#1}\def\org#1{#1}\def\st#1{\textit{#1}}

\bibitem[Arbabi \& Mezi{\'c}(2017)]{AR17}
{\sc \au{Arbabi, H.} \& \au{Mezi{\'c}, I.}} \yr{2017}  \at{Study of dynamics in
  post-transient flows using {K}oopman mode decomposition}.  \jt{Phys. Rev.
  Fluids}  \bvol{2},  \pg{124402}.

\bibitem[Barkley(2006)]{BA06}
{\sc \au{Barkley, D.}} \yr{2006}  \at{Linear analysis of the cylinder wake mean
  flow}.  \jt{Europhys. Lett.}  \bvol{75},  \pg{750}.

\bibitem[Beneddine {\em et~al.\/}(2016)Beneddine, Sipp, Arnault, Dandois \&
  Lesshafft]{BE16}
{\sc \au{Beneddine, S.}, \au{Sipp, D.}, \au{Arnault, A.}, \au{Dandois, J.} \&
  \au{Lesshafft, L.}} \yr{2016}  \at{Conditions for validity of mean flow
  stability analysis}.  \jt{J. Fluid Mech.}  \bvol{798},  \pg{485--504}.

\bibitem[Beneddine {\em et~al.\/}(2017)Beneddine, Yegavian, Sipp \&
  Leclaire]{BE17}
{\sc \au{Beneddine, S.}, \au{Yegavian, R.}, \au{Sipp, D.} \& \au{Leclaire, B.}}
  \yr{2017}  \at{Unsteady flow dynamics reconstruction from mean flow and point
  sensors: an experimental study}.  \jt{J. Fluid Mech.}  \bvol{824},
  \pg{174--201}.

\bibitem[Bengana {\em et~al.\/}(2019)Bengana, Loiseau, Robinet \&
  Tuckerman]{BE19b}
{\sc \au{Bengana, Y.}, \au{Loiseau, J.-Ch.}, \au{Robinet, J.-Ch.} \&
  \au{Tuckerman, L.~S.}} \yr{2019}  \at{Bifurcation analysis and frequency
  prediction in shear-driven cavity flow}.  \jt{J. Fluid Mech.}  \bvol{875},
  \pg{725--757}.

\bibitem[Bengana \& Tuckerman(2019)]{BE19}
{\sc \au{Bengana, Y.} \& \au{Tuckerman, L.~S.}} \yr{2019}  \at{Spirals and
  ribbons in counter-rotating {T}aylor--{C}ouette flow: Frequencies from mean
  flows and heteroclinic orbits}.  \jt{Phys. Rev. Fluids}  \bvol{4},
  \pg{044402}.

\bibitem[Bengana \& Tuckerman(2021)]{BE21}
{\sc \au{Bengana, Y.} \& \au{Tuckerman, L.~S}} \yr{2021}  \at{Frequency
  prediction from exact or self-consistent mean flows}.  \jt{Phys. Rev. Fluids}
   \bvol{6},  \pg{063901}.

\bibitem[Butler \& Farrell(1993)]{BU93}
{\sc \au{Butler, K.~M.} \& \au{Farrell, B.~F.}} \yr{1993}  \at{Optimal
  perturbations and streak spacing in wall-bounded turbulent shear flow}.
  \jt{Phys. Fluids A: Fluid Dynamics}  \bvol{5},  \pg{774--777}.

\bibitem[Conway(1990)]{CO90}
{\sc \au{Conway, J.~B.}} \yr{1990} {\em A course in functional analysis,
  2\textsuperscript{nd} edition\/}, ,  \vol{vol.~96}.  \publ{Springer}.

\bibitem[Cossu {\em et~al.\/}(2009)Cossu, Pujals \& Depardon]{CO09}
{\sc \au{Cossu, C.}, \au{Pujals, G.} \& \au{Depardon, S.}} \yr{2009}
  \at{Optimal transient growth and very large--scale structures in turbulent
  boundary layers}.  \jt{J. Fluid Mech.}  \bvol{619},  \pg{79--94}.

\bibitem[{\v{C}}rnjari{\'c}-{\v{Z}}ic {\em
  et~al.\/}(2019){\v{C}}rnjari{\'c}-{\v{Z}}ic, Ma{\'c}e{\v{s}}i{\'c} \&
  Mezi{\'c}]{VC19}
{\sc \au{{\v{C}}rnjari{\'c}-{\v{Z}}ic, N.}, \au{Ma{\'c}e{\v{s}}i{\'c}, S.} \&
  \au{Mezi{\'c}, I.}} \yr{2019}  \at{Koopman operator spectrum for random
  dynamical systems}.  \jt{Journal of Nonlinear Science}  \pg{pp. 1--50}.

\bibitem[Dahan {\em et~al.\/}(2012)Dahan, Morgans \& Lardeau]{DA12}
{\sc \au{Dahan, J.~A.}, \au{Morgans, A.~S.} \& \au{Lardeau, S.}} \yr{2012}
  \at{Feedback control for form-drag reduction on a bluff body with a blunt
  trailing edge}.  \jt{J. Fluid Mech.}  \bvol{704},  \pg{360--387}.

\bibitem[Dalla~Longa {\em et~al.\/}(2017)Dalla~Longa, Morgans \& Dahan]{DA17}
{\sc \au{Dalla~Longa, L.}, \au{Morgans, A.~S.} \& \au{Dahan, J.~A.}} \yr{2017}
  \at{Reducing the pressure drag of a d-shaped bluff body using linear feedback
  control}.  \jt{Theo. Comput. Fluid Dyn.}  \pg{pp. 1--11}.

\bibitem[Del~{\'A}lamo \& Jimenez(2006)]{DE06}
{\sc \au{Del~{\'A}lamo, J.~C.} \& \au{Jimenez, J.}} \yr{2006}  \at{Linear
  energy amplification in turbulent channels}.  \jt{J. Fluid Mech.}
  \bvol{559},  \pg{205--213}.

\bibitem[Deng {\em et~al.\/}(2020)Deng, Noack, Morzy{\'n}ski \& Pastur]{DE20}
{\sc \au{Deng, N.}, \au{Noack, B.~R.}, \au{Morzy{\'n}ski, M.} \& \au{Pastur,
  L.~R.}} \yr{2020}  \at{Low-order model for successive bifurcations of the
  fluidic pinball}.  \jt{J. Fluid Mech.}  \bvol{884}.

\bibitem[Evstafyeva {\em et~al.\/}(2017)Evstafyeva, Morgans \&
  Dalla~Longa]{EV17}
{\sc \au{Evstafyeva, O.}, \au{Morgans, A.~S.} \& \au{Dalla~Longa, L.}}
  \yr{2017}  \at{Simulation and feedback control of the ahmed body flow
  exhibiting symmetry breaking behaviour}.  \jt{J. Fluid Mech.}  \bvol{817}.

\bibitem[Franceschini {\em et~al.\/}(2022)Franceschini, Sipp, Marquet, Moulin
  \& Dandois]{franceschini2022identification}
{\sc \au{Franceschini, L.}, \au{Sipp, D.}, \au{Marquet, O.}, \au{Moulin, J.} \&
  \au{Dandois, J.}} \yr{2022}  \at{Identification and reconstruction of
  high-frequency fluctuations evolving on a low-frequency periodic limit cycle:
  application to turbulent cylinder flow}.  \jt{J. Fluid Mech.}  \bvol{942}.

\bibitem[Garnaud {\em et~al.\/}(2013)Garnaud, Lesshafft, Schmid \&
  Huerre]{GA13b}
{\sc \au{Garnaud, X.}, \au{Lesshafft, L.}, \au{Schmid, P.~J.} \& \au{Huerre,
  P.}} \yr{2013}  \at{The preferred mode of incompressible jets: linear
  frequency response analysis}.  \jt{J. Fluid Mech.}  \bvol{716},
  \pg{189--202}.

\bibitem[Gelb \& Vander~Velde(1968)]{GE68}
{\sc \au{Gelb, A.} \& \au{Vander~Velde, W.~E.}} \yr{1968} {\em Multiple-input
  describing functions and nonlinear system design\/}.  \publ{McGraw-Hill}.

\bibitem[Gohberg {\em et~al.\/}(2003)Gohberg, Kaashoek \& Spitkovsky]{GO03}
{\sc \au{Gohberg, I.}, \au{Kaashoek, M.s~A} \& \au{Spitkovsky, I.~M.}}
  \yr{2003}  \at{An overview of matrix factorization theory and operator
  applications}.  \jt{Factorization and integrable systems}  \pg{pp. 1--102}.

\bibitem[G{\'o}mez {\em et~al.\/}(2016)G{\'o}mez, Blackburn, Rudman, Sharma \&
  McKeon]{GO16}
{\sc \au{G{\'o}mez, F.}, \au{Blackburn, H.~M.}, \au{Rudman, M.}, \au{Sharma,
  A.~S.} \& \au{McKeon, B.~J.}} \yr{2016}  \at{A reduced-order model of
  three-dimensional unsteady flow in a cavity based on the resolvent operator}.
   \jt{J. Fluid Mech.}  \bvol{798},  \pg{R2}.

\bibitem[Hammond \& Redekopp(1997)]{HA97}
{\sc \au{Hammond, D.~A.} \& \au{Redekopp, L.~G.}} \yr{1997}  \at{Global
  dynamics of symmetric and asymmetric wakes}.  \jt{J. Fluid Mech.}
  \bvol{331},  \pg{231--260}.

\bibitem[Hecht(2012)]{MR3043640}
{\sc \au{Hecht, F.}} \yr{2012}  \at{New development in {F}ree{F}em++}.  \jt{J.
  Numer. Math.}  \bvol{20}~(3-4),  \pg{251--265}.

\bibitem[Herrmann {\em et~al.\/}(2021)Herrmann, Baddoo, Semaan, Brunton \&
  McKeon]{HE21}
{\sc \au{Herrmann, B.}, \au{Baddoo, P.~J.}, \au{Semaan, R.}, \au{Brunton,
  S.~L.} \& \au{McKeon, B.~J.}} \yr{2021}  \at{Data-driven resolvent analysis}.
   \jt{J. Fluid Mechanics}  \bvol{918},  \pg{A10}.

\bibitem[Herv{\'e} {\em et~al.\/}(2012)Herv{\'e}, Sipp, Schmid \&
  Samuelides]{HE12}
{\sc \au{Herv{\'e}, A.}, \au{Sipp, D.}, \au{Schmid, P.~J.} \& \au{Samuelides,
  M.}} \yr{2012}  \at{A physics-based approach to flow control using system
  identification}.  \jt{J. Fluid Mech.}  \bvol{702},  \pg{26--58}.

\bibitem[Hwang \& Cossu(2010)]{HW10}
{\sc \au{Hwang, Y.} \& \au{Cossu, C.}} \yr{2010}  \at{Linear non-normal energy
  amplification of harmonic and stochastic forcing in the turbulent channel
  flow}.  \jt{J. Fluid Mech.}  \bvol{664},  \pg{51--73}.

\bibitem[Illingworth {\em et~al.\/}(2018)Illingworth, Monty \& Marusic]{IL18}
{\sc \au{Illingworth, S.~J.}, \au{Monty, J.~P.} \& \au{Marusic, I.}} \yr{2018}
  \at{Estimating large-scale structures in wall turbulence using linear
  models}.  \jt{J. Fluid Mech.}  \bvol{842},  \pg{146--162}.

\bibitem[Iollo {\em et~al.\/}(2000)Iollo, Lanteri \& D{\'e}sid{\'e}ri]{IO00}
{\sc \au{Iollo, A.}, \au{Lanteri, S.} \& \au{D{\'e}sid{\'e}ri, J.-A.}}
  \yr{2000}  \at{Stability properties of pod--galerkin approximations for the
  compressible navier--stokes equations}.  \jt{Theo. Comput. Fluid Dyn.}
  \bvol{13}~(6),  \pg{377--396}.

\bibitem[Jeun {\em et~al.\/}(2016)Jeun, Nichols \& Jovanovi{\'c}]{JE16}
{\sc \au{Jeun, J.}, \au{Nichols, J.~W.} \& \au{Jovanovi{\'c}, M.~R.}} \yr{2016}
   \at{Input-output analysis of high-speed axisymmetric isothermal jet noise}.
  \jt{Phys. Fluids}  \bvol{28}~(4),  \pg{047101}.

\bibitem[Jovanovic \& Bamieh(2001)]{JO01}
{\sc \au{Jovanovic, M.} \& \au{Bamieh, B.}} \yr{2001} Modeling flow statistics
  using the linearized {N}avier--{S}tokes equations.  \bt{In {\em Proceedings
  of the 40th IEEE Conference on Decision and Control\/}}, ,  \vol{vol.~5},
  \pg{pp. 4944--4949}. IEEE.

\bibitem[Jovanovic \& Georgiou(2010)]{JO10}
{\sc \au{Jovanovic, M.} \& \au{Georgiou, T.}} \yr{2010} Reproducing second
  order statistics of turbulent flows using linearized {N}avier--{S}tokes
  equations with forcing.  \bt{In {\em 63rd Annual Meeting of the APS Division
  of Fluid Dynamics\/}}.

\bibitem[Jovanovi{\'c}(2021)]{JO21}
{\sc \au{Jovanovi{\'c}, M.~R.}} \yr{2021}  \at{From bypass transition to flow
  control and data-driven turbulence modeling: an input--output viewpoint}.
  \jt{Annu. Rev. Fluid Mech.}  \bvol{53},  \pg{311--345}.

\bibitem[Karban {\em et~al.\/}(2020)Karban, Bugeat, Martini, Towne, Cavalieri,
  Lesshafft, Agarwal, Jordan \& Colonius]{KA20}
{\sc \au{Karban, U.}, \au{Bugeat, B.}, \au{Martini, E.}, \au{Towne, A.},
  \au{Cavalieri, A.V.G.}, \au{Lesshafft, L.}, \au{Agarwal, A.}, \au{Jordan, P.}
  \& \au{Colonius, T.}} \yr{2020}  \at{Ambiguity in mean-flow-based linear
  analysis}.  \jt{J. Fluid Mech.}  \bvol{900},  \pg{R5}.

\bibitem[Khalil(2002)]{KH02}
{\sc \au{Khalil, H.~K.}} \yr{2002} {\em {Nonlinear systems; 3rd ed.}\/}.
  \publ{Prentice-Hall}.

\bibitem[Krishna~Kumar \& Kulkarni(2015)]{KU15}
{\sc \au{Krishna~Kumar, G.} \& \au{Kulkarni, S.~H.}} \at{ \yr{2015} } \jt{Ann.
  Funct. Anal.}  \bvol{6},  \pg{148--169}.

\bibitem[Leclercq {\em et~al.\/}(2019)Leclercq, Demourant, Poussot-Vassal \&
  Sipp]{LE19}
{\sc \au{Leclercq, C.}, \au{Demourant, F.}, \au{Poussot-Vassal, C.} \&
  \au{Sipp, D.}} \yr{2019}  \at{Linear iterative method for closed-loop control
  of quasiperiodic flows}.  \jt{J. Fluid Mech.}  \bvol{868},  \pg{26--65}.

\bibitem[Lee {\em et~al.\/}(1990)Lee, Kim \& Moin]{LE90}
{\sc \au{Lee, M.~J.}, \au{Kim, J.} \& \au{Moin, P.}} \yr{1990}  \at{Structure
  of turbulence at high shear rate}.  \jt{J. Fluid Mech.}  \bvol{216},
  \pg{561--583}.

\bibitem[Liu {\em et~al.\/}(2021)Liu, Sun, Yeh, Ukeiley, Cattafesta \&
  Taira]{LI21}
{\sc \au{Liu, Q.}, \au{Sun, Y.}, \au{Yeh, C.-A.}, \au{Ukeiley, L.~S.},
  \au{Cattafesta, L.~N.} \& \au{Taira, K.}} \yr{2021}  \at{Unsteady control of
  supersonic turbulent cavity flow based on resolvent analysis}.  \jt{J. Fluid
  Mech.}  \bvol{925},  \pg{A5}.

\bibitem[Luhar {\em et~al.\/}(2014)Luhar, Sharma \& McKeon]{LU14}
{\sc \au{Luhar, M.}, \au{Sharma, A.~S.} \& \au{McKeon, B.~J}} \yr{2014}
  \at{Opposition control within the resolvent analysis framework}.  \jt{J.
  Fluid Mech.}  \bvol{749},  \pg{597--626}.

\bibitem[Madhusudanan {\em et~al.\/}(2019)Madhusudanan, Illingworth \&
  Marusic]{MA19}
{\sc \au{Madhusudanan, A.}, \au{Illingworth, S.~J.} \& \au{Marusic, I.}}
  \yr{2019}  \at{Coherent large-scale structures from the linearized
  navier--stokes equations}.  \jt{J. Fluid Mech.}  \bvol{873},  \pg{89--109}.

\bibitem[Magruder {\em et~al.\/}(2018)Magruder, Gugercin \& Beattie]{MA18}
{\sc \au{Magruder, C.~C.}, \au{Gugercin, S.} \& \au{Beattie, C.~A}} \yr{2018}
  \at{Linear time-periodic dynamical systems: an ${H}_2$ analysis and a model
  reduction framework}.  \jt{Mathematical and Computer Modelling of Dynamical
  Systems}  \bvol{24},  \pg{119--142}.

\bibitem[Malkus(1956)]{MA56}
{\sc \au{Malkus, W.V.R.}} \yr{1956}  \at{Outline of a theory of turbulent shear
  flow}.  \jt{J. Fluid Mech.}  \bvol{1},  \pg{521--539}.

\bibitem[Manti{\v{c}}-Lugo {\em et~al.\/}(2014)Manti{\v{c}}-Lugo, Arratia \&
  Gallaire]{MA14}
{\sc \au{Manti{\v{c}}-Lugo, V.}, \au{Arratia, C.} \& \au{Gallaire, F.}}
  \yr{2014}  \at{Self-consistent mean flow description of the nonlinear
  saturation of the vortex shedding in the cylinder wake}.  \jt{Phys. Rev.
  Lett.}  \bvol{113},  \pg{084501}.

\bibitem[Manti{\v{c}}-Lugo {\em et~al.\/}(2015)Manti{\v{c}}-Lugo, Arratia \&
  Gallaire]{MA15}
{\sc \au{Manti{\v{c}}-Lugo, V.}, \au{Arratia, C.} \& \au{Gallaire, F.}}
  \yr{2015}  \at{A self-consistent model for the saturation dynamics of the
  vortex shedding around the mean flow in the unstable cylinder wake}.
  \jt{Phys. Fluids}  \bvol{27},  \pg{074103}.

\bibitem[McKeon \& Sharma(2010)]{MC10}
{\sc \au{McKeon, B.~J.} \& \au{Sharma, A.~S.}} \yr{2010}  \at{A critical-layer
  framework for turbulent pipe flow}.  \jt{J. Fluid Mech.}  \bvol{658},
  \pg{336--382}.

\bibitem[Meliga(2017)]{ME17}
{\sc \au{Meliga, P.}} \yr{2017}  \at{Harmonics generation and the mechanics of
  saturation in flow over an open cavity: a second-order self-consistent
  description}.  \jt{J. Fluid Mech.}  \bvol{826},  \pg{503--521}.

\bibitem[Mezi{\'c}(2013)]{ME13}
{\sc \au{Mezi{\'c}, I.}} \yr{2013}  \at{Analysis of fluid flows via spectral
  properties of the {K}oopman operator}.  \jt{Annu. Rev. Fluid Mech.}
  \bvol{45},  \pg{357--378}.

\bibitem[Mezi{\'c}(2020)]{ME20}
{\sc \au{Mezi{\'c}, I.}} \yr{2020}  \at{Spectrum of the {K}oopman operator,
  spectral expansions in functional spaces, and state-space geometry}.  \jt{J.
  Nonlin. Sci.}  \bvol{30},  \pg{2091--2145}.

\bibitem[Mezi\'c \& Surana(2016)]{ME16}
{\sc \au{Mezi\'c, I.} \& \au{Surana, A.}} \yr{2016}  \at{Koopman mode
  decomposition for periodic/quasi-periodic time dependence}.
  \jt{IFAC-PapersOnLine}  \bvol{49}~(18),  \pg{690--697}.

\bibitem[Mittal(2008)]{MI08}
{\sc \au{Mittal, S.}} \yr{2008}  \at{Global linear stability analysis of
  time-averaged flows}.  \jt{Int. J. Num. Meth. Fluids}  \bvol{58},
  \pg{111--118}.

\bibitem[Moarref \& Jovanovi{\'c}(2012)]{MO12}
{\sc \au{Moarref, R.} \& \au{Jovanovi{\'c}, M.~R.}} \yr{2012}  \at{Model-based
  design of transverse wall oscillations for turbulent drag reduction}.  \jt{J.
  Fluid Mech.}  \bvol{707},  \pg{205--240}.

\bibitem[Moarref {\em et~al.\/}(2014)Moarref, Jovanovi{\'c}, Tropp, Sharma \&
  McKeon]{MO14}
{\sc \au{Moarref, R.}, \au{Jovanovi{\'c}, M.~R.}, \au{Tropp, J.~A.},
  \au{Sharma, A.~S.} \& \au{McKeon, B.~J.}} \yr{2014}  \at{A low-order
  decomposition of turbulent channel flow via resolvent analysis and convex
  optimization}.  \jt{Phys. Fluids}  \bvol{26}~(5),  \pg{051701}.

\bibitem[Morra {\em et~al.\/}(2019)Morra, Semeraro, Henningson \& Cossu]{MO19}
{\sc \au{Morra, P.}, \au{Semeraro, O.}, \au{Henningson, D.~S.} \& \au{Cossu,
  C.}} \yr{2019}  \at{On the relevance of reynolds stresses in resolvent
  analyses of turbulent wall-bounded flows}.  \jt{J. Fluid Mech.}  \bvol{867},
  \pg{969--984}.

\bibitem[Noack {\em et~al.\/}(2003)Noack, Afanasiev, Morzy{\'n}ski, Tadmor \&
  Thiele]{NO03}
{\sc \au{Noack, B.~R.}, \au{Afanasiev, K.}, \au{Morzy{\'n}ski, M.}, \au{Tadmor,
  G.} \& \au{Thiele, F.}} \yr{2003}  \at{A hierarchy of low-dimensional models
  for the transient and post-transient cylinder wake}.  \jt{J. Fluid Mech.}
  \bvol{497},  \pg{335--363}.

\bibitem[Noiray {\em et~al.\/}(2008)Noiray, Durox, Schuller \& Candel]{NO08}
{\sc \au{Noiray, N.}, \au{Durox, D.}, \au{Schuller, T.} \& \au{Candel, S.}}
  \yr{2008}  \at{A unified framework for nonlinear combustion instability
  analysis based on the flame describing function}.  \jt{J. Fluid Mech.}
  \bvol{615},  \pg{139--167}.

\bibitem[Oteski {\em et~al.\/}(2015)Oteski, Duguet, Pastur \&
  Le~Qu{\'e}r{\'e}]{OT15}
{\sc \au{Oteski, L.}, \au{Duguet, Y.}, \au{Pastur, L.} \& \au{Le~Qu{\'e}r{\'e},
  P.}} \yr{2015}  \at{Quasiperiodic routes to chaos in confined two-dimensional
  differential convection}.  \jt{Phys. Rev. E}  \bvol{92},  \pg{043020}.

\bibitem[Padovan {\em et~al.\/}(2020)Padovan, Otto \& Rowley]{PA20}
{\sc \au{Padovan, A.}, \au{Otto, S.~E.} \& \au{Rowley, C.~W.}} \yr{2020}
  \at{Analysis of amplification mechanisms and cross-frequency interactions in
  nonlinear flows via the harmonic resolvent}.  \jt{J. Fluid Mech.}
  \bvol{900},  \pg{A14}.

\bibitem[Pickering {\em et~al.\/}(2021)Pickering, Rigas, Schmidt, Sipp \&
  Colonius]{PI21}
{\sc \au{Pickering, E.}, \au{Rigas, G.}, \au{Schmidt, O.~T.}, \au{Sipp, D.} \&
  \au{Colonius, T.}} \yr{2021}  \at{Optimal eddy viscosity for resolvent-based
  models of coherent structures in turbulent jets}.  \jt{J. Fluid Mech.}
  \bvol{917}.

\bibitem[Pier(2002)]{PI02}
{\sc \au{Pier, B.}} \yr{2002}  \at{On the frequency selection of
  finite-amplitude vortex shedding in the cylinder wake}.  \jt{J. Fluid Mech.}
  \bvol{458},  \pg{407--417}.

\bibitem[Pujals {\em et~al.\/}(2009)Pujals, Garc{\'\i}a-Villalba, Cossu \&
  Depardon]{PU09}
{\sc \au{Pujals, G.}, \au{Garc{\'\i}a-Villalba, M.}, \au{Cossu, C.} \&
  \au{Depardon, S.}} \yr{2009}  \at{A note on optimal transient growth in
  turbulent channel flows}.  \jt{Phys. Fluids}  \bvol{21},  \pg{015109}.

\bibitem[Sartor {\em et~al.\/}(2015)Sartor, Mettot \& Sipp]{SA15b}
{\sc \au{Sartor, F.}, \au{Mettot, C.} \& \au{Sipp, D.}} \yr{2015}
  \at{Stability, receptivity, and sensitivity analyses of buffeting transonic
  flow over a profile}.  \jt{AIAA Journal}  \bvol{53},  \pg{1980--1993}.

\bibitem[Schmid(2010)]{SC10}
{\sc \au{Schmid, P.~J.}} \yr{2010}  \at{Dynamic mode decomposition of numerical
  and experimental data}.  \jt{J. Fluid Mech.}  \bvol{656},  \pg{5--28}.

\bibitem[Schmidt {\em et~al.\/}(2018)Schmidt, Towne, Rigas, Colonius \&
  Br{\`e}s]{SC18}
{\sc \au{Schmidt, O.~T.}, \au{Towne, A.}, \au{Rigas, G.}, \au{Colonius, T.} \&
  \au{Br{\`e}s, G.~A.}} \yr{2018}  \at{Spectral analysis of jet turbulence}.
  \jt{J. Fluid Mech.}  \bvol{855},  \pg{953--982}.

\bibitem[Sell(1978)]{SE78}
{\sc \au{Sell, G.~R.}} \yr{1978}  \at{The structure of a flow in the vicinity
  of an almost periodic motion}.  \jt{Journal of Differential Equations}
  \bvol{27}~(3),  \pg{359--393}.

\bibitem[Semeraro {\em et~al.\/}(2016{\natexlab{{\em a\/}}})Semeraro, Jaunet,
  Jordan, Cavalieri \& Lesshafft]{SE16a}
{\sc \au{Semeraro, O.}, \au{Jaunet, V.}, \au{Jordan, P.}, \au{Cavalieri, A.~V.}
  \& \au{Lesshafft, L.}} \yr{2016{\natexlab{{\em a\/}}}}  \bt{Stochastic and
  harmonic optimal forcing in subsonic jets}.  \pg{p. 2935}.

\bibitem[Semeraro {\em et~al.\/}(2016{\natexlab{{\em b\/}}})Semeraro,
  Lesshafft, Jaunet \& Jordan]{SE16b}
{\sc \au{Semeraro, O.}, \au{Lesshafft, L.}, \au{Jaunet, V.} \& \au{Jordan, P.}}
  \yr{2016{\natexlab{{\em b\/}}}}  \at{Modeling of coherent structures in a
  turbulent jet as global linear instability wavepackets: Theory and
  experiment}.  \jt{Int. J. Heat Fluid Flow}  \bvol{62},  \pg{24--32}.

\bibitem[Sipp \& Lebedev(2007)]{SI07}
{\sc \au{Sipp, D.} \& \au{Lebedev, A.}} \yr{2007}  \at{Global stability of base
  and mean flows: a general approach and its applications to cylinder and open
  cavity flows}.  \jt{J. Fluid Mech.}  \bvol{593},  \pg{333--358}.

\bibitem[Sipp \& Schmid(2016)]{SI16}
{\sc \au{Sipp, D.} \& \au{Schmid, P.~J.}} \yr{2016}  \at{Linear closed-loop
  control of fluid instabilities and noise-induced perturbations: A review of
  approaches and tools}.  \jt{Appl. Mech. Rev.}  \bvol{68},  \pg{020801}.

\bibitem[Suzuki(1976)]{SU76}
{\sc \au{Suzuki, N.}} \yr{1976}  \at{On the convergence of {N}eumann series in
  {B}anach space}.  \jt{Mathematische Annalen}  \bvol{220},  \pg{143--146}.

\bibitem[Symon {\em et~al.\/}(2019)Symon, Sipp \& McKeon]{SY19}
{\sc \au{Symon, S.}, \au{Sipp, D.} \& \au{McKeon, B.~J.}} \yr{2019}  \at{A tale
  of two airfoils: resolvent-based modelling of an oscillator versus an
  amplifier from an experimental mean}.  \jt{J. Fluid Mech.}  \bvol{881},
  \pg{51--83}.

\bibitem[Toedtli {\em et~al.\/}(2019)Toedtli, Luhar \& McKeon]{TO19}
{\sc \au{Toedtli, S.~S.}, \au{Luhar, M.} \& \au{McKeon, B.~J.}} \yr{2019}
  \at{Predicting the response of turbulent channel flow to varying-phase
  opposition control: resolvent analysis as a tool for flow control design}.
  \jt{Phys. Rev. Fluids}  \bvol{4}~(7),  \pg{073905}.

\bibitem[Towne {\em et~al.\/}(2020)Towne, Lozano-Dur{\'a}n \& Yang]{TO20}
{\sc \au{Towne, A.}, \au{Lozano-Dur{\'a}n, A.} \& \au{Yang, X.}} \yr{2020}
  \at{Resolvent-based estimation of space--time flow statistics}.  \jt{J. Fluid
  Mech.}  \bvol{883},  \pg{A17}.

\bibitem[Towne {\em et~al.\/}(2018)Towne, Schmidt \& Colonius]{TO18}
{\sc \au{Towne, A.}, \au{Schmidt, O.~T.} \& \au{Colonius, T.}} \yr{2018}
  \at{Spectral proper orthogonal decomposition and its relationship to dynamic
  mode decomposition and resolvent analysis}.  \jt{J. Fluid Mech.}  \bvol{847},
   \pg{821--867}.

\bibitem[Turton {\em et~al.\/}(2015)Turton, Tuckerman \& Barkley]{TU15}
{\sc \au{Turton, S.~E.}, \au{Tuckerman, L.~S.} \& \au{Barkley, D.}} \yr{2015}
  \at{Prediction of frequencies in thermosolutal convection from mean flows}.
  \jt{Phys. Rev. E}  \bvol{91},  \pg{043009}.

\bibitem[Vigo(1998)]{VI98}
{\sc \au{Vigo, G.}} \yr{1998}  \at{The proper orthogonal decomposition applied
  to unsteady compressible {N}avier--{S}tokes equation}.  \jt{INRIA Rapport de
  Recherche}  \bvol{3385}.

\bibitem[Wereley \& Hall(1990)]{WE90}
{\sc \au{Wereley, N.~M.} \& \au{Hall, S.~R.}} \yr{1990} Frequency response of
  linear time periodic systems.  \bt{In {\em 29th IEEE conference on decision
  and control\/}},  \pg{pp. 3650--3655}. IEEE.

\bibitem[Wereley \& Hall(1991)]{WE91}
{\sc \au{Wereley, N.~M.} \& \au{Hall, S.~R.}} \yr{1991} Linear time periodic
  systems: transfer function, poles, transmission zeroes and directional
  properties.  \bt{In {\em 1991 American Control Conference\/}},  \pg{pp.
  1179--1184}. IEEE.

\bibitem[Williams {\em et~al.\/}(2015)Williams, Kevrekidis \& Rowley]{WI15}
{\sc \au{Williams, M.~O.}, \au{Kevrekidis, I.~G.} \& \au{Rowley, C.~W.}}
  \yr{2015}  \at{A data--driven approximation of the {K}oopman operator:
  Extending dynamic mode decomposition}.  \jt{J. Nonlinear Science}  \bvol{25},
   \pg{1307--1346}.

\bibitem[Yeh \& Taira(2019)]{YE19}
{\sc \au{Yeh, C.-A.} \& \au{Taira, K.}} \yr{2019}  \at{Resolvent-analysis-based
  design of airfoil separation control}.  \jt{J. Fluid Mech.}  \bvol{867},
  \pg{572--610}.

\bibitem[Zare {\em et~al.\/}(2017)Zare, Jovanovi{\'c} \& Georgiou]{ZA17}
{\sc \au{Zare, A.}, \au{Jovanovi{\'c}, M.~R.} \& \au{Georgiou, T.~T.}}
  \yr{2017}  \at{Colour of turbulence}.  \jt{J. Fluid Mech.}  \bvol{812},
  \pg{636--680}.

\bibitem[Zhou(2008)]{ZH08}
{\sc \au{Zhou, J.}} \yr{2008}  \at{Zeros and poles of linear continuous-time
  periodic systems: Definitions and properties}.  \jt{IEEE transactions on
  automatic control}  \bvol{53}~(9),  \pg{1998--2011}.

\bibitem[Zhou \& Hagiwara(2002)]{ZH02}
{\sc \au{Zhou, J.} \& \au{Hagiwara, T.}} \yr{2002}  \at{{$H_2$} and
  {$H_\infty$} norm computations of linear continuous-time periodic systems via
  the skew analysis of frequency response operators}.  \jt{Automatica}
  \bvol{38},  \pg{1381--1387}.

\end{thebibliography}

\end{document}